\documentstyle[11pt,epsf,fleqn]{article}
\begin{document}
\title{Radiative CP violation in the Higgs sector of the Next-to-minimal supersymmetric model}
\author{S.W. Ham$^{(1)}$, Y.S. Jeong$^{(2)}$, S.K. Oh$^{(1,2)}$
\\
\\
{\it $^{\rm (1)}$ Center for High Energy Physics, Kyungpook National University}\\
{\it Daegu 702-701, Korea} \\
{\it $^{\rm (2)}$ Department of Physics, Konkuk University, Seoul 143-701, Korea}
\\
\\
}
\date{}
\maketitle
\begin{abstract}
We investigate the neutral Higgs sector in the next-to-minimal supersymmetric standard model (NMSSM)
with explicit CP violation at the one-loop level by using the effective potential method.
In general, explicit CP violation is possible at the tree-level in the Higgs potential of the NMSSM,
which may possess a complex phase.
The tree-level Higgs potential can be made CP conserving by assuming that all the relevant parameters are real.
However, the CP-conserving Higgs potential at the tree level may still develop complex phases
at the one-loop level through radiative corrections.
These complex phases exhibit explicit CP violation through the mixings between the scalar and pseudoscalar Higgs bosons.
One of the complex phases arise in the neutralino sector.
For a particular choice of the relevant parameter values, the scalar-pseudoscalar mixings are found
to strongly dependent on this phase.
Meanwhile, for a wide region in the parameter space, the ratios of CP mixing elements of
the neutral Higgs boson mass matrix to CP non-mixing ones increase weakly against this phase.
Also, the neutral Higgs boson masses are roughly stable against the variation of this phase.
\end{abstract}
\vfil
\eject

\section{Introduction}

Supersymmetry (SUSY) must be broken in order to be realized in nature.
In various supersymmetric models, the SUSY breaking may be accomplished by introducing soft SUSY breaking terms.
The introduction of soft SUSY breaking terms may also invoke irremovable complex phases.
These complex phases may contribute to the mixing between the scalar and pseudoscalar Higgs bosons, thus account for explicit CP violation.
In the minimal supersymmetric standard model (MSSM),
there can be as many as eleven non-trivial phases associated with the scalar fermions, the charginos, and the neutralinos.
However, at the tree level, the MSSM can avoid explicit CP violation
by adjusting the complex phases away via redefinition of various participating fields in the model.
It has been found that, at the one-loop level the MSSM can accommodate explicit CP violation in its neutral Higgs sector [1].
Many authors have investigated the explicit CP violation in the MSSM in the radiatively corrected Higgs potential [1-4],
where five complex phases (three from the scalar fermion sector of the third generation,
one from the charginos, and one from the neutralinos) can participate in the scalar-pseudoscalar mixing.

Unlike in the case of the MSSM, explicit CP violation can be realized
in the next-to-minimal supersymmetric standard model (NMSSM) even at the tree level.
At the tree level in the Higgs potential of the NMSSM, a nontrivial complex phase emerges after
redefining the three Higgs fields [5].
This complex phase from the tree-level Higgs potential may persist in the one-loop effective potential if the
scalar top quarks are degenerate in mass.
In Ref.[5], assuming the degeneracy of the scalar top quark masses,
the scalar pseudoscalar mixing is investigated in the NMSSM with explicit CP violation.
It has been shown that large mixings between the scalar and pseudoscalar Higgs bosons may be realized as
the vacuum expectation value (VEV) of the neutral Higgs singlet approaches to the electroweak scale [5].

We have elsewhere investigated the implications of  the mass splitting between the scalar quarks of the third generation
in the NMSSM at the one-loop level with explicit CP violation [6].
In this case, two additional complex phase are induced from the scalar top and scalar bottom quark masses.
We have then extended our investigations to the chargino sector in the NMSSM with explicit CP violation [7].
An additional complex phase other than the tree-level phase appearing from  the mixing between the chargino masses
is shown to contribute the scalar-pseudoscalar mixing, too, at the one-loop level.

In this paper, we investigate the possibility where the tree-level Higgs potential is CP conserving
while explicit CP violation occurs in the Higgs sector of the NMSSM through the one-loop radiative corrections.
We assume that the tree-level complex phase is zero.
All the parameters in the NMSSM Higgs potential at the tree level are assumed to be real.
Then, the possible sources of complex phases contributing to the scalar-pseudoscalar mixing are
the radiative corrections from the scalar top, the scalar bottom, the scalar tau lepton, the chargino,
and the neutralino sectors.
In fact, in the one-loop effective potential for each sector may be embedded a complex phase.
We investigate the effects of these five complex phases in the radiatively corrected Higgs potential of the NMSSM.
One of the complex phases comes from the neutralino sector.
We pay our attention on this phase.
The dependence on this phase of the scalar-pseudoscalar mixings are investigated, while the values
of other parameters are fixed.
The scalar-pseudoscalar mixings are strongly affected by this phase, at least for a specific set of parameter values.

Our paper is organized as follows.
In Sec. II we briefly describe the Higgs sector of the NMSSM at the tree level.
In Sec. III, for the scenario of explicit CP violation, radiative corrections from the scalar top quark sector,
the scalar bottom quark sector, the scalar tau lepton sector,
the chargino sector, and the neutralino sector are explicitly calculated using the effective potential method.
Numerical analysis and results for the explicit CP violation at the one-loop level are presented in Sec. IV.
Conclusions and some remarks are given in Sec. V.
Appendices supply the formulae for the elements of the mass matrix of the neutral Higgs bosons,
which contain the radiative corrections from each sector of the scalar top quarks, the scalar bottom quarks,
the scalar tau leptons, the charginos, and the neutralinos.

\section{The Higgs sector}

The NMSSM has two SU(2) doublet superfields $H_1 = (H_1^0, H^-)$ and $H_2 = (H^+, H_2^0)$ with
hypercharges $- 1/2$ and $1/2$, respectively,  and a neutral SU(2) singlet superfield $N$ with zero-hypercharge.
For the fermion matter fields, we take into account only the third generation of quarks and leptons.
The general form of the NMSSM superpotential may be expressed as
\begin{equation}
W = h_t Q H_2 t_R^c + h_b Q H_1 b_R^c + h_{\tau} L H_1 {\tau}_R^c + \lambda N H_1 H_2 - {k \over 3} N^3 \ ,
\end{equation}
where $H_1 H_2 = H_1^0 H_2^0 - H_1^- H_2^+$, and all the coupling coefficients are dimensionless.
The chiral superfields $Q$ and $L$ contain the left-handed quark and lepton doublets,
and $t_R^c$, $b_R^c$ and ${\tau}_R^c$ are the charge conjugate of the right-handed $t$ quark, $b$ quark, and $\tau$ lepton,
respectively, and the parameters $h_t$, $h_b$, and $h_{\tau}$ are their respective Yukawa coupling coefficients.

In the NMSSM [8, 9], the tree-level Higgs potential may be decomposed into three parts as
\[
    V^0 = V_D + V_F + V_{\rm S} \ ,
\]
where
\begin{eqnarray}
V_D & = & {g_2^2 \over 8} (H_1^{\dag} \vec\sigma H_1 + H_2^{\dag} \vec\sigma H_2)^2 + {g_1^2\over 8}(|H_2|^2 - |H_1|^2)^2  \ , \cr
V_F & = & |\lambda|^2[(|H_1|^2+|H_2|^2)|N|^2+|H_1 H_2|^2] + |k|^2|N|^4 -(\lambda k^*H_1H_2N^{*2}+ {\rm H.c.}) \ , \cr
V_{\rm S} & = & m_{H_1}^2|H_1|^2 + m_{H_2}^2|H_2|^2 + m_N^2|N|^2 - (\lambda A_{\lambda} H_1 H_2 N + {1\over 3} k A_k N^3 + {\rm H.c.})\ ,
\end{eqnarray}
with $g_1$ and $g_2$ being the U(1) and SU(2) gauge coupling constants,
respectively, and $\vec \sigma = (\sigma^1, \sigma^2, \sigma^3)$ are the Pauli matrices.
The soft SUSY breaking $V_{\rm S}$ has two additional parameters $A_{\lambda}$ and $A_k$,
both of mass dimension, and three soft masses $m_{H_1}$, $m_{H_2}$, and $m_N$.
Generally, $\lambda$, $k$, $A_{\lambda}$, and $A_k$ can be complex.
Among them, $\lambda A_{\lambda}$ and $k A_k$ can be adjusted to be real and positive by redefining the phases of $H_1H_2$ and $N$.
Thus the tree-level Higgs potential may have at most only one physical phase.
It may be chosen to be $\phi$ in $\lambda k^*$ = $\lambda k e^{i \phi}$.
This phase may allow for explicit CP violation in the NMSSM.
It has been found that the scalar-pseudoscalar mixing in tree-level Higgs potential of the NMSSM stems
from either the triple coupling term of the Higgs doublets and the Higgs singlet or the cubic term of the Higgs singlet itself,
but not from the coupling between two Higgs doublets [5-7].

In this paper, we impose the restriction of $\phi = 0$ on the tree-level Higgs potential of the NMSSM.
We start with the tree-level Higgs potential which is phase-free, by assuming that $\lambda$, $A_{\lambda}$, $k$, and $A_k$  are real.
Thus, $\phi$ = 0 in our tree-level Higgs potential.
Now, the two Higgs doublets and the Higgs singlet may be expressed via a unitary transformation as
\begin{eqnarray}
\begin{array}{lll}
        H_1 & = & \left ( \begin{array}{c}
          v_1 + S_1 + i \sin \beta A   \cr
          \sin \beta C^{+ *}
  \end{array} \right )  \ ,  \cr
        H_2 & = & \left ( \begin{array}{c}
          \cos \beta C^+           \cr
          v_2 + S_2 + i \cos \beta A
  \end{array} \right )   \ ,  \cr
        N & = & \left ( \begin{array}{c}
          x + X + i Y
  \end{array} \right )   \ ,
\end{array}
\end{eqnarray}
where $S_1$, $S_2$, $X$ are the scalar Higgs fields, $A$, $Y$ are the pseudoscalar Higgs fields,
and $C^+$ is the charged Higgs field.

>From the tree-level Higgs potential, the symmetric 3 $\times$ 3 mass matrix $M_{S}$ for the scalar Higgs bosons
is obtained in the basis $(S_1, S_2, X)$ as
\begin{eqnarray}
(M_S)_{11} & = & m_Z^2 \cos^2 \beta + \lambda x (A_{\lambda} + kx) \tan \beta \ , \cr
(M_S)_{22} & = & m_Z^2 \sin^2 \beta + \lambda x (A_{\lambda} + kx) \cot \beta \ , \cr
(M_S)_{33} & = & (2 k x)^2 - k x A_k + {\lambda \over 2x} v^2 A_{\lambda} \sin 2 \beta \ , \cr
(M_S)_{12} & = & (\lambda^2 v^2 - {1 \over 2} m_Z^2) \sin 2 \beta - \lambda x (A_{\lambda} + k x)  \ , \cr
(M_S)_{13} & = & 2 \lambda^2 x v \cos \beta - \lambda v \sin \beta (A_{\lambda} + 2 k x) \ , \cr
(M_S)_{23} & = & 2 \lambda^2 x v \sin \beta - \lambda v \cos \beta (A_{\lambda} + 2 k x) \ ,
\end{eqnarray}
where the gauge boson masses are given as $m_Z^2 = (g_1^2 + g_2^2) v^2/2$ and $m_W^2 = g_2^2 v^2/2$ for
$v = \sqrt{v_1^2 + v_2^2}$ = 175 GeV after the electroweak symmetry breaking.
In the above mass matrix, the soft SUSY breaking masses $m_{H_1}^2$, $m_{H_2}^2$, and $m_N^2$ already are
eliminated by minimization conditions with respect to $S_1$, $S_2$, and $X$.
Thus the mass matrix for the scalar Higgs bosons is a function of $\tan \beta = v_2/v_1$, $\lambda$,
$A_{\lambda}$, $k$, $A_k$ and $x$.
The mass matrix $M_S$ is then diagonalized in the basis of mass eigenstates.
Let us denote the mass eigenstates of $M_S$ as $S_1, S_2, S_3$, which are linear combinations of the weak eigenstates,
namely, $S_1, S_2, X$.
(Notations are a little bit confusing but hereafter $S_i$ ($i$ = 1,2,3) are the mass eigenstates.)
The eigenvalues of the mass matrix $M_S$ are obtained and sorted as
\begin{eqnarray}
m_{S_j}^2 & = & {1 \over 3} {\rm Tr}(M_S) + 2 \sqrt{W}
\cos \left \{{\Theta + 2 j \pi \over 3} \right \} \  \  (j = 1, \ 2, \ 3)  \ ,
\end{eqnarray}
where
\begin{equation}
\Theta = \cos^{-1} \left ({U \over \sqrt{W^3}} \right)  \ ,
\end{equation}
with
\begin{eqnarray}
W & = & - {1 \over 18} \{{\rm Tr}(M_S)\}^2 + {1 \over 6} {\rm Tr}(M_S M_S) \ , \cr
  &   &      \cr
U & = & - {5 \over 108} \{{\rm Tr}(M_S)\}^3 + {1 \over 12} {\rm Tr}(M_S) {\rm Tr}(M_S M_S)
+ {1 \over 2} \det(M_S) \ .
\end{eqnarray}
Thus, $m_{S_1}^2$ is the squared mass of the mass eigenstate $S_1$, and so on.

For the pseudoscalar Higgs bosons, the symmetric 2 $\times$ 2 mass matrix $M_P$ is obtained in the basis $(A, Y)$ as
\begin{eqnarray}
(M_P)_{11} & = & {2 \lambda x (A_{\lambda} + kx) \over \sin 2 \beta}  \ , \cr
(M_P)_{22} & = & 3 k x A_k + 2 \lambda k v^2 \sin 2 \beta + {\lambda v^2 \over 2x} A_{\lambda} \sin 2 \beta  \ , \cr
(M_P)_{12} & = & \lambda v (A_{\lambda} - 2 k x)  \ .
\end{eqnarray}
The pseudoscalar Higgs boson masses are given by the eigenvalues of $M_P$ as
\begin{equation}
m_{P_1, P_2}^2 = {1 \over 2} \left \{ {\rm Tr} (M_P) \mp \sqrt{({\rm Tr} M_P)^2 - 4 {\rm det} (M_P)  } \right \}     \ .
\end{equation}
Finally, the charged Higgs boson mass at the tree level is given as
\begin{equation}
m_C^2 = m_W^2 - \lambda^2 v^2 + {2 \lambda x \over \sin 2 \beta} (A_{\lambda} + k x) \ .
\end{equation}
Unlike in the case of the MSSM, the tree-level mass of the charged Higgs boson can be either heavier or
lighter than the $W$ boson
according to whether the second is smaller than the third term or not.

Next, we calculate tree-level masses of other relevant fields.
These will be used later for evaluating the radiative corrections to the neutral Higgs sector.
The fermion matter fields obtain their masses after the electroweak symmetry breaking
as $m_t^2 = (h_t v_2)^2$ for top quark, $m_b^2 = (h_b v_1)^2$ for bottom quark,
and $m_{\tau}^2 = (h_{\tau} v_1)^2$ for tau lepton.
The scalar top quarks, the scalar bottom quarks, the scalar tau leptons,
and the charginos have their tree-level masses as
\begin{eqnarray}
m_{{\tilde t}_{1, 2}}^2 & = & m_t^2 + {1 \over 2}(m_Q^2 + m_T^2) + {m_Z^2 \over 4} \cos 2 \beta
\mp \left [ \left \{ {1 \over 2} (m_Q^2 - m_T^2) \right. \right. \cr
& &\mbox{}\left. \left. + \left ( {2 \over 3} m_W^2 - {5 \over 12} m_Z^2 \right ) \cos 2 \beta \right \}^2   \right. \cr
& &\left. \mbox{} + m_t^2 (A_t^2  + \lambda^2 x^2 \cot^2 \beta + 2 A_t \lambda x \cot \beta \cos \phi_t) \rule{0mm}{5.6mm} \right ]^{1 \over 2}    \ , \cr
m_{{\tilde b}_{1, 2}}^2 & = & m_b^2 + {1 \over 2} (m_Q^2 + m_B^2) - {m_Z^2 \over 4} \cos 2 \beta
\mp \left [ \left \{ {1 \over 2} (m_Q^2 - m_B^2) \right. \right. \cr
& &\mbox{}\left. \left. + \left ( {1 \over 12} m_Z^2 - {1 \over 3} m_W^2 \right ) \cos 2 \beta \right \}^2 \right. \cr
& &\left. \mbox{} + m_b^2 (A_b^2 + \lambda^2 x^2 \tan^2 \beta + 2 A_b \lambda x \tan \beta \cos \phi_b) \rule{0mm}{5.6mm}  \right ]^{1 \over 2}  \ , \cr
m_{{\tilde \tau}_{1, 2}}^2 & = & m_{\tau}^2 + {1 \over 2}(m_L^2 + m_E^2)
    - {m_Z^2 \over 4} \cos 2 \beta \mp \left [ \left \{ {1 \over 2} (m_L^2 - m_E^2) \right. \right. \cr
& &\mbox{}\left. \left. + \left ( {3 \over 8} m_Z^2 - {1 \over 2} m_W^2 \right ) \cos 2 \beta \right \}^2 \right. \cr
& &\left. \mbox{} + m_{\tau}^2 (A_{\tau}^2 + \lambda^2 x^2 \tan^2 \beta
    + 2 A_{\tau} \lambda x \tan \beta \cos \phi_{\tau}) \rule{0mm}{5.4mm}  \right ]^{1 \over 2}   \ , \cr
m_{{\tilde \chi}_{1, 2}}^2 & = & {1 \over 2} (M_2^2 + \lambda^2 x^2) + m_W^2
\mp \left [ \left \{ {1 \over 2} (M_2^2 - \lambda^2 x^2 ) - m_W^2 \cos 2 \beta \right \}^2 \right. \cr
& &\left. \mbox{} + 2 m_W^2 \cos^2 \beta (M_2^2 + \lambda^2 x^2 \tan^2 \beta
    + 2 M_2 \lambda x \tan \beta \cos \phi_c) \rule{0mm}{5.4mm} \right ]^{1\over 2}      \ ,
\end{eqnarray}
where $m_Q$, $m_T$, $m_B$, $m_L$, and $m_E$ are the soft SUSY breaking breaking masses, and $A_t$, $A_b$, and
$A_{\tau}$ are the trilinear soft SUSY breaking parameters of mass dimension.
One can find that there are four phases  $\phi_t$, $\phi_b$, $\phi_{\tau}$, and $\phi_c$ in the above expressions:
They are originated from the trilinear soft SUSY breaking parameters $A_t$, $A_b$, $A_{\tau}$ and the SU(2) gaugino
mass $M_2$ which are complex in general whereas $\lambda$ is assumed to be real and positive.
Note that the scalar fermion masses are not symmetric under $m_Q \leftrightarrow m_T$,  $m_Q \leftrightarrow m_B$,
or $m_L \leftrightarrow m_E$ since $D$-terms are included in them.

In the NMSSM, there are five neutralinos.
Their tree-level masses are obtained from the 5 $\times$ 5 mass matrix
\begin{equation}
\begin{array}{ccc}
{\cal M}_{{\tilde \chi}^0} & = & \left ( \begin{array}{ccccc}
- M_1 e^{i \phi_1} & 0 & - {{\displaystyle g_1} \over {\displaystyle \sqrt{2}}} H_1^0 &
{{\displaystyle g_1} \over {\displaystyle \sqrt{2}}} H_2^0 & 0  \cr
 & & & & \cr
0  & - M_2 & {{\displaystyle g_2} \over {\displaystyle \sqrt{2}}} H_1^0 &
- {{\displaystyle g_2} \over {\displaystyle \sqrt{2}}} H_2^0 & 0  \cr
 & & & & \cr
- {{\displaystyle g_1} \over {\displaystyle \sqrt{2}}} H_1^0  &
{{\displaystyle g_2} \over {\displaystyle \sqrt{2}}} H_1^0 & 0 & \lambda N & \lambda H_2^0   \cr
 & & & & \cr
{{\displaystyle g_1} \over {\displaystyle \sqrt{2}}} H_2^0  &
- {{\displaystyle g_2} \over {\displaystyle \sqrt{2}}} H_2^0 & \lambda N &  0 & \lambda H_1^0 \cr
 & & & & \cr
0 & 0 & \lambda H_2^0 & \lambda H_1^0 & - 2 k N \cr
\end{array} \right )  \ ,
\end{array}
\end{equation}
where $M_1$ is the U(1) gaugino mass.
Note that we have already assumed that $\lambda$ and $k$ are real.
Thus, the neutralino mass matrix may have two complex numbers, $M_1$ and $M_2$.
Let the relative phase between them be $\phi_1$.
This is therefore the only non-trivial phase in the neutralino mass matrix in our case.

The above neutralino mass matrix is complex and symmetric, but not Hermitian.
In other to calculate the neutralino masses, the Hermitian matrix ${\cal M}_{{\tilde \chi}^0} {\cal M}_{{\tilde \chi}^0}^{\dagger}$ is
diagonalized through a similarity transformation.
The squared neutralino masses at the tree level are denoted as $m_{{\tilde \chi}^0}^2$ ($i$ = 1 to 5),
and they are sorted such that $m_{{\tilde \chi}_i^0}^2 < m_{{\tilde \chi}_j^0}^2$ for $i < j$.
If they are all different from each other, they are obtained as the solutions of
\begin{equation}
 {\rm det} ( {\cal M}_{{\tilde \chi}^0} {\cal M}_{{\tilde \chi}^0}^{\dagger}
    - m_{{\tilde \chi}_i^0}^2 1_{5 \times 5} ) = 0  \ ,
\end{equation}
which may be rewritten as
\begin{equation}
m_{{\tilde \chi}_i^0}^{10} + {\tilde A} m_{{\tilde \chi}_i^0}^8 + {\tilde B} m_{{\tilde \chi}_i^0}^6 + {\tilde C} m_{{\tilde \chi}_i^0}^4 + {\tilde D} m_{{\tilde \chi}_i^0}^2 + {\tilde E} = 0 \ ,
\end{equation}
where the field-dependent coefficients ${\tilde A}$, ${\tilde B}$, ${\tilde C}$, ${\tilde D}$, and ${\tilde E}$ are
very complicated and obtainable from Appendix A.

\section{Radiative corrections}

Now let us consider the radiative corrections.
In order to describe the explicit CP violation scenario, it would be convenient to decompose the neutral Higgs fields as
\begin{eqnarray}
\begin{array}{lll}
        H_1^0 & = & \left ( \begin{array}{c}
          v_1 + h_1 + i \sin \beta h_3    \cr
  \end{array} \right )  \ ,  \cr
        H_2^0 & = & \left ( \begin{array}{c}
          v_2 + h_2 + i \cos \beta h_3
  \end{array} \right )   \ ,  \cr
        N & = & \left ( \begin{array}{c}
          x + h_4 + i h_5
\end{array} \right )   \ ,
\end{array}
\end{eqnarray}
where $h_i$ ($i$ = 1 to 5) are the neutral Higgs fields which do not have definite CP parities.
In our case with  explicit CP violation scenario, they are mixtures of CP-even  and CP-odd components.
Through radiative corrections, there emerges CP mixing among $h_i$.
On the other hand, the charged Higgs bosons does not mix because both of them are CP-even.

Now, we turn to the radiative corrections.
The incomplete cancellation between ordinary particles and their superpartners give the one-loop corrections
to the tree-level Higgs boson masses.
The full Higgs potential up to the one-loop level is the sum of the tree-level Higgs potential $V^0$
and the radiative corrections $V^1$ as
\[
            V = V^0 + V^1 \ ,
\]
where $V^1$ may be decomposed into
\begin{equation}
            V^1 = V^t +V^b +V^{\tau} +V^{\tilde \chi} +V^{{\tilde \chi}^0} \ ,
\end{equation}
with superscripts indicate the radiative corrections due to the relevant particles and their superpartners.
In fact, the neutral part and the charged one are related to each other.
In the NMSSM with explicit CP violation, all the neutral Higgs boson masses can be expressed
in terms of the charged Higgs boson mass.
Note that the charged Higgs boson mass should be used as input.
And $V^{\tilde \chi}$ and $V^{{\tilde \chi}^0}$ should be evaluated simultaneously.

We employ the effective potential method in order to calculate radiative corrections [10].
The general expression for the one-loop effective potential is  given by
\[
V^1 = \sum_k {c_k \over 64 \pi^2} (-1)^{2 J_k} (2 J_k + 1) \{{\cal M}_k (h_i) \}^4 \left [\log { \{{\cal M}_k (h_i) \}^2 \over \Lambda^2} - {3\over 2} \right ]  \ ,
\]
where $J_k$ is the spin of particle or superparticle,  $\Lambda$ is the renormalization scale,
and $c_k = c_{\rm color} c_{\rm charge}$ with $c_{\rm color}$ being the color factor
and $c_{\rm charge}$ the charge factor of the corresponding  particle or superparticle.
We have $c_{\rm color}$ = 3 (1) for colored (uncolored) particles and $c_{\rm charge}$ = 2 (1) for
charged (neutral) particles.
Note that ${\cal M}_k^2$ depends on the neutral Higgs fields $h_i$ ($i$ = 1 to 5).

Explicitly, we write down the formulae for the relevant one-loop effective potentials:
The one-loop effective potential for the contributions of top quark and scalar top quarks is given by
\[
V^t = \sum_{i = 1}^2 {3 {\cal M}_{{\tilde t}_i}^4 \over 32 \pi^2}
 \left (\log {{\cal M}_{{\tilde t}_i}^2 \over \Lambda^2} - {3 \over 2} \right )
  - {3 {\cal M}_t^4 \over 16 \pi^2} \left (\log {{\cal M}_t^2 \over \Lambda^2}
  - {3\over 2} \right ) \ ,
\]
the one for bottom and scalar bottom quarks by
\[
V^b = \sum_{i = 1}^2 {3 {\cal M}_{{\tilde b}_i}^4 \over 32 \pi^2}
 \left (\log {{\cal M}_{{\tilde b}_i}^2 \over \Lambda^2} - {3\over 2} \right )
 -  {3 {\cal M}_b^4 \over 16 \pi^2} \left (\log {{\cal M}_b^2 \over \Lambda^2}
 - {3\over 2} \right ) \ ,
\]
the one tau lepton and scalar tau leptons by
\[
V^{\tau} = \sum_{i = 1}^2 {{\cal M}_{{\tilde \tau}_i}^4 \over 32 \pi^2}
 \left (\log {{\cal M}_{{\tilde \tau}_i}^2 \over \Lambda^2} - {3\over 2} \right )
 - {{\cal M}_{\tau}^4 \over 16 \pi^2} \left (\log {{\cal M}_{\tau}^2 \over \Lambda^2} - {3\over 2} \right ) \ ,
\]
the one for $W$ boson, the charged Higgs boson, and the charginos by
\[
V^{{\tilde \chi}} = {3 {\cal M}_W^4  \over 32 \pi^2} \left (\log {{\cal M}_W^2 \over \Lambda^2} - {3\over 2} \right )
 + {{\cal M}_{C^+}^4 \over 32 \pi^2} \left ( \log  {{\cal M}_{C^+}^2 \over \Lambda^2} - {3\over 2} \right )
 - \sum_{i = 1}^2 {{\cal M}_{{\tilde \chi}_i}^4 \over 16 \pi^2}
 \left (\log {{\cal M}_{{\tilde \chi}_i}^2 \over \Lambda^2} - {3\over 2} \right )   \ ,
\]
and lastly, the one for $Z$ boson, the neutral Higgs bosons and the neutralinos is given by
\[
V^{{\tilde \chi}^0} = {3 {\cal  M}_Z^4 \over  64 \pi^2}
  \left (\log  {{\cal M}_Z^2 \over  \Lambda^2} - {3 \over 2}\right )
  + \sum_{i  = 1}^5 {{\cal  M}_{h_i}^4 \over 64 \pi^2}
 \left( \log {{\cal M}_{h_i}^2 \over \Lambda^2} - {3 \over 2} \right )
 - \sum_{i = 1}^5 {{\cal M}_{{\tilde \chi}^0_i}^4 \over 32 \pi^2}
 \left (\log {{\cal M}_{{\tilde \chi}^0_i}^2 \over \Lambda^2} - {3\over 2}
 \right ) \ .
\]
Note that in the above one-loop effective potentials fermions enter with a negative sign while bosons
with a positive sign.

If CP be conserved in the Higgs sector, $h_3$ and $h_5$ would be the pseudoscalar components of the Higgs fields,
and the minimum conditions for them would be trivial.
In our case of explicit CP violation, the minimum conditions for them are non-trivial.
Thus, we obtain two CP-odd tadpole minimum conditions for $h_3$ and $h_5$, respectively, as
\begin{eqnarray}
0 & = & {3 m_t^2 A_t \lambda x \sin \phi_t \over 16 \pi^2 v^2 \sin^2 \beta}
f (m_{{\tilde t}_1}^2, m_{{\tilde t}_2}^2)
+ {3 m_b^2 A_b \lambda x \sin \phi_b \over 16 \pi^2 v^2 \cos^2 \beta}
f (m_{{\tilde b}_1}^2, m_{{\tilde b}_2}^2) \cr
  & & \cr
  & &\mbox{} + {m_{\tau}^2 A_{\tau} \lambda x \sin \phi_{\tau} \over 16 \pi^2 v^2 \cos^2 \beta}
f (m_{{\tilde \tau}_1}^2, m_{{\tilde \tau}_2}^2)
- {m_W^2 \lambda x M_2 \sin \phi_c \over 4 \pi^2 v^2}
f (m_{{\tilde \chi}_1}^2, m_{{\tilde \chi}_2}^2) \cr
  & & \cr
& &\mbox{} + \sum_{k = 1}^5 {m_{{\tilde \chi}^0_k}^2  \over 32 \pi^2 v}
\left \{\log \left ({m_{{\tilde \chi}^0_k}^2 \over \Lambda^2} \right ) - 1 \right \} {{\tilde B}_3 m_{{\tilde \chi}^0_k}^6 + {\tilde C}_3 m_{{\tilde \chi}^0_k}^4 + {\tilde D_3} m_{{\tilde \chi}^0_k}^2 + {\tilde E}_3 \over \prod\limits_{a \not= k} (m_{{\tilde \chi}^0_k}^2 - m_{{\tilde \chi}^0_a}^2)} \ ,
\end{eqnarray}
\begin{eqnarray}
0 & = & {3 m_t^2 A_t \lambda \sin \phi_t \over 16 \pi^2 v \sin^2 \beta}
f (m_{{\tilde t}_1}^2, m_{{\tilde t}_2}^2)
+ {3 m_b^2 A_b \lambda \sin \phi_b \over 16 \pi^2 v \cos^2 \beta}
f (m_{{\tilde b}_1}^2, m_{{\tilde b}_2}^2) \cr
  & & \cr
  & &\mbox{} + {m_{\tau}^2 A_{\tau} \lambda \sin \phi_{\tau} \over 16 \pi^2 v \cos^2 \beta}
f (m_{{\tilde \tau}_1}^2, m_{{\tilde \tau}_2}^2)
+ {m_W^2 \lambda M_2 \sin \phi_c \over 4 \pi^2 v}
f (m_{{\tilde \chi}_1}^2, m_{{\tilde \chi}_2}^2) \cr
  & & \cr
& &\mbox{} + \sum_{k = 1}^5 {m_{{\tilde \chi}^0_k}^2 \over 16 \pi^2 v \sin 2 \beta}
\left \{\log \left ({m_{{\tilde \chi}^0_k}^2 \over \Lambda^2} \right ) - 1 \right \} {{\tilde B}_5 m_{{\tilde \chi}^0_k}^6 + {\tilde C}_5 m_{{\tilde \chi}^0_k}^4 + {\tilde D}_5 m_{{\tilde \chi}^0_k}^2
    + {\tilde E}_5 \over \prod\limits_{a \not= k} (m_{{\tilde \chi}^0_k}^2 - m_{{\tilde \chi}^0_a}^2)} \ ,
\end{eqnarray}
where the five terms come respectively from the scalar top quark, the scalar bottom quark, the scalar tau lepton,
the chargino, and the neutralino contributions.
The dimensionless function arising from radiative corrections is defined by
\[
f(m_x^2, \  m_y^2) = {1 \over (m_y^2 - m_x^2)} \left \{m_x^2 \log  {m_x^2 \over \Lambda^2} - m_y^2 \log {m_y^2 \over \Lambda^2} \right \} + 1 \ ,
\]
and we need to introduce ${\tilde A}_i = (\partial {\tilde A} / \partial h_i)$,
${\tilde B}_i = (\partial {\tilde B} / \partial h_i)$, ${\tilde C}_i = (\partial {\tilde C} / \partial h_i)$,
${\tilde D}_i = (\partial {\tilde D} / \partial h_i)$, and ${\tilde E}_i = (\partial {\tilde E} / \partial h_i)$:
In the above minimum conditions, just ${\tilde B}_i$, ${\tilde C}_i$, ${\tilde D}_i$,
and  ${\tilde E}_i$ ($i$ = 3 and 5) enter.
Note that no term associated with the tree-level is present in the above minimum conditions since $\phi$ = 0.

By differentiating the Higgs potential $V$ at the one-loop level with respect to the neutral Higgs fields, a $5 \times 5$
symmetric mass matrix $M$ for them is obtained in the $(h_1, h_2, h_3, h_4, h_5)$-basis, which may be decomposed as
\[
M_{ij} = M^0_{ij} + {\bar M}_{ij}^1 \ ,
\]
where $M^0_{ij}$ is obtained from $V^0$ while ${\bar M}_{ij}^1$ from $V^1$.
Explicitly, $M^0_{ij}$ may be expressed in terms of the scalar and pseudoscalar mass matrix elements as
\begin{eqnarray}
    && M_{11}^0 = (M_S)_{11}  \cr
    && M_{22}^0 = (M_S)_{22}  \cr
    && M_{44}^0 = (M_S)_{33}  \cr
    && M_{12}^0 = (M_S)_{12}  \cr
    && M_{14}^0 = (M_S)_{13}  \cr
    && M_{24}^0 = (M_S)_{23}
\nonumber
\end{eqnarray}
and
\begin{eqnarray}
    && M_{33}^0 = (M_P)_{11}  \cr
    && M_{55}^0 = (M_P)_{22}  \cr
    && M_{35}^0 = (M_P)_{12}
\nonumber
\end{eqnarray}
and the remaining elements of $M_{ij}^0$ are explicitly zero since we assume that $\phi = 0$.

Thus, the elements of the neutral Higgs boson mass matrix at the one-loop level may be written as
\begin{eqnarray}
M_{11} & = & (M_S)_{11} + m_A^2 \sin^2 \beta + M_{11}^1 \ , \cr
M_{22} & = & (M_S)_{22} + m_A^2 \cos^2 \beta + M_{22}^1 \ , \cr
M_{33} & = & (M_P)_{11} + m_A^2 + M_{33}^1 \ , \cr
M_{44} & = & (M_S)_{33} + m_B^2 + M_{44}^1 \ , \cr
M_{55} & = & (M_P)_{22} + m_B^2 + M_{55}^1 \ , \cr M_{12} & = & (M_S)_{12} - m_A^2 \cos \beta \sin \beta + M_{12}^1 \ , \cr M_{13} & = & M_{13}^1 \ , \cr
M_{14} & = & (M_S)_{13} - {m_B^2 x \over v \cos \beta} + M_{14}^1
\ , \cr M_{15} & = & M_{15}^1  \ , \cr
M_{23} & = & M_{23}^1  \ , \cr
M_{24} & = & (M_S)_{23} - {m_B^2 x \over v \sin \beta} + M_{24}^1 \ , \cr
M_{25} & = & M_{25}^1   \ , \cr
M_{34} & = & M_{34}^1  \ , \cr
M_{35} & = & (M_P)_{12} + {m_A^2 v \cos \beta \sin \beta \over x} + M_{35}^1 \ , \cr
M_{45} & = & M_{45}^1 \ ,
\end{eqnarray}
where two expressions are introduced for convenience as
\begin{eqnarray}
m_A^2 & = & {3 m_t^2 A_t \lambda x \cos \phi_t \over 16 \pi^2 v^2 \sin^3 \beta \cos \beta} f (m_{{\tilde t}_1}^2, m_{{\tilde t}_2}^2)
+ {3 m_b^2 A_b \lambda x \cos \phi_b \over 16 \pi^2 v^2 \cos^3 \beta \sin \beta} f (m_{{\tilde b}_1}^2, m_{{\tilde b}_2}^2) \cr
  & & \cr
  & &\mbox{} + {m_{\tau}^2 A_{\tau} \lambda x \cos \phi_{\tau} \over 16 \pi^2 v^2 \cos^3 \beta \sin \beta}
f (m_{{\tilde \tau}_1}^2, m_{{\tilde \tau}_2}^2)
- {m_W^2 \lambda x M_2 \cos \phi_c \over 2 \pi^2 v^2 \sin 2 \beta}
f (m_{{\tilde \chi}_1}^2, m_{{\tilde \chi}_2}^2) \cr
  & & \cr
& &\mbox{} + \sum_{k = 1}^5 {m_{{\tilde \chi}^0_k}^2 \over 32 \pi^2}
\left \{\log \left ({m_{{\tilde \chi}^0_k}^2 \over \Lambda^2} \right ) - 1 \right \}
{{\tilde A}_{33} m_{{\tilde \chi}^0_k}^8  + {\tilde B}_{33} m_{{\tilde \chi}^0_k}^6
+ {\tilde C}_{33} m_{{\tilde \chi}^0_k}^4 + {\tilde D}_{33} m_{{\tilde \chi}^0_k}^2
+ {\tilde E}_{33} \over \prod\limits_{a \not= k} (m_{{\tilde \chi}^0_k}^2 - m_{{\tilde \chi}^0_a}^2)}
\nonumber
\end{eqnarray}
and
\begin{eqnarray}
m_B^2 & = & {3 m_t^2 A_t \lambda \cos \phi_t \over 16 \pi^2 x \tan \beta} f (m_{{\tilde t}_1}^2, m_{{\tilde t}_2}^2)
+ {3 m_b^2 A_b \lambda \cos \phi_b \over 16 \pi^2 x \cot \beta} f (m_{{\tilde b}_1}^2, m_{{\tilde b}_2}^2) \cr
  & & \cr
  & &\mbox{} + {m_{\tau}^2 A_{\tau} \lambda \cos \phi_{\tau} \over 16 \pi^2 x \cot \beta}
f (m_{{\tilde \tau}_1}^2, m_{{\tilde \tau}_2}^2)
- {m_W^2 \lambda M_2 \sin 2 \beta \cos \phi_c \over 8 \pi^2 x}
f (m_{{\tilde \chi}_1}^2, m_{{\tilde \chi}_2}^2) \cr
  & & \cr
& &\mbox{} + \sum_{k = 1}^5 {m_{{\tilde \chi}^0_k}^2 \over 32 \pi^2}
\left \{\log \left ({m_{{\tilde \chi}^0_k}^2 \over \Lambda^2} \right ) - 1 \right \}
    {{\tilde A}_{55} m_{{\tilde \chi}^0_k}^8  + {\tilde B}_{55} m_{{\tilde \chi}^0_k}^6
    + {\tilde C}_{55} m_{{\tilde \chi}^0_k}^4 + {\tilde D}_{55} m_{{\tilde \chi}^0_k}^2 + {\tilde E}_{55} \over
\prod\limits_{a \not = k} (m_{{\tilde \chi}^0_k}^2 - m_{{\tilde \chi}^0_a}^2)}
\nonumber
\end{eqnarray}
In each expression, the first term comes from the scalar top quark contributions, the second term from
the scalar bottom quarks,
the third term from the scalar tau leptons, the fourth one from charginos, and the last term from neutralinos.
In $m_A$ and $m_B$, the coefficients ${\tilde A}_{ii}$, ${\tilde B}_{ii}$, ${\tilde C}_{ii}$, ${\tilde D}_{ii}$,
and ${\tilde E}_{ii}$ ($i$ = 1 to 5) are given by
${\tilde A}_{ij} = \bigtriangledown_{ij} {\tilde A}$,
${\tilde B}_{ij} = \bigtriangledown_{ij} {\tilde B}$,
${\tilde C}_{ij} = \bigtriangledown_{ij} {\tilde C}$,
${\tilde D}_{ij} = \bigtriangledown_{ij} {\tilde D}$,
and ${\tilde E}_{ij} = \bigtriangledown_{ij} {\tilde E}$, where the second-order differential operator is defined as
\begin{eqnarray}
\bigtriangledown_{ij} & = & {\partial^2 \over \partial h_i \partial h_j}
    - {1 \over v_1} {\partial \over \partial h_1} \delta_{1j}- {1 \over v_2} {\partial \over \partial h_2} \delta_{2j}
    - \left ({\sin^2 \beta \over v_1} {\partial \over \partial h_1}
    + {\cos^2 \beta \over v_2} {\partial \over \partial h_2}\right) \delta_{3j}  \cr
& &\mbox{} - {1 \over x} {\partial \over \partial h_4} \delta_{4j}
- {1 \over x} {\partial \over \partial h_4} \delta_{5j} \ ,
\end{eqnarray}
for $j \ge i$ ($i,j$ = 1 to 5).
In the above differential operator, all the terms except the first one are caused by the shift in the vacuum
due to the radiative corrections.

We may further decompose $M_{ij}^1$ as
\[
M_{ij}^1 = M^t_{ij} + M^b_{ij} + M^{\tau}_{ij} + M_{ij}^{\tilde \chi} + M_{ij}^{{\tilde \chi}^0} \ ,
\]
where $M^t_{ij}$ is obtained from $V^t$, $M^b_{ij}$ from $V^b$, $M^{\tau}_{ij}$ from $V^{\tau}$,
$M^{\tilde \chi}_{ij}$ from $V^{\tilde \chi}$, and $M^{{\tilde \chi}^0}_{ij}$ from $V^{{\tilde \chi}^0}$,
after imposing  the CP-odd tadpole minimum conditions for $h_3$ and $h_5$.
We note that $M^{\tilde \chi}_{ij}$ and $M^{{\tilde \chi}^0}_{ij}$ are matrices for squared masses.
The elements of $M^{\tau}$ have no color factor.
In Appendix B are presented the complicated formulae for $M^t_{ij}$.
In Appendix C for $M^b_{ij}$, in Appendix D for $M^{\tau}_{ij}$, and in Appendix E for $M_{ij}^{\tilde \chi}$.

Now, we concentrate on $M_{ij}^{{\tilde \chi}^0}$, which represents the radiative corrections due to the contributions
from the neutralino sector:
It consists of  the contributions from $Z$ boson, tree-level scalar and pseudoscalar Higgs bosons, and neutralinos.
Thus, we may decompose it further again as
\[
M_{ij}^{{\tilde \chi}^0} = \Delta^Z_{ij} +\Delta^S_{ij} +\Delta^P_{ij} +\Delta^{{\tilde \chi}^0}_{ij}
\]
where each term represents the relevant contribution.
Let us describe one by one.

The contribution of $Z$ boson affects the four elements $M_{11}^{{\tilde \chi}^0}$, $M_{22}^{{\tilde \chi}^0}$,
and $M_{12}^{{\tilde \chi}^0} = M_{21}^{{\tilde \chi}^0}$ only, since the Higgs singlet $N$ does not couple to $Z$ boson.
Calculations yield the contribution of $Z$ boson as
\begin{equation}
\begin{array}{cccc}
\left ( \begin{array}{c}
\Delta_{11}^Z \cr \Delta_{22}^Z \cr \Delta_{12}^Z
\end{array} \right )
& = & \left ( \begin{array}{c}
\cos^2 \beta   \cr \sin^2 \beta \cr \cos \beta \sin \beta
\end{array} \right )
& {{\displaystyle 3 m_Z^4} \over {\displaystyle 16 \pi^2 v^2}} {\displaystyle \log} \left ({{\displaystyle m_Z^2} \over {\displaystyle \Lambda^2}} \right ) \ ,
\end{array}
\end{equation}
with $\Delta_{21}^Z = \Delta_{12}^Z$.

The contribution of the tree-level scalar Higgs bosons is obtained as
\begin{equation}
\Delta_{i j}^S  = \sum_{l = 1}^3 {m_{S_l}^2 \over 64 \pi^2}
\left (\log {m_{S_l}^2 \over \Lambda^2} - 1 \right )
{\partial^2 m_{S_l}^2 \over \partial h_i \partial h_j}
+ \sum_{l = 1}^3 {\log (m_{S_l}^2 / \Lambda^2) \over 64 \pi^2}
\left ({\partial m_{S_l}^2 \over \partial h_i} \right ) \left ({\partial m_{S_l}^2 \over \partial h_j} \right )  \ ,
\end{equation}
where $m^2_{S_l}$ ($l$ = 1, 2, 3) are the tree-level masses of the scalar Higgs bosons of Eq. (5), and
\begin{equation}
{\partial m_{S_a}^2 \over \partial h_i} = -
{A_i m_{S_a}^4 + B_i m_{S_a}^2 + C_i \over (m_{S_a}^2 - m_{S_b}^2) (m_{S_a}^2 - m_{S_c}^2)}  \ ,
\end{equation}
and
\begin{eqnarray}
{\partial^2 m_{S_a}^2 \over \partial h_i \partial h_j} & = &\mbox{}
- { (\bigtriangledown_{ij} A) m_{S_a}^4 + (\bigtriangledown_{ij} B) m_{S_a}^2
    + (\bigtriangledown_{ij} C) \over (m_{S_a}^2 - m_{S_b}^2) (m_{S_a}^2 - m_{S_c}^2)} \cr
& &\mbox{} + {1 \over (m_{S_a}^2 - m_{S_b}^2)} \left (
{\partial m_{S_a}^2 \over \partial \phi_i}
{\partial m_{S_b}^2 \over \partial \phi_j}
+ {\partial m_{S_b}^2 \over \partial \phi_i}
{\partial m_{S_a}^2 \over \partial \phi_j} \right ) \cr
& &\mbox{}+ {1 \over (m_{S_a}^2 - m_{S_c}^2)} \left (
{\partial m_{S_a}^2 \over \partial \phi_i}
{\partial m_{S_c}^2 \over \partial \phi_j}
+ {\partial m_{S_c}^2 \over \partial \phi_i}
{\partial m_{S_a}^2 \over \partial \phi_j} \right )   \ ,
\end{eqnarray}
for $i$ = 1 to 5, and $m^2_{S_a} \ne m^2_{S_b} \ne m^2_{S_c}$.
Here, we introduce
\begin{eqnarray}
A & = &\mbox{} - {\rm Tr} ({\cal M}_S) \ , \cr
B & = & {\cal M}_{S_{11}} {\cal M}_{S_{22}} + {\cal M}_{S_{11}} {\cal M}_{S_{33}}
    + {\cal M}_{S_{22}} {\cal M}_{S_{33}} - {\cal M}_{S_{12}}^2 - {\cal M}_{S_{23}}^2
    - {\cal M}_{S_{13}}^2  \ , \cr
C & = &\mbox{} -  {\rm det} ({\cal M}_S) \ ,
\end{eqnarray}
where ${\cal M}_{S_{ij}}$ are given in Appendix F,
and $A_i = (\partial A / \partial h_i)$, $B_i = (\partial B / \partial h_i)$, and $C_i = (\partial C / \partial h_i)$.

As aforementioned, all terms in $\bigtriangledown_{ij}$ except the first one are caused by the shifts in the vacuum
due to the radiative corrections.
Note that the shifts take place along $h_1$, $h_2$, and $h_4$.
Those shifts occur actually on only the diagonal elements of the mass matrix of the neutral Higgs boson through
$\bigtriangledown_{ij}$:
$M_{11}$ by the second term of $\bigtriangledown$, $M_{22}$ by the third term, $M_{33}$ by the fourth term,
$M_{44}$ by the fifth term, and $M_{55}$ by the sixth term, respectively.
Since the tree-level Higgs potential does not possess the scalar-pseudoscalar mixing,
both $h_3$ and $h_5$ become the pseudoscalar Higgs fields and
$\partial V^0 / \partial h_3 = \partial V^0 / \partial h_5 = 0$.
At the one-loop level the non-trivial solutions for $h_3$ and $h_5$ already are obtained
by the CP-odd tadpole minimum conditions.

Next, the contribution of the tree-level pseudoscalar Higgs bosons is obtained as
\begin{eqnarray}
\Delta_{ij}^P & = & {(\bigtriangledown_{ij} A_P) \over 64 \pi^2}
\left \{m_{P_1}^2 \left (\log {m_{P_1}^2 \over \Lambda^2} - 1 \right )
    + m_{P_2}^2 \left (\log {m_{P_2}^2 \over \Lambda^2} - 1 \right ) \right \} \cr
& &\mbox{} + {1 \over 64 \pi^2} \left ({\partial A_P \over \partial h_i} {\partial A_P \over \partial h_j}
    + {\partial A_P \over \partial h_j}
    {\partial W_P \over \partial h_i} \right ) {\log (m_{P_1}^2/ m_{P_2}^2) \over (m_{P_1}^2 - m_{P_2}^2) }
- {(\bigtriangledown_{ij} W_P) \over 64 \pi^2} f (m_{P_1}^2, \ m_{P_2}^2) \cr
& &\mbox{} + {1 \over 64 \pi^2} \left ({\partial W_P \over \partial h_i} {\partial W_P \over \partial h_j} \right )
    {g (m_{P_1}^2, \ m_{P_2}^2) \over (m_{P_1}^2 - m_{P_2}^2)^2}
    + {1 \over 64 \pi^2} \left ({\partial A_P \over \partial h_i}
    {\partial A_P \over \partial h_j } \right ) \log \left ({m_{P_1}^2 m_{P_2}^2 \over \Lambda^4 } \right )    \ ,
\end{eqnarray}
for $i,j$ = 1 to 5, and $m^2_{P_1}$ and $m^2_{P_2}$ are the tree-level masses of pseudoscalar Higgs bosons.
Here, we have introduced $A_P$ and $W_P$ as
\begin{eqnarray}
    A_P & = & {\left ( {\cal M}_{P_{44}} + {\cal M}_{P_{55}} \right ) \over 2 }    \ , \cr
    W_P & = & {{\left ( {\cal M}_{P_{44}} - {\cal M}_{P_{55}} \right )^2 \over 4 } + {\cal M}_{P_{45}}^2}  \ .
\end{eqnarray}
where ${\cal M}_{P_{ij}}$ are given in Appendix G.

Lastly, the contribution of the neutralinos is obtained as
\begin{equation}
    \Delta_{i j}^{{\tilde \chi}^0}  = - \sum_{l = 1}^5
{m_{{\tilde \chi}^0_l}^2 \over 32 \pi^2}
\left (\log {m_{{\tilde \chi}^0_l}^2 \over \Lambda^2} - 1 \right )
{\partial^2 m_{{\tilde \chi}^0_l}^2 \over \partial h_i \partial h_j}
- \sum_{l = 1}^5 {1 \over 32 \pi^2} \log{m_{{\tilde \chi}^0_l}^2 \over \Lambda^2}
    \left ({\partial m_{{\tilde \chi}^0_l}^2 \over \partial h_i} \right )
    \left ({\partial m_{{\tilde \chi}^0_l}^2 \over \partial h_j} \right )  \ .
\end{equation}
where
\begin{equation}
{\partial m_a^2 \over \partial h_i} = -
{{\tilde A}_i m_a^8 + {\tilde B}_i m_a^6 + {\tilde C}_i m_a^4 + {\tilde D}_i m_a^2
    + {\tilde E}_i \over (m_a^2 - m_b^2) (m_a^2 - m_c^2) (m_a^2 - m_d^2) (m_a^2 - m_e^2)} \ ,
\end{equation}
and
\begin{eqnarray}
{\partial^2 m_a^2 \over \partial h_i \partial h_j}
    & = &\mbox{} - {{\tilde A}_{ij} m_a^8 + {\tilde B}_{ij} m_a^6 + {\tilde C}_{ij} m_a^4 + {\tilde D}_{ij} m_a^2
    + {\tilde E}_{ij} \over (m_a^2 - m_b^2) (m_a^2 - m_c^2) (m_a^2 - m_d^2) (m_a^2 - m_e^2) }  \cr
& &\mbox{} + {1 \over (m_a^2 - m_b^2)} \left (
{\partial m_a^2 \over \partial h_i}
{\partial m_b^2 \over \partial h_j}
+ {\partial m_b^2 \over \partial h_i}
{\partial m_a^2 \over \partial h_j} \right ) \cr
& &\mbox{} + {1 \over (m_a^2 - m_c^2)} \left (
{\partial m_a^2 \over \partial h_i}
{\partial m_c^2 \over \partial h_j}
+ {\partial m_c^2 \over \partial h_i}
{\partial m_a^2 \over \partial h_j} \right ) \cr
& &\mbox{} + {1 \over (m_a^2 - m_d^2)} \left (
{\partial m_a^2 \over \partial h_i}
{\partial m_d^2 \over \partial h_j}
+ {\partial m_d^2 \over \partial h_i}
{\partial m_a^2 \over \partial h_j} \right ) \cr
& &\mbox{} + {1 \over (m_a^2 - m_e^2)} \left (
{\partial m_a^2 \over \partial h_i}
{\partial m_e^2 \over \partial h_j}
+ {\partial m_e^2 \over \partial h_i}
{\partial m_a^2 \over \partial h_j} \right ) \ .
\end{eqnarray}
for $i,j$ = 1 to 5.
In the above two expressions, we use the shorthand notations for the neutralino masses:
$m^2_a = m^2_{{\tilde \chi}^0_a}$ and so on. We assume that none of the neutralino masses are equal one another.

We note that $\Delta_{i j}^{{\tilde \chi}^0}$ in above equation is not in its final form, in the sense that
the two CP-odd tadpole minimum conditions should be applied to obtain the explicit results.
The minimum conditions on $h_1$, $h_2$, and $h_4$ are already imposed through the differential operators
$\bigtriangledown_{ij}$.

\section{Numerical Results}

Experimental lower bounds on the Higgs boson masses in various models have been set by the LEP2 data.
They have excluded at the 95 \% confidence level a Higgs boson in the Standard Model,
which is lighter than 114.4 GeV [11].
For the MSSM without explicit CP violation, the lower bounds on the scalar and pseudoscalar Higgs boson masses
have been set
as about 84 GeV and 86 GeV, respectively, by the data [12].
For the MSSM with explicit CP violation, on the other hand, the data have allowed
the possibility of the lightest neutral Higgs boson with a mass of 60-70 GeV [13].

Compared to the MSSM, the Higgs sector of the NMSSM have seemed not utilized the LEP2 data in full.
Some years ago, the analysis of the LEP2 data have shown that a massless scalar Higgs boson might exist
in the NMSSM without explicit CP violation [9].
This analysis have been done when the center of mass energy and the luminosity of the LEP2 have not reached
their final capacity [9].
In the NMSSM with explicit CP violation, the mass of the lightest neutral Higgs boson has been roughly
evaluated with LEP2 data [6].

If CP symmetry be conserved in the Higgs sector of the NMSSM, the five Higgs fields decouple
such that $h_1$, $h_2$, and $h_4$ would become the scalar Higgs bosons while $h_3$ and
$h_5$ the pseudoscalar Higgs bosons.
There would be no mixing between them.
The CP violation is dictated by the six off-diagonal elements of the mass matrix,
$M_{13}$, $M_{23}$, $M_{34}$, $M_{15}$, $M_{25}$, and $M_{45}$, of the neutral Higgs bosons
in the basis $(h_1, h_2, h_3, h_4, h_5)$, which describe the scalar-pseudoscalar mixing.
The remaining off-diagonal elements describe the scalar-scalar mixing or the pseudoscalar-pseudoscalar mixing,
and have nothing to do with the CP violation.

For numerical analysis, we set the fermion masses as $m_t$ = 175 GeV, $m_b$ = 4 GeV, and $m_{\tau}$ = 1.7 GeV,
the gauge boson masses as $m_Z$ = 91.1 GeV and $m_W$ = 80.4 GeV,
and the weak mixing angle as $\sin^2 \theta_W$ = 0.23.
By assuming universal gaugino masses at GUT scale, the low-energy values
of the U(1) and SU(2) gaugino masses has a mass relation of $M_1 = 5/3 \tan^2 \theta_W M_2$.
We take the assumption.

Now, to be specific, we take a representative point in the parameter space:
We choose $\tan \beta$ = 5, $\lambda$ = 0.5, $k$ = 0.4, $A_{\lambda}$ = 590 GeV,
$A_k$ = 5 GeV, $x$ = 178 GeV, $m_Q = m_L$ = 967 GeV, $m_T = m_B = m_E$ = 794 GeV,
$A_t = A_b = A_{\tau}$ = 907 GeV, and $M_2$ = 357 GeV.
The complex phases except the one from neutralino contributions are fixed as
$\phi_t = \phi_b = \phi_{\tau} = \phi_c = 10 \pi/19$.
Then, we are left essentially with two free parameters: $\Lambda$ and $\phi_1$.
The phase $\phi_1$ from the neutralino contributions is allowed to vary as $0 \le \phi_1 < 2 \pi$.

We would like to mention about the dependence on the renormalization scale of the neutral Higgs boson masses.
Physically observable quantities such as the neutral Higgs boson masses should be scale-independent.
However, Fig. 1 shows a weak scale-dependence of the neutral Higgs boson masses,
where $m_{h_i}$ ($i$ = 1 to 5) are plotted against $\Lambda$ for $\phi_1 = 10 \pi/19$, and the other parameter values are
set as above.
This is because we consider only up to the one-loop correction:
If all of the higher-order contributions are further taken into account, the scale-dependence would disappear.
Hereafter, in our analysis, we fix $\Lambda$ as 300 GeV in the one-loop effective potential.

We are interested in the effect of $\phi_1$ on the $5\times5$ neutral Higgs boson mass matrix $M$ and
the $5\times5$ orthogonal mass matrix $O$ that diagonalizes it.
Among the 10 off-diagonal elements of the symmetric $M$, six of them describe the scalar-pseudoscalar mixing:
$M_{13}$, $M_{23}$, $M_{34}$, $M_{15}$, $M_{25}$, and $M_{45}$. The scalar-pseudoscalar mixing,
which is responsible for CP violation in the NMSSM, is triggered by five phases in these matrix elements,
namely, $\phi_t$, $\phi_b$, $\phi_{\tau}$, $\phi_c$, and $\phi_1$.
If $\sin (\phi_t) = \sin (\phi_b) = \sin (\phi_{\tau}) = \sin (\phi_c) = \sin (\phi_1)$ = 0,
these matrix elements would be zero, and the scalar-pseudoscalar mixings are said to be minimal.
Thus, the scalar-pseudoscalar mixings would disappear to restore the CP symmetry, if the Higgs sector
contains no complex phases.
In this case, $O$ would also be simplified as $O_{13}$, $O_{23}$, $O_{34}$, $O_{15}$, $O_{25}$,
and $O_{45}$ would be zero.

In Fig. 2, we first plot the numerical result for the neutral Higgs boson masses, as functions of $\phi_1$,
where $\Lambda$ = 300 GeV and the values of the other parameters are the same as Fig. 1.
The mass of the lightest neutral Higgs boson is about 80 GeV, and roughly stable, for our choice of parameter values.
The mass of the next-to-lightest neutral Higgs boson, on the other hand, is found to vary by 40 \%
for $\phi_1$ between 0 and $2\pi$.
This indicates that the neutralino contributions are not negligible for some parameter values.

We then plot the dependence on $\phi_1$ of $M_{13}$, $M_{23}$, $M_{34}$, $M_{15}$, $M_{25}$, and $M_{45}$
in Figs. 3a and 3b, for the same parameter values as Fig. 2.
The unit is (GeV)$^2$.
In Fig. 3a, three curves for $M_{13}$ (solid curve), $M_{23}$ (dashed curve), and $M_{34}$ (dotted curve)
are displayed, while in Fig. 3b those for $M_{15}$ (solid curve), $M_{25}$ (dashed curve), $M_{45}$ (dotted curve).
It can be seen that the scalar-pseudoscalar mixing elements for the neutral Higgs bosons fluctuate widely
in the NMSSM with explicit CP violation, by the one-loop neutralino contributions.

These results may be compared to the case of the MSSM.
In the case of  the MSSM with explicit CP violation, there is a complex phase and only
two scalar-pseudoscalar mixing elements.
As the relevant complex phase varies between 0 and $2 \pi$,
it is found that one of the scalar-pseudoscalar mixing elements in the MSSM increases as much as twice
its value at zero complex phase,
due to the neutralino contribution [3].

In order to obtain a more comprehensive picture on the amount of mixing between scalar Higgs bosons
and pseudoscalar Higgs bosons,
which is directly linked to the amount of CP violation the the Higgs sector of the NMSSM,
we would not only examine the magnitudes of $M_{13}$, $M_{23}$, $M_{34}$, $M_{15}$, $M_{25}$, and $M_{45}$,
but also examine the relative sizes of them against the other matrix elements.
The ratios of the mixing matrix elements to the non-mixing matrix elements would tell us the relative importance of the
scalar-pseudoscalar mixing elements.

We define the ratios of mixing matrix elements to non-mixing ones by
\[
R_{(p,q)}^{(i,j)} = {|M_{ij}| \over |M_{pq}|} \ ,
\]
where $(i,j)$ are (1,3), (2,3), (3,4), (1,5), (2,5), and (4,5) and $(p, q)$ are (1,2), (1,4), (2,4), and (3,5).
The number of matrix elements that do not mix the CP states is 4: $M_{12}$, $M_{14}$, $M_{24}$, and $M_{35}$.
Since we have 6 mixing matrix elements, the number of possible ratios in the NMSSM is 24,
whereas in the MSSM there are just two ratios.

In Figs. 4a to 4h, we plot all of these 24 ratios of the mixing matrix elements to the non-mixing ones,
as functions of $\phi_1$, for the same parameter values as Fig. 2.
We check that the absolute values of non-mixing elements $M_{12}$, $M_{14}$, $M_{24}$, and $M_{35}$ are much
larger than 1 (GeV)$^2$ for Figs. 4.
These ratios vary largely as $\phi_1$ varies from 0 to $2 \pi$.
On all Figs. 4a to 4h, these ratios reach their peak values at the values of $\phi_1$
where the mixing matrix elements themselves
become extremum values.
>From Figs. 4, one may deduce that for a particular set of parameter values
the scalar-pseudoscalar mixing matrix elements are comparable in size to the non-mixing matrix elements.

On the other hand, we can randomly explore the parameter space to check the values of $R_{(p,q)}^{(i,j)}$.
This is because the results in Figs. 4 might not faithfully represent the whole parameter space of the NMSSM,
since, up to now, our analysis is carried out for a specific set of parameter values (except $\phi_1$).
Thus,  we would like to search in more detail the meaningful region of the parameter space in order to see
if the mass matrix elements can be still comparable in size there.

The boundaries of the region of the parameter space we search are defined by
\[
\begin{array}{l}
\Lambda = 300 \ {\rm GeV} \cr
0 < \phi_t = \phi_b = \phi_{\tau} = \phi_c < 2 \pi \cr
2 \le \tan \beta \le 40 \cr
0 < \lambda, k \le 0.7 \cr
0 < A_{\lambda}, A_k, x, M_2 \le 1000 \ {\rm GeV} \cr
0 < A_t = A_b = A_{\tau} \le 1000 \ {\rm GeV} \cr
0 < m_Q = m_L \le 1000 \ {\rm GeV} \cr
0 < m_T = m_B = m_E \le 1000 \ {\rm GeV}
\end{array}
\]
and
\[
    0 \le \phi_1 < 2 \pi \ .
\]
We evaluate $R_{(p,q)}^{(i,j)}$ at a point in this region, which represent naturally a given set of parameter values.
We take three values of $\phi_1$: $\phi_1 = 0$, $\phi_1 = \pi/3$, and $\phi_1 = 2 \pi/3$.
For each value of $\phi_1$, we choose 20,000 points by Monte Carlo method in the above region,
and at each point 24 $R_{(p,q)}^{(i,j)}$ are evaluated.
Thus, for each value of $\phi_1$, we obtain 480,000 numbers for $R_{(p,q)}^{(i,j)}$.

We put these 480,000 numbers into four categories for convenience:
$R_{(p,q)}^{(i,j)} \ge 1$, $1 > R_{(p,q)}^{(i,j)} \ge 0.1$,  $0.1 > R_{(p,q)}^{(i,j)} \ge 0.01$,
and $0.01 > R_{(p,q)}^{(i,j)}$.
In Table 1, we count them and list the result.
One can see in Table 1 that, for example, for $\phi_1 = 0$, 188 of 480,000 $R_{(p,q)}^{(i,j)}$ are larger than 1.

It can roughly be said that the larger $R_{(p,q)}^{(i,j)}$ are the larger mixings between CP states become.
>From Table 1, it is apparent that we have clearly more large $R_{(p,q)}^{(i,j)}$ for $\phi_1 \ne 0$
than for $\phi_1 =0 $.
The number of large $R_{(p,q)}^{(i,j)}$, larger than 0.1, increases as $\phi_1$ changes from 0 to $2\pi/3$.
This may be interpreted as the effect of  $\phi_1$, the complex phase in the neutralino sector.

\begin{table}[ht]
\caption{
For each of  $\phi_1$ = 0, $\pi/3$, and $2 \pi/3$, we evaluate 24  $R_{(p,q)}^{(i,j)}$ at 20,000 points
in the region of the parameter space defined in the text.
The 480,000 values of  $R_{(p,q)}^{(i,j)}$ are classified into four categories.
We list the counts in each category. }

\begin{center}
\begin{tabular}{c|c|c|c|c}
\hline
\hline
       & $R_{(p,q)}^{(i,j)} \ge 1$ & $1 > R_{(p,q)}^{(i,j)} \ge 0.1$ &  $0.1 > R_{(p,q)}^{(i,j)} \ge 0.01$ & $0.01 > R_{(p,q)}^{(i,j)}$   \\
\hline
\hline
$\phi_1 = 0$ &  188 &  1,525  & 13,614 & 464,673    \\
\hline
$\phi_1 = \pi/3$ &  1,051 & 9,056 & 66,312 & 403,581  \\
\hline
$\phi_1 = 2 \pi/3$ & 1,417  & 12,807 & 73,584 & 392,192 \\
\hline
\hline
\end{tabular}
\end{center}
\end{table}

Now, we turn our attention to the $5\times5$ orthogonal matrix that diagonalizes the mass matrix
of the neutral Higgs bosons.
This orthogonal matrix also contains pieces of information about the CP violation and the mixing
between scalar and pseudoscalar Higgs bosons.
The elements $O_{ij}$ determine the couplings of the physical neutral Higgs bosons to other fields in the NMSSM.
In Figs. 5a and 5b, we plot as a function of $\phi_1$, for the same parameter values as Fig. 2,
the squares of  $O_{13}$, $O_{23}$, $O_{34}$, $O_{15}$, $O_{25}$, and $O_{45}$, which link the CP-even and CP-odd states.
One can see that $O_{34}^2$, $O_{15}^2$, and $O_{25}^2$ fluctuate widely for $0 \le \phi_1 < 2\pi$,
while the remaining ones remain nearly zero.
Therefore, for the particular set of parameter values,
it is possible to expect that neutralino contributions may affect seriously the scalar-pseudoscalar mixings in the NMSSM.
Note that, in our analysis, $\phi_1 = 0$ does not imply the absence of CP violation,
because there are other non-zero complex phases, i.e., $\phi_t = \phi_b = \phi_{\tau} = \phi_c = 10 \pi/19$.

Meanwhile, one can evaluate effectively the size of the CP violation by measuring the dimensionless parameter [6]
\[
\rho = 5^5 O_{11}^2 O_{21}^2 O_{31}^2 O_{41}^2 O_{51}^2 \ ,
\]
The range of $\rho$ is from 0 to 1 since
the elements of the transformation matrix satisfy the orthogonality condition of $\sum_{j = 1}^{5} O_{j1}^2$ = 1.
If $\rho = 0$, there is no CP violation in the Higgs sector of the NMSSM.
On the other hand, if $\rho = 1$, CP symmetry is maximally violated.
The maximal CP violation that leads to $\rho$ = 1 takes place
when $O_{11}^2 = O_{21}^2 = O_{31}^2 = O_{41}^2 = O_{51}^2$ = 1/5.
In Fig. 6, we plot $\rho$, as a function of $\phi_1$, for the same parameter values as Fig. 2.
One can observe that $\rho$ reaches to $\sim 10^{-2}$ for certain values of $\phi_1$.
This shows that it is possible for the neutralino sector to play a dominant role in the CP violation
in the NMSSM at the one-loop level,
through the scalar-pseudoscalar mixings, at least for the particular set of parameter values.

Here, too, we explore the parameter space randomly and check the value of $\rho$.
The regions of the parameter space is the same as the aforementioned one for $R_{(p,q)}^{(i,j)}$.
We take 19 values of $\phi_1$ from 0 to $2\pi$.
For each value of $\phi_1$, we choose 20,000 points in the region and evaluate $\rho$, and take
the largest value among them.
The result is shown in Fig. 7, where the maximum value of $\rho$, say $\rho_{\rm max}$, for each $\phi_1$ is plotted.
We see that $\rho_{\rm max}$ is considerably large for a wide range of $\phi_1$.
Thus, the CP violation can be significantly affected by the presence of $\phi_1$.
Without the neutralino contributions in the NMSSM, the explicit CP violation
at the one-loop level might be considerably reduced.

We now estimate the dependence on $\phi_1$ of the mass of the lightest neutral Higgs boson $m_{h_1}$ in more detail.
We first choose randomly a point in the region of the parameter space defined above.
We check at this point if the condition of $m_{h_{(n+1)}} - m_{h_{(n)}} > 10$ GeV, ($n$ = 1 to 4) is satisfied.
Then, we change only the value of $\phi_1$ arbitrarily while fixing other parameter values of the point:
This defines another point.
We check at the new point too if the condition of $m_{h_{(n+1)}} - m_{h_{(n)}} > 10$ GeV, ($n$ = 1 to 4) is satisfied.
If so, we keep the two points as a pair.
Let the values of $\phi_1$ of a given pair of the two points be $c_1$ and $c_2$.

Now, at each of the the two points, we evaluate the mass of the lightest neutral Higgs boson, $m_{h_1}$, and
introduce the ratio of the mass values as
\[
    R_1 = {m_{h_1} (\phi_1 = c_1) \over m_{h_1} (\phi_1 = c_2)}
\]
for $m_{h_1} (\phi_1 = c_1) > m_{h_i} (\phi_1 = c_2)$, and
\[
        R_1 = {m_{h_1} (\phi_1 = c_2) \over m_{h_1} (\phi_1 = c_1)}
\]
for $m_{h_1} (\phi_1 = c_1) \le m_{h_1} (\phi_1 = c_2)$.
In this way, we always have $R_1 \ge 1$.
Note that $R_1 = 1$ implies the independence of the lightest neutral Higgs boson mass on $\phi_1$, i.e.,
on the neutralino contributions.

We select $20,000$ pairs of points randomly in the aforementioned region of the parameter space and
evaluate $R_1$ for each pair.
We plot the 20,000 points of $R_1$ against $C_3 = |c_2 - c_1|$ in Fig. 8.
One can see in Fig. 8 that only a few of the points are scattered away from the line of $R_1 =1$.
This indicates that  for most of the parameter region the $\phi_1$ dependence of the lightest neutral
Higgs boson mass is weak.
In other words, the neutralino contributions to the lightest neutral Higgs boson mass in the NMSSM are rather small.
The masses of other neutral Higgs bosons behave similarly.
Thus, one might say that the masses of the five neutral Higgs bosons are roughly independent on the CP phase
of the neutralino sector.

\section{Conclusions}

We have investigated the possibility of explicit CP violation at the one-loop level in the neutral Higgs sector
of the NMSSM, where
the tree-level Higgs potential is assumed to possess CP symmetry.
The mixing between the scalar and pseudoscalar Higgs bosons, which induce the CP violation,
occurs via the radiative corrections at the one-loop level.
For the radiative corrections, we consider the loop contributions due to top quark, bottom quark,
tau lepton, $W$ boson,
$Z$ boson, the charged Higgs boson, five the neutral Higgs bosons and their superpartners.
With those contributions, five phases appear. They are: $\phi_t$ from the scalar top quark sector,
$\phi_b$ from the scalar bottom quark
sector, $\phi_{\tau}$ from the tau lepton sector, $\phi_c$ from the chargino sector, and $\phi_1$
from the neutralino sector.

The effects of CP violation in the NMSSM at the one-loop level may manifest themselves in many places:
The mass matrix of the neutral Higgs bosons contain some elements that mix CP-even and CP-odd fields.
Some elements of the orthogonal matrix that diagonalizes the mass matrix of the neutral Higgs bosons have
also carry information
about the CP violation. They will be zero if CP is conserved.
The five complex phases are all responsible for, and contribute to, the CP violation.
We pay attention to $\phi_1$, and examine the size of contributions from the neutralino sector
to the matrix elements relevant for the CP violation in the NMSSM at the one-loop level.

As $\phi_1$ is allowed to vary from 0 to $2\pi$ while the other complex phases are fixed,
we can see the dependence on $\phi_1$ of
the CP violation. From Figs. 2 to Fig. 6, we observe that for a certain set of parameter values most
of the relevant quantities
that we examine depend  significantly on $\phi_1$.
One of the scalar-pseudoscalar mixing matrix elements change from nearly zero to as large as more
than 10,000 (GeV)$^2$, depending on $\phi_1$.
Actually, the absolute values of these scalar-pseudoscalar mixing matrix elements do not convey
more useful information than the
relative sizes among the mixing matrix elements and non-mixing matrix elements.

The ratios of the mixing matrix elements and non-mixing matrix elements are examined.
They also depend clearly on $\phi_1$, exhibiting the neutralino contributions.
Some of them approach to 1, for certain value of $\phi_1$, and many of them are larger than 0.01
for a wide range of $\phi_1$.
The results imply that the scalar-pseudoscalar mixing matrix elements may be comparable
in sizes to the non-mixing matrix elements,
for a certain set of parameter values in the NMSSM, due to the neutralino contributions at the one-loop level.
The masses themselves of the five neutral Higgs bosons are not so affected by $\phi_1$,
at least for the certain set of parameter values.

The elements of the orthogonal matrix that diagonalizes the mass matrix of the neutral Higgs bosons
carry more direct information on the CP violation
than the mass matrix itself.
As can be seen in Figs. 5, some of the matrix elements $O_{ij}^2$ fluctuate very widely between 0 and 1,
depending on $\phi_1$.
Moreover, $\rho$, which is a quantity that measures directly the amount of the CP violation, is also checked.
It is found that $\rho$ is larger than 0.01 for some values of $\phi_1$, whereas $\rho$ is nearly zero when $\phi_1 =0$.
Thus, it is reasonable to expect that the neutralino sector at the one-loop level dominantly contributes
the CP violation in the NMSSM,  for a certain set of parameter values.

We also explore a meaningful region of the parameter space by Monte Carlo method in order to
check the above observations are
not confined to the particular set of parameter values.
In the parameter region, we pick randomly 20,000 points for given $\phi_1$ to evaluate $\rho$ at them.
We choose the largest of them and denote it $\rho_{\rm max}$ for given $\phi_1$.
It is found that $\rho_{\rm max}$ shows a clear dependence on $\phi_1$.
Moreover, $\rho_{\rm max}$ at $\phi_1 =0$ is definitely smaller than $\rho_{\rm max}$ at non-zero $\phi_1$ values.
Actually, for some values of $\phi_1$, $\rho_{\rm max}$ close to as large as 0.29.
To conclude, we find that for some parameter values in the NMSSM with explicit CP violation at the one-loop level
the neutralino contributions to the scalar-pseudoscalar mixing exhibit relatively significant effects.

\vskip 0.3 in

\noindent
{\large {\bf Acknowledgments}}
\vskip 0.2 in
\noindent
This work was supported by Korea Research Foundation Grant (2001-050-D00005).

\vskip 0.2 in

\vfil\eject
\vskip 0.2 in
\noindent
{\large {\bf Appendix A}}
\vskip 0.2 in
\noindent
The elements for the field-dependent mass matrix ${\cal M}_{{\tilde \chi}^0} {\cal M}_{{\tilde \chi}^0}^{\dagger}$
of the neutralinos at the tree level are
\begin{eqnarray}
{\cal M}_{{{\tilde \chi}^0}_{11}} & = &
{ { {g_1^2 \over 2}\left( |H_1^0|^2 + |H_2^0|^2 \right)}
+ {1\over 4} M_2^2}  \ , \cr
 & & \cr
{\cal M}_{{{\tilde \chi}^0}_{22}} & = &
{ {g_2^2\over 2}\left( |H_1^0|^2 + |H_2^0|^2 \right) + {M_2^2 }} \ , \cr
 & & \cr
{\cal M}_{{{\tilde \chi}^0}_{33}} & = &
{{1\over 2}{\left( g_1^2 + g_2^2 \right) |H_1^0|^2}
+ \lambda^2 \left(|N|^2 + |H_2^0|^2 \right)} \ , \cr
 & & \cr
{\cal M}_{{{\tilde \chi}^0}_{44}} & = &
{{1 \over 2}{\left( g_1^2 + g_2^2 \right) |H_2^0|^2 }
+ \lambda^2 \left( |N|^2 + |H_1^0|^2 \right)} \ , \cr
 & & \cr
{\cal M}_{{{\tilde \chi}^0}_{55}} & = &
{4 k^2 |N|^2
+\lambda^2 \left( |H_1^0|^2 + |H_2^0|^2  \right)} \ , \cr
 & & \cr
{\cal M}_{{{\tilde \chi}^0}_{12}} & = &
{ -{g_1 g_2 \over 2 }\left (|H_1^0|^2 + |H_2^0|^2  \right )} \ , \cr
 & & \cr
{\cal M}_{{{\tilde \chi}^0}_{13}} & = &
{ {g_1 \over 2\sqrt{2}}
\left( 2 \lambda H_2^0 N^* + {H_1^0}^* M_1 e^{i{\phi_1}} \right)} \ , \cr
& & \cr
{\cal M}_{{{\tilde \chi}^0}_{14}} & = &
{ -{g_1 \over 2\sqrt{2}}
\left( 2 \lambda H_1^0 N^* + {H_2^0}^* M_1 e^{i{\phi_1}} \right)} \ , \cr
 & & \cr
{\cal M}_{{{\tilde \chi}^0}_{15}} & = &
{ {g_1\over \sqrt{2}} \lambda
\left( {H_1^0}^* H_2^0 - H_1^0 {H_2^0}^*  \right)}
\ , \cr
 & & \cr
{\cal M}_{{{\tilde \chi}^0}_{23}} & = &
{-{g_2\over \sqrt{2}}
\left( {H_1^0}^* M_2 + \lambda N^* H_2^0 \right)} \ , \cr
 & & \cr
{\cal M}_{{{\tilde \chi}^0}_{24}} & = &
{{g_2 \over \sqrt{2}} \left( {H_2^0}^* M_2 + \lambda N^* H_1^0 \right)} \ , \cr
 & & \cr
{\cal M}_{{{\tilde \chi}^0}_{25}} & = &
{{g_2 \over \sqrt{2}} \lambda
\left( H_1^0 {H_2^0}^* - {H_1^0}^* H_2^0 \right)} \ , \cr
 & & \cr
{\cal M}_{{{\tilde \chi}^0}_{34}} & = &
{- {1 \over 2}{\left( g_1^2 + g_2^2 \right) H_1^0 {H_2^0}^* }
+ \lambda^2 {H_1^0}^* H_2^0} \ , \cr
 & & \cr
{\cal M}_{{{\tilde \chi}^0}_{35}} & = &
{\lambda \left( \lambda N {H_1^0}^* - 2 k N^* H_2^0\right)} \ , \cr
 & & \cr
{\cal M}_{{{\tilde \chi}^0}_{45}} & = &
{\lambda \left( \lambda N {H_2^0}^* - 2 k N^* H_1^0 \right)} \ .
\end{eqnarray}
\noindent
{\large {\bf Appendix B}}
\vskip 0.2 in
\noindent
The elements for the mass matrix of the neutral Higgs bosons due to the radiative contributions
of the top quark and scalar top quarks are
\begin{eqnarray}
M_{11}^t & = & {3 \over 8 \pi^2 v^2}
\left ( {m_t^2 \lambda x \Delta_{{\tilde t}_1} \over \sin \beta}
+ {\cos \beta \Delta_{\tilde t} \over 2} \right )^2
{g(m_{{\tilde t}_1}^2, \ m_{{\tilde t}_2}^2)
\over (m_{{\tilde t}_2}^2 - m_{{\tilde t}_1}^2)^2}
+ {3 m_Z^4 \cos^2 \beta \over 128 \pi^2 v^2}
\log \left ({m_{{\tilde t}_1}^2  m_{{\tilde t}_2}^2 \over \Lambda^4} \right ) \cr
& & \cr
& &\mbox{} - {3 \cos^2 \beta \over 16 \pi^2 v^2}
\left( {4 m_W^2 \over 3} - {5 m_Z^2 \over 6} \right)^2
f(m_{{\tilde t}_1}^2, \ m_{{\tilde t}_2}^2) \cr
& & \cr
& &\mbox{}
+ {3 m_Z^2 \cos \beta \over 16 \pi^2 v^2}
\left ( {m_t^2 \lambda x \Delta_{{\tilde t}_1} \over \sin \beta}
+ {\cos \beta \Delta_{\tilde t} \over 2} \right )
{\displaystyle \log (m_{{\tilde t}_2}^2 / m_{{\tilde t}_1}^2)
 \over (m_{{\tilde t}_2}^2 - m_{{\tilde t}_1}^2)}
 \ , \cr
& & \cr
M_{22}^t & = & {3 \over 8 \pi^2 v^2}
\left ( {m_t^2 A_t \Delta_{{\tilde t}_2} \over \sin \beta}
- {\sin \beta \Delta_{\tilde t} \over 2} \right )^2
{g(m_{{\tilde t}_1}^2, \ m_{{\tilde t}_2}^2)
\over(m_{{\tilde t}_2}^2 - m_{{\tilde t}_1}^2)^2}
- {3 m_t^4 \over 4 \pi^2 v^2 \sin^2 \beta} \log \left ({m_t^2 \over \Lambda^2} \right ) \cr
& & \cr
& &\mbox{} - {3 \sin^2 \beta \over 16 \pi^2 v^2}
\left( {4 m_W^2 \over 3} - {5 m_Z^2 \over 6} \right)^2
f(m_{{\tilde t}_1}^2, \ m_{{\tilde t}_2}^2) \cr
& & \cr
& &\mbox{} + {3 \sin \beta \over 16 \pi^2 v^2}
\left ({4 m_t^2 \over \sin^2 \beta} - m_Z^2 \right)
\left ( {m_t^2 A_t \Delta_{{\tilde t}_2} \over \sin \beta}
- {\sin \beta \Delta_{\tilde t} \over 2} \right )
{\displaystyle \log (m_{{\tilde t}_2}^2 / m_{{\tilde t}_1}^2)
 \over (m_{{\tilde t}_2}^2 - m_{{\tilde t}_1}^2)} \cr
& & \cr
& &\mbox{}
 + {3 \over 32 \pi^2 v^2}
\left ({2 m_t^2 \over \sin \beta} - {m_Z^2 \sin \beta \over 2} \right)^2
\log \left ({m_{{\tilde t}_1}^2  m_{{\tilde t}_2}^2 \over \Lambda^4} \right ) \ , \cr
& & \cr
M_{33}^t & = & {3 m_t^4 A_t^2 \lambda^2 x^2 \sin^2 \phi_t \over 8 \pi^2 v^2 \sin^4 \beta}
{g(m_{{\tilde t}_1}^2, \ m_{{\tilde t}_2}^2) \over (m_{{\tilde t}_2}^2
- m_{{\tilde t}_1}^2 )^2} \ , \cr
& & \cr
M_{44}^t & = & {3 m_t^4 \lambda^2 {\Delta_{{\tilde t}_1}^2}
\over 8 \pi^2 \tan^2 \beta}
{g(m_{{\tilde t}_1}^2, \ m_{{\tilde t}_2}^2) \over (m_{{\tilde t}_2}^2 - m_{{\tilde t}_1}^2)^2 }  \ , \cr
 & & \cr
M_{55}^t & = & {3 m_t^4 A_t^2 \lambda^2 \sin^2 \phi_t \over 8 \pi^2 \tan^2 \beta}
{g(m_{{\tilde t}_1}^2, \ m_{{\tilde t}_2}^2) \over (m_{{\tilde t}_2}^2 - m_{{\tilde t}_1}^2 )^2}  \ , \cr
& & \cr
M_{12}^t & = & {3 \over 8 \pi^2 v^2}
\left ( {m_t^2 \lambda x \Delta_{{\tilde t}_1}
\over \sin \beta} + {\cos \beta \Delta_{\tilde t} \over 2} \right )
\left ( {m_t^2 A_t \Delta_{{\tilde t}_2}\over \sin \beta}
- {\sin \beta \Delta_{\tilde t} \over 2} \right )
{g(m_{{\tilde t}_1}^2, \ m_{{\tilde t}_2}^2)
\over (m_{{\tilde t}_2}^2 - m_{{\tilde t}_1}^2)^2} \cr
& & \cr
& & \mbox{} + {3 \sin 2 \beta \over 32 \pi^2 v^2}
\left ({4 m_W^2 \over 3} - {5 m_Z^2 \over 6} \right)^2
f(m_{{\tilde t}_1}^2, \ m_{{\tilde t}_2}^2)
\cr & &
\cr & &\mbox{} + {3 \sin \beta \over 32 \pi^2 v^2}
\left ({4 m_t^2 \over \sin^2 \beta} - m_Z^2 \right)
\left ({m_t^2 \lambda x \Delta_{{\tilde t}_1} \over \sin \beta}
+ {\cos \beta \Delta_{\tilde t} \over 2} \right )
{\displaystyle \log (m_{{\tilde t}_2}^2 / m_{{\tilde t}_1}^2) \over (m_{{\tilde t}_2}^2 - m_{{\tilde t}_1}^2)} \cr
& & \cr
& &\mbox{} + {3 m_Z^2 \cos \beta \over 32 \pi^2 v^2}
\left ({m_t^2 A_t \Delta_{{\tilde t}_2} \over \sin \beta}
- {\sin \beta \Delta_{\tilde t} \over 2} \right )
{\displaystyle \log (m_{{\tilde t}_2}^2 / m_{{\tilde t}_1}^2)
 \over (m_{{\tilde t}_2}^2 - m_{{\tilde t}_1}^2)} \cr
& & \cr
& &\mbox{}
+ {3 m_Z^2 \sin 2 \beta \over 256 \pi^2 v^2}
\left ({4 m_t^2 \over \sin^2 \beta} - m_Z^2 \right)
\log \left ({m_{{\tilde t}_1}^2 m_{{\tilde t}_2}^2 \over \Lambda^4} \right ) \ , \cr
& & \cr
M_{13}^t & = & \mbox{} - {3 m_t^2 A_t \lambda x \sin \phi_t \over 8 \pi^2 v^2 \sin^2 \beta}
\left ({m_t^2 \lambda x \Delta_{{\tilde t}_1} \over \sin \beta}
+ {\cos \beta \Delta_{\tilde t} \over 2} \right )
{g(m_{{\tilde t}_1}^2, \ m_{{\tilde t}_2}^2) \over (m_{{\tilde t}_2}^2 - m_{{\tilde t}_1}^2)^2 }  \cr
& & \cr
& &\mbox{} - {3 m_Z^2 m_t^2 A_t \lambda x \sin \phi_t \over 32 \pi^2 v^2 \sin \beta \tan \beta}
{\log ({m_{{\tilde t}_2}^2 / m_{{\tilde t}_1}^2}) \over (m_{{\tilde t}_2}^2 - m_{{\tilde t}_1}^2) } \ , \cr
& & \cr
M_{14}^t & = & {3 m_t^2 \lambda \Delta_{{\tilde t}_1}\over 8 \pi^2 v \tan \beta} \left ( {m_t^2 \lambda x \Delta_{{\tilde t}_1} \over \sin \beta}
+ {\cos \beta \Delta_{\tilde t} \over 2} \right )
{g(m_{{\tilde t}_1}^2, \ m_{{\tilde t}_2}^2) \over (m_{{\tilde t}_2}^2 - m_{{\tilde t}_1}^2)^2 }
- {3 m_t^2 \lambda^2 x \cot \beta \over 8 \pi^2 v \sin \beta}
f(m_{{\tilde t}_1}^2, \ m_{{\tilde t}_2}^2)  \cr
& & \cr
& &\mbox{} + {3 m_Z^2 m_t^2 \lambda \cos \beta\Delta_{{\tilde t}_1}
\over 32 \pi^2 v \tan \beta} {\log( {m_{{\tilde t}_2}^2 / m_{{\tilde t}_1}^2})
\over (m_{{\tilde t}_2}^2 - m_{{\tilde t}_1}^2)} \ , \cr
& & \cr
M_{15}^t & = & \mbox{} - {3 m_t^2 A_t \lambda \sin \phi_t \over 8 \pi^2 v \tan \beta}
\left ({m_t^2 \lambda x \Delta_{{\tilde t}_1} \over \sin \beta}
+ {\cos \beta \Delta_{\tilde t} \over 2} \right )
{g(m_{{\tilde t}_1}^2, \ m_{{\tilde t}_2}^2) \over (m_{{\tilde t}_2}^2 - m_{{\tilde t}_1}^2)^2 }  \cr
& & \cr
& &\mbox{} - {3 m_Z^2 m_t^2 A_t \lambda \cos \beta \sin \phi_t \over 32 \pi^2 v \tan \beta}
\ {\log ({m_{{\tilde t}_2}^2 / m_{{\tilde t}_1}^2}) \over (m_{{\tilde t}_2}^2 - m_{{\tilde t}_1}^2) } \ , \cr
& & \cr
M_{23}^t & = & \mbox{} - {3 m_t^2 A_t \lambda x \sin \phi_t \over 8 \pi^2 v^2 \sin^2 \beta}
\left ({ m_t^2 A_t \Delta_{{\tilde t}_2} \over \sin \beta} - {\sin \beta \Delta_{\tilde t} \over 2} \right )
{g(m_{{\tilde t}_1}^2, \ m_{{\tilde t}_2}^2) \over (m_{{\tilde t}_2}^2 - m_{{\tilde t}_1}^2)^2 } \cr
& & \cr
& &\mbox{} - {3 m_t^2 \lambda x A_t \sin \phi_t \over 16 \pi^2 v^2 \sin \beta}
\left ( {2 m_t^2 \over \sin^2 \beta} - {m_Z^2 \over 2}\right )
{\log (m_{{\tilde t}_2}^2 / m_{{\tilde t}_1}^2)
\over (m_{{\tilde t}_2}^2 - m_{{\tilde t}_1}^2)} \ , \cr
& & \cr
M_{24}^t & = & {3 m_t^2 \lambda \Delta_{{\tilde t}_1} \over 8 \pi^2 v \tan \beta} \left( {m_t^2 A_t \Delta_{{\tilde t}_2} \over \sin \beta}
 - {\sin \beta \Delta_{\tilde t} \over 2} \right )
{g(m_{{\tilde t}_1}^2, \ m_{{\tilde t}_2}^2) \over (m_{{\tilde t}_2}^2 - m_{{\tilde t}_1}^2)^2} \cr
& & \cr
& &\mbox{} + {3 m_t^2 \lambda \cos \beta \Delta_{{\tilde t}_1} \over 16 \pi^2 v} \left ({2 m_t^2 \over \sin^2 \beta} - {m_Z^2 \over 2} \right)
{\log (m_{{\tilde t}_2}^2 / m_{{\tilde t}_1}^2) \over (m_{{\tilde t}_2}^2 - m_{{\tilde t}_1}^2)} , \cr
& & \cr
M_{25}^t & = & \mbox{} - {3 m_t^2 \lambda A_t \sin \phi_t \over 8 \pi^2 v \tan \beta}
\left ({m_t^2 A_t \Delta_{{\tilde t}_2} \over \sin \beta} - {\sin \beta \Delta_{\tilde t} \over 2} \right )
{g(m_{{\tilde t}_1}^2, \ m_{{\tilde t}_2}^2) \over (m_{{\tilde t}_2}^2 - m_{{\tilde t}_1}^2)^2 }  \cr
& & \cr
& &\mbox{}
- {3 m_t^2 \lambda A_t \sin \phi_t \over 16 \pi^2 v \sin \beta \tan \beta}
\left (2 m_t^2 - {m_Z^2 \sin^2 \beta \over 2} \right )
{\log (m_{{\tilde t}_2}^2 / m_{{\tilde t}_1}^2)
\over (m_{{\tilde t}_2}^2 - m_{{\tilde t}_1}^2)} \ , \cr
& & \cr
M_{34}^t & = & \mbox{} - {3 m_t^4 \lambda^2 x A_t \sin \phi_t \Delta_{{\tilde t}_1} \over 8 \pi^2 v \sin^2 \beta \tan \beta}
{g(m_{{\tilde t}_1}^2, \ m_{{\tilde t}_2}^2)
\over (m_{{\tilde t}_2}^2 - m_{{\tilde t}_1}^2)^2 } \ , \cr
& & \cr
M_{35}^t & = & \mbox{} {3 m_t^4 \lambda^2 x A_t^2 \sin^2 \phi_t \over 8 \pi^2 v \sin^2 \beta \tan \beta}
{g(m_{{\tilde t}_1}^2, \ m_{{\tilde t}_2}^2) \over (m_{{\tilde t}_2}^2 - m_{{\tilde t}_1}^2)^2 }  \ , \cr
M_{45}^t & = & \mbox{} - {3 m_t^4 \lambda^2 A_t \sin \phi_t \Delta_{{\tilde t}_1} \over 8 \pi^2 \tan^2 \beta}
{g(m_{{\tilde t}_1}^2, \ m_{{\tilde t}_2}^2) \over (m_{{\tilde t}_2}^2 - m_{{\tilde t}_1}^2)^2 } \ ,
\end{eqnarray}
where
\begin{eqnarray}
\Delta_{{\tilde t}_1} & = & A_t \cos \phi_t + \lambda x \cot \beta  \  , \cr
\Delta_{{\tilde t}_2} & = & A_t + \lambda x \cot \beta \cos \phi_t \ , \cr
\Delta_{\tilde t} & = & \left ( {4 m_W^2 \over 3} - {5 m_Z^2 \over 6} \right)
        \left \{(m_Q^2 - m_T^2) + \left ( {4 m_W^2 \over 3} - {5 m_Z^2 \over 6} \right) \cos 2 \beta \right \}  \ ,
\end{eqnarray}
\[
g(m_1^2,m_2^2) = {m_2^2 + m_1^2 \over m_1^2 - m_2^2} \log {m_2^2 \over m_1^2} + 2 \ .
\]
\vskip 0.2 in
\noindent
{\large {\bf Appendix C}}
\vskip 0.2 in
\noindent
The elements for the mass matrix of the neutral Higgs bosons due to the radiative contributions of the bottom quark
and scalar bottom quarks are
\begin{eqnarray}
M_{11}^b & = & {3 \over 8 \pi^2 v^2}
\left ( {m_b^2 A_b \Delta_{{\tilde b}_1}
\over \cos \beta} + {\cos \beta \Delta_{\tilde b} \over 2} \right )^2
{ g(m_{{\tilde b}_1}^2,\ m_{{\tilde b}_2}^2) \over (m_{{\tilde b}_2}^2 - m_{{\tilde b}_1}^2)^2}
- {3 m_b^4 \over 4 \pi^2 v^2 \cos^2 \beta} \log \left ({m_b^2 \over \Lambda^2} \right) \cr
& & \cr
& &\mbox{} - {3 \cos^2 \beta \over 16 \pi^2 v^2}
\left( { m_Z^2 \over 6} - {2 m_W^2 \over 3} \right)^2
f(m_{{\tilde b}_1}^2, \ m_{{\tilde b}_2}^2) \cr
& & \cr
& & \mbox{} + {3 \cos \beta \over 16 \pi^2 v^2}
\left ({4 m_b^2 \over \cos^2 \beta} - m_Z^2 \right)
\left ( {m_b^2 A_b \Delta_{{\tilde b}_1} \over \cos \beta} + {\cos \beta \Delta_{\tilde b} \over 2} \right )
{\displaystyle \log (m_{{\tilde b}_2}^2 / m_{{\tilde b}_1}^2)
\over (m_{{\tilde b}_2}^2 - m_{{\tilde b}_1}^2)} \cr
& & \cr
& &\mbox{} + {3 \over 32 \pi^2 v^2}
\left ({2 m_b^2 \over \cos  \beta} - {m_Z^2 \cos \beta \over 2} \right)^2
\log \left ( {m_{{\tilde b}_1}^2  m_{{\tilde b}_2}^2 \over \Lambda^4} \right )\ , \cr
& & \cr
M_{22}^b & = & {3 \over 8 \pi^2 v^2}
\left \{ {m_b^2 \lambda x \Delta_{{\tilde b}_2} \over \cos \beta}
- {\sin \beta \Delta_{\tilde b} \over 2} \right \}^2
{g(m_{{\tilde b}_1}^2, \ m_{{\tilde b}_2}^2) \over (m_{{\tilde b}_2}^2 - m_{{\tilde b}_1}^2)^2}
+ {3 m_Z^4 \sin^2 \beta \over 128 \pi^2 v^2}
\log \left ({m_{{\tilde b}_1}^2  m_{{\tilde b}_2}^2 \over \Lambda^4} \right ) \cr
& & \cr
& &\mbox{} - {3 \sin^2 \beta \over 16 \pi^2 v^2}
\left( { m_Z^2 \over 6} - {2 m_W^2 \over 3} \right)^2
f(m_{{\tilde b}_1}^2, \ m_{{\tilde b}_2}^2)   \cr
& & \cr
& &\mbox{} + {3 m_Z^2 \sin \beta \over 16 \pi^2 v^2} \left \{ {m_b^2 \lambda x \Delta_{{\tilde b}_2} \over \cos \beta} - {\sin \beta \Delta_{\tilde b} \over 2} \right \} {\displaystyle \log (m_{{\tilde b}_2}^2 / m_{{\tilde b}_1}^2)
 \over (m_{{\tilde b}_2}^2 - m_{{\tilde b}_1}^2)}   \ , \cr
& & \cr
M_{33}^b & = & {3 m_b^4 \lambda^2 x^2 A_b^2 \sin^2 \phi_b \over 8 \pi^2 v^2 \cos^4 \beta} {g(m_{{\tilde b}_1}^2, \ m_{{\tilde b}_2}^2) \over (m_{{\tilde b}_2}^2 - m_{{\tilde b}_1}^2 )^2} \ , \cr
& & \cr
M_{44}^b & = & {3 m_b^4 \lambda^2 {\Delta_{{\tilde b}_2}^2} \over 8 \pi^2 \cot^2 \beta} {g(m_{{\tilde b}_1}^2, \ m_{{\tilde b}_2}^2) \over (m_{{\tilde b}_2}^2 - m_{{\tilde b}_1}^2)^2}   \ , \cr
& & \cr
M_{55}^b & = & {3 m_b^4 \lambda^2 A_b^2 \sin^2 \phi_b \over 8 \pi^2 \cot^2 \beta} {g(m_{{\tilde b}_1}^2, \ m_{{\tilde b}_2}^2) \over (m_{{\tilde b}_2}^2 - m_{{\tilde b}_1}^2 )^2}   \ , \cr
& & \cr
M_{12}^b & = & {3 \over 8 \pi^2 v^2} \left \{ {m_b^2 \lambda x \Delta_{{\tilde b}_2} \over \cos \beta} - {\sin \beta \Delta_{\tilde b} \over 2} \right \} \left \{ {m_b^2 A_b \Delta_{{\tilde b}_1} \over \cos \beta} + {\cos \beta \Delta_{\tilde b}\over 2} \right \}
{g(m_{{\tilde b}_1}^2, \ m_{{\tilde b}_2}^2) \over (m_{{\tilde b}_2}^2 - m_{{\tilde b}_1}^2)^2}  \cr
& & \cr
& & \mbox{} + {3 \sin 2 \beta \over 32 \pi^2 v^2}
\left ({ m_Z^2 \over 6} - {2 m_W^2 \over 3} \right)^2
f(m_{{\tilde b}_1}^2, \ m_{{\tilde b}_2}^2)  \cr
& & \cr
& &\mbox{} + {3 \cos \beta \over 32 \pi^2 v^2} \left ({4 m_b^2 \over \cos^2 \beta} - m_Z^2 \right)
\left \{{m_b^2 \lambda x \Delta_{{\tilde b}_2} \over \cos \beta} - {\sin \beta \Delta_{\tilde b} \over 2} \right \} {\displaystyle \log (m_{{\tilde b}_2}^2 / m_{{\tilde b}_1}^2) \over (m_{{\tilde b}_2}^2 - m_{{\tilde b}_1}^2)} \cr
& & \cr
& &\mbox{} + {3 m_Z^2 \sin 2 \beta \over 64 \pi^2 v^2}
\left \{{m_b^2 A_b \Delta_{{\tilde b}_1} \over \cos^2 \beta} + {\Delta_{\tilde b} \over 2} \right \}
{\displaystyle \log (m_{{\tilde b}_2}^2 / m_{{\tilde b}_1}^2) \over (m_{{\tilde b}_2}^2 - m_{{\tilde b}_1}^2)} \cr
& & \cr
& &\mbox{} + {3 m_Z^2 \sin 2 \beta \over 256 \pi^2 v^2}
\left ({4 m_b^2 \over \cos^2 \beta} - m_Z^2 \right) \log \left ({m_{{\tilde b}_1}^2 m_{{\tilde b}_2}^2 \over \Lambda^4} \right ) \ , \cr
& & \cr
M_{13}^b & = & \mbox{} - {3 m_b^2 \lambda x A_b \sin \phi_b \over 8 \pi^2 v^2 \cos^2 \beta}\left \{ {m_b^2 \ A_b \Delta_{{\tilde b}_1 } \over \cos \beta}
+ {\cos \beta \Delta_{\tilde b} \over 2} \right \}
{g(m_{{\tilde b}_1}^2, \ m_{{\tilde b}_2}^2) \over (m_{{\tilde b}_2}^2 - m_{{\tilde b}_1}^2)^2}  \cr
& & \cr
& &\mbox{} - {3 m_b^2 \lambda x A_b \sin \phi_b \over 16 \pi^2 v^2 \cos \beta}
\left ({2m_b^2 \over \cos^2 \beta} - {m_Z^2 \over 2} \right )
{\displaystyle \log (m_{{\tilde b}_2}^2 / m_{{\tilde b}_1}^2) \over (m_{{\tilde b}_2}^2 - m_{{\tilde b}_1}^2)} \ , \cr
& & \cr
M_{14}^b & = & {3 m_b^2 \lambda \Delta_{{\tilde b}_2}\over 8 \pi^2 v \cot \beta} \left \{ {m_b^2 A_b \Delta_{{\tilde b}_1} \over \cos \beta}
+ {\cos \beta \Delta_{\tilde b} \over 2} \right \}
{g(m_{{\tilde b}_1}^2, \ m_{{\tilde b}_2}^2) \over (m_{{\tilde b}_2}^2 - m_{{\tilde b}_1}^2)^2} \cr
& & \cr
& &\mbox{} + {3 m_b^2 \lambda \cos \beta \Delta_{{\tilde b}_2}\over 32 \pi^2 v \cot \beta}
\left ( {4 m_b^2 \over \cos^2 \beta} - m_Z^2 \right ) {\log ({m_{{\tilde b}_2}^2 / m_{{\tilde b}_1}^2}) \over ( { m_{{\tilde b}_2}^2 - m_{{\tilde b}_1}^2}) } \ , \cr
& & \cr
M_{15}^b & = & \mbox{} - {3 m_b^2 \lambda A_b \sin \phi_b \over 8 \pi^2 v \cot \beta}
\left ({ m_b^2 A_b \Delta_{{\tilde b}_1}\over \cos \beta} + {\cos \beta\Delta_{\tilde b} \over 2} \right )
{g(m_{{\tilde b}_1}^2, \ m_{{\tilde b}_2}^2) \over (m_{{\tilde b}_2}^2 - m_{{\tilde b}_1}^2)^2} \cr
& & \cr
& &\mbox{} -{3 m_b^2 \lambda A_b \cos \beta \sin \phi_b \over 32 \pi^2 v \cot \beta}
\left ({4 m_b^2 \over \cos^2 \beta} - m_Z^2\right ) {\log ({m_{{\tilde b}_2}^2 / m_{{\tilde b}_1}^2}) \over ( { m_{{\tilde b}_2}^2 - m_{{\tilde b}_1}^2}) } \ , \cr
& & \cr
M_{23}^b & = & \mbox{} - {3 m_b^2 \lambda x A_b \sin \phi_b \over 8 \pi^2 v^2 \cos^2 \beta}
\left ({m_b^2 \lambda x \Delta_{{\tilde b}_2} \over \cos \beta} - {\sin \beta \Delta_{\tilde b} \over 2} \right )
{g(m_{{\tilde b}_1}^2, \ m_{{\tilde b}_2}^2) \over (m_{{\tilde b}_2}^2 - m_{{\tilde b}_1}^2)^2 } \cr
& & \cr
& &\mbox{}
- {3 m_Z^2 m_b^2 \lambda x A_b \tan \beta \sin \phi_b \over 32 \pi^2 v^2 \cos \beta}
{\log ({m_{{\tilde b}_2}^2 / m_{{\tilde b}_1}^2}) \over ( { m_{{\tilde b}_2}^2 - m_{{\tilde b}_1}^2}) } \ , \cr
& & \cr
M_{24}^b & = & {3 m_b^2 \lambda \Delta_{{\tilde b}_2} \over 8 \pi^2 v \cot \beta} \left ( {m_b^2 \lambda x \Delta_{{\tilde b}_2} \over \cos \beta}
 - {\sin \beta \Delta_{\tilde b} \over 2} \right )
{ g(m_{{\tilde b}_1}^2, \ m_{{\tilde b}_2}^2)
\over(m_{{\tilde b}_2}^2 - m_{{\tilde t}_1}^2)^2}
- {3 m_b^2 \lambda^2 x \tan \beta \over 8 \pi^2 v \cos \beta}
f(m_{{\tilde b}_1}^2, \ m_{{\tilde b}_2}^2) \cr
& &\mbox{} + {3 m_Z^2 m_b^2  \lambda \sin \beta \Delta_{{\tilde b}_2}
\over 32 \pi^2 v \cot\beta}
\ {\log ({m_{{\tilde b}_2}^2 / m_{{\tilde b}_1}^2})
\over ( { m_{{\tilde b}_2}^2 - m_{{\tilde b}_1}^2}) }
\ , \cr
& & \cr
M_{25}^b & = & \mbox{} - {3 m_b^2 \lambda A_b \sin \phi_b \over 8 \pi^2 v \cot \beta}
\left ({m_b^2 \lambda x \Delta_{\tilde{b}_2} \over \cos \beta}
- {\sin \beta \Delta_{{\tilde b}} \over 2} \right )
{g(m_{{\tilde b}_1}^2, \ m_{{\tilde b}_2}^2) \over (m_{{\tilde b}_2}^2 - m_{{\tilde b}_1}^2)^2 }   \cr
& &\mbox{} - {3 m_Z^2 m_b^2 A_b \lambda \sin \beta \sin \phi_b  \over 32 \pi^2 v \cot \beta}
  {\log (m_{{\tilde b}_2}^2 / m_{{\tilde b}_1}^2) \over (m_{{\tilde b}_2}^2 - m_{{\tilde b}_1}^2)} \ , \cr
& & \cr
M_{34}^b & = & \mbox{} - {3 m_b^4 \lambda^2 x A_b \sin \phi_b \Delta_{{\tilde b}_2} \over 8 \pi^2 v \cos^2 \beta \cot \beta}
{g(m_{{\tilde b}_1}^2, \ m_{{\tilde b}_2}^2)
\over(m_{{\tilde b}_2}^2 - m_{{\tilde b}_1}^2)^2 } \ , \cr
& & \cr
M_{35}^b & = & \mbox{} {3 m_b^4 \lambda^2 x A_b^2 \sin^2 \phi_b \over 8 \pi^2 v \cos^2 \beta \cot \beta }
{g(m_{{\tilde b}_1}^2, \ m_{{\tilde b}_2}^2) \over (m_{{\tilde b}_2}^2 - m_{{\tilde b}_1}^2)^2 }
\ , \cr
M_{45}^b & = & \mbox{} - {3 m_b^4 \lambda^2 A_b \sin \phi_b \Delta_{{\tilde b}_2} \over 8 \pi^2 \cot^2 \beta}
{g(m_{{\tilde b}_1}^2, \ m_{{\tilde b}_2}^2) \over (m_{{\tilde b}_2}^2 - m_{{\tilde b}_1}^2)^2 } \ ,
\end{eqnarray}
where
\begin{eqnarray}
\Delta_{{\tilde b}_1} & = & A_b + \lambda x \tan \beta \cos \phi_b \  , \cr
\Delta_{{\tilde b}_2} & = & A_b \cos \phi_b + \lambda x \tan \beta  \ , \cr
\Delta_{\tilde b} & = & \left ( { m_Z^2 \over 6} - {2 m_W^2 \over 3} \right)
\left \{ m_Q^2 - m_B^2 + \left ( { m_Z^2 \over 6} - {2 m_W^2 \over 3} \right) \cos 2 \beta \right \}  \ .
\end{eqnarray}
\vskip 0.2 in
\noindent
{\large {\bf Appendix D}}
\vskip 0.2 in
\noindent
The elements for the mass matrix of the neutral Higgs bosons due to the radiative contributions
of the tau lepton and scalar tau leptons are
\begin{eqnarray}
M_{11}^{\tau} & = & {1 \over 8 \pi^2 v^2}
\left \{ {m_{\tau}^2 A_{\tau} \Delta_{{\tilde \tau}_1} \over \cos \beta}
+ {\cos \beta \Delta_{\tilde \tau} \over 2}\right \}^2
{g(m_{{\tilde \tau}_1}^2, \ m_{{\tilde \tau}_2}^2)
\over (m_{{\tilde \tau}_2}^2 - m_{{\tilde \tau}_1}^2)^2} - { m_{\tau}^4 \over 4 \pi^2 v^2 \cos^2 \beta}
\log \left ({m_{\tau}^2 \over \Lambda^2} \right )\cr
& & \cr
& &\mbox{} - {\cos^2 \beta \over 16 \pi^2 v^2}
\left(  {3 m_Z^2\over 4 } - {m_W^2} \right)^2
f(m_{{\tilde \tau}_1}^2, \ m_{{\tilde \tau}_2}^2)
\cr & &
\cr & &\mbox{} + {\cos \beta \over 16 \pi^2 v^2}
\left ({4 m_{\tau}^2 \over \cos^2 \beta} - m_Z^2 \right)
\left \{ {m_{\tau}^2 A_{\tau} \Delta_{{\tilde \tau}_1} \over \cos
\beta} + {\cos \beta \Delta_{\tilde \tau} \over 2} \right \}
{\displaystyle \log (m_{{\tilde \tau}_2}^2 / m_{{\tilde \tau}_1}^2)
 \over (m_{{\tilde \tau}_2}^2 - m_{{\tilde \tau}_1}^2)} \cr
& & \cr
& &\mbox{} + {1 \over 32 \pi^2 v^2} \left ({2 m_{\tau}^2 \over \cos  \beta} - {m_Z^2 \cos \beta \over 2} \right)^2
\log \left ({m_{{\tilde \tau}_1}^2  m_{{\tilde \tau}_2}^2 \over \Lambda^4} \right ) \ , \cr
& & \cr
M_{22}^{\tau} & = & {1 \over 8 \pi^2 v^2} \left \{ {m_{\tau}^2
\lambda x \Delta_{{\tilde \tau}_2} \over \cos \beta} - {\sin \beta
\Delta_{\tilde \tau} \over 2} \right \}^2 {g(m_{{\tilde \tau}_1}^2,
\ m_{{\tilde \tau}_2}^2) \over (m_{{\tilde \tau}_2}^2 - m_{{\tilde \tau}_1}^2)^2} + { m_Z^4 \sin^2 \beta \over 128 \pi^2 v^2}
\log \left ({m_{{\tilde \tau}_1}^2  m_{{\tilde \tau}_2}^2 \over \Lambda^4} \right ) \cr
& & \cr
& &\mbox{} - {\sin^2 \beta  \over 16 \pi^2 v^2}
\left( {3 m_Z^2 \over 4 } - {m_W^2} \right)^2
f(m_{{\tilde \tau}_1}^2, \ m_{{\tilde \tau}_2}^2)   \cr
& & \cr
& &\mbox{} + { m_Z^2 \sin \beta \over 16 \pi^2 v^2}
\left \{ {m_{\tau}^2 \lambda x \Delta_{{\tilde \tau}_2} \over \cos \beta} -
{\sin \beta \Delta_{\tilde \tau} \over 2} \right \}
{\displaystyle \log (m_{{\tilde \tau}_2}^2 / m_{{\tilde \tau}_1}^2)
 \over (m_{{\tilde \tau}_2}^2 - m_{{\tilde \tau}_1}^2)}   \ , \cr
& & \cr
M_{33}^{\tau} & = & { m_{\tau}^4 \lambda^2 x^2 A_{\tau}^2 \sin^2 \phi_\tau \over 8 \pi^2 v^2 \cos^4 \beta} {g(m_{\tilde{\tau}_1}^2, \ m_{\tilde{\tau}_2}^2) \over (m_{\tilde{\tau}_2}^2 - m_{\tilde{\tau}_1}^2 )^2}  \ , \cr
& & \cr
M_{44}^{\tau} & = & { m_{\tau}^4 \lambda^2 \Delta_{{\tilde \tau}_2}^2 \over 8 \pi^2 \cot^2 \beta} {g(m_{{\tilde \tau}_1}^2, \ m_{{\tilde \tau}_2}^2) \over (m_{{\tilde \tau}_2}^2 - m_{\tilde{\tau}_1}^2)^2 }   \ , \cr
& & \cr
M_{55}^{\tau} & = & { m_{\tau}^4 \lambda^2 A_{\tau}^2 \sin^2 \phi_\tau \over 8 \pi^2 \cot^2 \beta} {g(m_{{\tilde \tau}_1}^2, \ m_{{\tilde \tau}_2}^2) \over (m_{{\tilde \tau}_2}^2 - m_{{\tilde \tau}_1}^2 )^2}   \ , \cr
& & \cr
M_{12}^{\tau} & = & {1 \over 8 \pi^2 v^2} \left \{ {m_{\tau}^2 \lambda x \Delta_{{\tilde \tau}_2} \over \cos \beta} - {\sin \beta \Delta_{\tilde \tau} \over 2} \right \} \left \{ {m_{\tau}^2 A_{\tau} \Delta_{{\tilde \tau}_1} \over \cos \beta}
+ {\cos \beta \Delta_{\tilde \tau}\over 2} \right \} {g(m_{{\tilde \tau}_1}^2, \ m_{{\tilde \tau}_2}^2) \over (m_{{\tilde \tau}_2}^2 - m_{{\tilde \tau}_1}^2)^2} \cr
& & \cr
& & \mbox{} + {\sin 2 \beta \over 32 \pi^2 v^2}
\left ({3 m_Z^2\over 4 } - {m_W^2} \right)^2
f(m_{{\tilde \tau}_1}^2, \ m_{{\tilde \tau}_2}^2)  \cr
& & \cr
& &\mbox{} + { \cos \beta \over 32 \pi^2 v^2} \left ({4 m_\tau^2 \over \cos^2 \beta} - m_Z^2 \right) \left \{{m_\tau^2 \lambda x \Delta_{{\tilde \tau}_2} \over \cos \beta}
- {\sin \beta \Delta_{\tilde \tau} \over 2} \right \} {\displaystyle \log (m_{{\tilde \tau}_2}^2 / m_{{\tilde \tau}_1}^2)  \over (m_{{\tilde \tau}_2}^2 - m_{{\tilde \tau}_1}^2)} \cr
& & \cr
& &\mbox{} + { m_Z^2 \sin 2 \beta \over 64 \pi^2 v^2}
\left \{{m_{\tau}^2 A_{\tau} \Delta_{{\tilde \tau}_1} \over \cos^2 \beta} + {\Delta_{\tilde \tau}\over 2} \right \} {\displaystyle \log (m_{{\tilde \tau}_2}^2 / m_{{\tilde \tau}_1}^2) \over (m_{{\tilde \tau}_2}^2 - m_{{\tilde \tau}_1}^2)} \cr
& & \cr
& &\mbox{} + { m_Z^2 \sin 2 \beta \over 256 \pi^2 v^2} \left ({4 m_{\tau}^2 \over \cos^2 \beta} - m_Z^2 \right) \log \left ({m_{{\tilde \tau}_1}^2 m_{{\tilde \tau}_2}^2 \over \Lambda^4} \right ) \ , \cr
& & \cr
M_{13}^{\tau} & = & \mbox{} - {m_{\tau}^2 \lambda x A_{\tau} \sin \phi_{\tau} \over 8 \pi^2 v^2 \cos^2 \beta}\left \{ {m_{\tau}^2 \ A_{\tau} \Delta_{{\tilde \tau}_1 } \over \cos \beta}
+ {\cos \beta \Delta_{\tilde \tau} \over 2}\right \} {g(m_{{\tilde \tau}_1}^2, \ m_{{\tilde \tau}_2}^2) \over (m_{{\tilde \tau}_2}^2 - m_{{\tilde \tau}_1}^2)^2}  \cr
& & \cr
& &\mbox{} - {m_{\tau}^2 \lambda x A_\tau \sin \phi_{\tau} \over 16 \pi^2 v^2 \cos \beta} \left ({2 m_{\tau}^2 \over \cos^2 \beta} - {m_Z^2 \over 2} \right )
{\displaystyle \log (m_{{\tilde \tau}_2}^2 / m_{{\tilde \tau}_1}^2) \over (m_{{\tilde \tau}_2}^2 - m_{{\tilde \tau}_1}^2)}  \ , \cr
& & \cr
M_{14}^{\tau} & = & {m_{\tau}^2 \lambda \Delta_{{\tilde \tau}_2}\over 8 \pi^2 v \cot\beta} \left \{ {m_{\tau}^2 A_{\tau} \Delta_{{\tilde \tau}_1} \over \cos \beta} + {\cos \beta \Delta_{\tilde \tau} \over 2} \right \}
{g(m_{{\tilde \tau}_1}^2, \ m_{{\tilde \tau}_2}^2) \over (m_{{\tilde \tau}_2}^2 - m_{{\tilde \tau}_1}^2)^2} \cr
& & \cr
& &\mbox{} + {m_{\tau}^2 \lambda \cos \beta \Delta_{{\tilde \tau}_2}\over 32 \pi^2 v \cot \beta} \left ( {4 m_{\tau}^2\over \cos^2 \beta} - m_Z^2 \right ) \ {\log ({m_{{\tilde \tau}_2}^2 / m_{{\tilde \tau}_1}^2}) \over ( { m_{{\tilde \tau}_2}^2 - m_{{\tilde \tau}_1}^2}) } \ , \cr
& & \cr
M_{15}^{\tau} & = & \mbox{} - {m_{\tau}^2 \lambda A_{\tau} \sin \phi_{\tau} \over 8 \pi^2 v \cot \beta} \left ({ m_{\tau}^2 A_{\tau} \Delta_{{\tilde \tau}_1}\over \cos \beta} + {\cos \beta \Delta_{\tilde \tau} \over 2} \right )
{g(m_{{\tilde \tau}_1}^2, \ m_{{\tilde \tau}_2}^2) \over (m_{{\tilde \tau}_2}^2 - m_{{\tilde \tau}_1}^2)^2}  \cr
& & \cr
& &\mbox{} - {m_{\tau}^2 \lambda A_{\tau} \cos \beta \sin \phi_{\tau} \over 32 \pi^2 v \cot \beta} \left ({4 m_{\tau}^2 \over \cos^2 \beta} - m_Z^2 \right ) \ {\log ({m_{{\tilde \tau}_2}^2 / m_{{\tilde \tau}_1}^2}) \over ( { m_{{\tilde \tau}_2}^2 - m_{{\tilde \tau}_1}^2}) } \ , \cr
& & \cr
M_{23}^{\tau} & = & \mbox{} - {m_{\tau}^2 \lambda x A_{\tau} \sin \phi_{\tau} \over 8 \pi^2 v^2 \cos^2 \beta} \left ({m_{\tau}^2 \lambda x \Delta_{{\tilde \tau}_2} \over \cos \beta} - {\sin \beta \Delta_{\tilde \tau} \over 2} \right )
{g(m_{{\tilde \tau}_1}^2, \ m_{{\tilde \tau}_2}^2) \over (m_{{\tilde \tau}_2}^2 - m_{{\tilde \tau}_1}^2)^2 } \cr
& & \cr
& &\mbox{} - {m_Z^2 m_{\tau}^2 \lambda x A_{\tau} \tan \beta \sin \phi_{\tau} \over 32 \pi^2 v^2 \cos \beta} {\log ({m_{{\tilde \tau}_2}^2 / m_{{\tilde \tau}_1}^2}) \over ( {m_{{\tilde \tau}_2}^2 - m_{{\tilde \tau}_1}^2}) } \ , \cr
& & \cr
M_{24}^{\tau} & = & {m_{\tau}^2 \lambda \Delta_{{\tilde \tau}_2} \over 8 \pi^2 v \cot \beta} \left ( {m_{\tau}^2 \lambda x \Delta_{{\tilde \tau}_2} \over \cos \beta} - {\sin \beta \Delta_{\tilde \tau} \over 2} \right )
{ g(m_{{\tilde \tau}_1}^2, \ m_{{\tilde \tau}_2}^2) \over(m_{{\tilde \tau}_2}^2 - m_{{\tilde \tau}_1}^2)^2}
- {m_{\tau}^2 \lambda^2 x \tan \beta \over 8 \pi^2 v \cos \beta}
f(m_{{\tilde \tau}_1}^2, \ m_{{\tilde \tau}_2}^2) \cr
& & \cr
& &\mbox{} + {m_Z^2 m_{\tau}^2  \lambda \sin \beta \Delta_{{\tilde \tau}_2} \over 32 \pi^2 v \cot \beta} \ {\log ({m_{{\tilde \tau}_2}^2 / m_{{\tilde \tau}_1}^2}) \over ( { m_{{\tilde \tau}_2}^2 - m_{{\tilde \tau}_1}^2}) } \ , \cr
& & \cr
M_{25}^{\tau} & = & \mbox{} -{m_{\tau}^2 \lambda A_{\tau} \sin \phi_{\tau} \over 8 \pi^2 v \cot \beta} \left ({m_{\tau}^2 \lambda x \Delta_{{\tilde \tau}_2} \over \cos \beta} - {\sin \beta \Delta_{\tilde \tau} \over 2} \right )
{g(m_{{\tilde \tau}_1}^2, \ m_{{\tilde \tau}_2}^2) \over (m_{{\tilde \tau}_2}^2 - m_{{\tilde \tau}_1}^2)^2 } \cr
& & \cr
& &\mbox{} - {m_{\tau}^2 m_Z^2 A_{\tau} \lambda \sin \beta \sin \phi_{\tau}
 \over 32 \pi^2 v \cot \beta}
  {\log (m_{{\tilde \tau}_2}^2 / m_{{\tilde \tau}_1}^2) \over (m_{{\tilde \tau}_2}^2 - m_{{\tilde \tau}_1}^2)} \ , \cr
& & \cr
M_{34}^{\tau} & = & \mbox{} - { m_{\tau}^4 \lambda^2 x A_{\tau} \sin \phi_{\tau} \Delta_{{\tilde \tau}_2} \over 8 \pi^2 v \cos^2 \beta \cot \beta} {g(m_{{\tilde \tau}_1}^2, \ m_{{\tilde \tau}_2}^2) \over(m_{{\tilde \tau}_2}^2 - m_{{\tilde \tau}_1}^2)^2 }  \ , \cr
& & \cr
M_{35}^{\tau} & = & \mbox{} { m_{\tau}^4 \lambda^2 x A_{\tau}^2 \sin^2 \phi_{\tau} \over 8 \pi^2 v \cos^2 \beta \cot \beta }
{g(m_{{\tilde \tau}_1}^2, \ m_{{\tilde \tau}_2}^2) \over (m_{{\tilde \tau}_2}^2 - m_{{\tilde \tau}_1}^2)^2 }  \ , \cr
& & \cr
M_{45}^{\tau} & = & \mbox{} - { m_{\tau}^4 \lambda^2 A_{\tau} \sin \phi_{\tau} \Delta_{{\tilde \tau}_2} \over 8 \pi^2 \cot^2 \beta} {g(m_{{\tilde \tau}_1}^2, \ m_{{\tilde \tau}_2}^2) \over (m_{{\tilde \tau}_2}^2 - m_{{\tilde \tau}_1}^2)^2 } \ ,
\end{eqnarray}
where
\begin{eqnarray}
\Delta_{{\tilde \tau}_1} & = & A_\tau + \lambda x \tan \beta \cos \phi_{\tau} \ , \cr
\Delta_{{\tilde \tau}_2} & = & A_{\tau} \cos \phi_{\tau} + \lambda x \tan \beta  \  , \cr
\Delta_{\tilde \tau} & = & \left ( {3 m_Z^2 \over 4 } - {m_W^2} \right )
 \left \{ m_L^2 - m_E^2 + \left ( {3 m_Z^2 \over 4 } - {m_W^2} \right) \cos 2 \beta \right \}   \ .
\end{eqnarray}
\vskip 0.2 in
\noindent
{\large {\bf Appendix E}}
\vskip 0.2 in
\noindent
The elements for the mass  matrix of the neutral Higgs bosons due to the radiative
contributions of the $W$ boson, charged Higgs boson, and charginos are
\begin{eqnarray}
M_{11}^{\chi} & = & \mbox{} - {m_W^4 \cos^2 \beta \Delta _{{\tilde \chi}_1}^2 \over 4 \pi^2 v^2}
{g(m_{{\tilde \chi}_1}^2, \ m_{{\tilde \chi}_2}^2)
\over (m_{{\tilde \chi}_2}^2 - m_{{\tilde \chi}_1}^2)^2}
+ {m_W^4 \cos^2 \beta  \over 2 \pi^2 v^2}
f(m_{{\tilde \chi}_1}^2, \ m_{{\tilde \chi}_2}^2) \cr
& & \cr
& &\mbox{} - {m_W^4 \cos^2 \beta \over 2 \pi^2 v^2}
{\Delta_{{\tilde \chi}_1} \log ({m_{{\tilde \chi}_2}^2 / m_{{\tilde \chi}_1}^2}) \over (m_{{\tilde \chi}_2}^2 - m_{{\tilde \chi}_1}^2)}
+ {m_W^4 \cos^2 \beta \over 8 \pi^2 v^2} \log \left ({m_W^6 m_{C^+}^2 \over
m_{{\tilde \chi}_1}^4 m_{{\tilde \chi}_2}^4} \right ) \cr
& &\mbox{} + {\sin^2 \beta \over 16 \pi^2 v^2} (2 \lambda^2 v^2 - m_W^2) m_{C^+}^2 \left \{\log \left ({m_{C^+}^2 \over \Lambda^2} \right ) - 1 \right \}  \ , \cr
 & & \cr
M_{22}^{\chi} & = & \mbox{} - {m_W^4 \sin^2 \beta \Delta _{{\tilde \chi}_2}^2 \over 4 \pi^2 v^2}
{g(m_{{\tilde \chi}_1}^2, \ m_{{\tilde \chi}_2}^2)
\over (m_{{\tilde \chi}_2}^2 - m_{{\tilde \chi}_1}^2)^2}
+ {m_W^4 \sin^2 \beta \over 2 \pi^2 v^2}
f(m_{{\tilde \chi}_1}^2, \ m_{{\tilde \chi}_2}^2) \cr
& & \cr
& &\mbox{} - {m_W^4 \sin^2 \beta \over 2 \pi^2 v^2}
{\Delta_{{\tilde \chi}_2} \log ({m_{{\tilde \chi}_2}^2 / m_{{\tilde \chi}_1}^2}) \over (m_{{\tilde \chi}_2}^2 - m_{{\tilde \chi}_1}^2)}
+ {m_W^4 \sin^2 \beta \over 8 \pi^2 v^2} \log \left ({m_W^6 m_{C^+}^2 \over
m_{{\tilde \chi}_1}^4 m_{{\tilde \chi}_2}^4} \right ) \cr
& &\mbox{} + {\cos^2 \beta \over 16 \pi^2 v^2} (2 \lambda^2 v^2 - m_W^2) m_{C^+}^2 \left \{\log \left ({m_{C^+}^2 \over \Lambda^2} \right ) - 1 \right \}  \ , \cr
 & & \cr
M_{33}^{\chi} & = & \mbox{} - {m_W^4 \over \pi^2 v^2}
{M_2^2 \lambda^2 x^2 \sin^2 \phi_c \over (m_{{\tilde \chi}_2}^2 - m_{{\tilde \chi}_1}^2)^2}
 g(m_{{\tilde \chi}_1}^2, \ m_{{\tilde \chi}_2}^2) \cr
& & \cr
& &\mbox{} + {(2 \lambda^2 v^2 - m_W^2) m_{C^+}^2 \over 16 \pi^2 v^2}
\left \{\log \left ({m_{C^+}^2 \over \Lambda^2} \right ) - 1 \right \} \ , \cr
 & & \cr
M_{44}^{\chi} & = &\mbox{} - {\lambda^2 \over 16 \pi^2}
{\Delta_{\tilde \chi}^2
g(m_{{\tilde \chi}_1}^2, \ m_{{\tilde \chi}_2}^2) \over
(m_{{\tilde \chi}_2}^2 - m_{{\tilde \chi}_1}^2)^2}
+ {\lambda^4 x^2 \over 8 \pi^2}
f(m_{{\tilde \chi}_1}^2, \ m_{{\tilde \chi}_2}^2)
 - {\lambda^4 x^2 \over 16 \pi^2} \log \left ({m_{{\tilde \chi}_1}^2 m_{{\tilde \chi}_2}^2 \over \Lambda^4} \right ) \cr
& & \cr
& &\mbox{} - {\lambda^3 x \over 8 \pi^2} {\Delta_{\tilde \chi}
\log (m_{{\tilde \chi}_2}^2 / m_{{\tilde \chi}_1}^2)
\over (m_{{\tilde \chi}_2}^2 - m_{{\tilde \chi}_1}^2)}
- {\lambda A_{\lambda} \sin 2 \beta m_{C^+}^2 \over 32 \pi^2 x}
\left \{ \log \left ({m_{C^+}^2 \over \Lambda^2} \right ) - 1 \right \} \ , \cr
 & & \cr
M_{55}^{\chi} & = &\mbox{} - {m_W^4 \over 4 \pi^2}
{M_2^2 \lambda^2 \sin^2 2 \beta \sin^2 \phi_c \over
(m_{{\tilde \chi}_2}^2 - m_{{\tilde \chi}_1}^2)^2}
g(m_{{\tilde \chi}_1}^2, \ m_{{\tilde \chi}_2}^2) \cr
& & \cr
& &\mbox{} - {\lambda \sin 2 \beta \over 32 \pi^2 x}
(A_{\lambda} + 8 k x) m_{C^+}^2
\left \{\log \left ({m_{C^+}^2 \over \Lambda^2} \right ) - 1 \right \}
 \ ,  \cr
 & & \cr
M_{12}^{\chi} & = &\mbox{} - {m_W^4 \over 8 \pi^2 v^2}
{\sin 2 \beta \Delta_{{\tilde \chi}_1} \Delta_{{\tilde \chi}_2}
\over (m_{{\tilde \chi}_2}^2 - m_{{\tilde \chi}_1}^2)^2}
g(m_{{\tilde \chi}_1}^2, \ m_{{\tilde \chi}_2}^2)
- {m_W^4 \sin 2 \beta \over 4 \pi^2 v^2}
f(m_{{\tilde \chi}_1}^2, \ m_{{\tilde \chi}_2}^2) \cr
& & \cr
& &\mbox{} - {m_W^4 \over 8 \pi^2 v^2}
{\sin 2 \beta (\Delta_{{\tilde \chi}_1} + \Delta_{{\tilde \chi}_2})
\over (m_{{\tilde \chi}_2}^2 - m_{{\tilde \chi}_1}^2)}
\log \left ({m_{{\tilde \chi}_2}^2 \over m_{{\tilde \chi}_1}^2} \right)
+ {m_W^4 \sin 2 \beta \over 16 \pi^2 v^2}
\log \left ( {m_W^6 m_{C^+}^2 \over m_{{\tilde \chi}_1}^4 m_{{\tilde \chi}_2}^4} \right ) \cr
& & \cr
& &\mbox{} - {\lambda^2 m_W^2 \sin 2 \beta \over 8 \pi^2}
\log \left( {m_{C^+}^2 \over \Lambda^2}\right)
+ {\sin 2 \beta \over 32 \pi^2 v^2} (m_W^2 - 2 \lambda^2 v^2) m_{C^+}^2
\left \{\log \left ({m_{C^+}^2 \over \Lambda^2} \right ) - 1 \right \}  \ , \cr
 & & \cr
M_{13}^{\chi} & = &\mbox{} - {m_W^4 \over 2 \pi^2 v^2}
{M_2 \lambda x \cos \beta \sin \phi_c \Delta_{{\tilde \chi}_1} \over
 (m_{{\tilde \chi}_2}^2 - m_{{\tilde \chi}_1}^2)^2}
g(m_{{\tilde \chi}_1}^2, \ m_{{\tilde \chi}_2}^2) \cr
& & \cr
& &\mbox{} - {m_W^4 \over 2 \pi^2 v^2}
{M_2 \lambda x \cos \beta \sin \phi_c \over
(m_{{\tilde \chi}_2}^2 - m_{{\tilde \chi}_1}^2)}
\log \left ( {m_{{\tilde \chi}_2}^2 \over m_{{\tilde \chi}_1}^2} \right)
 \ , \cr
 & & \cr
M_{14}^{\chi} & = &\mbox{} - {m_W^2 \over 8 \pi^2 v}
{\lambda \cos \beta \Delta_{{\tilde \chi}_1} \Delta_{\tilde \chi}
\over (m_{{\tilde \chi}_2}^2 - m_{{\tilde \chi}_1}^2)^2}
g(m_{{\tilde \chi}_1}^2, \ m_{{\tilde \chi}_2}^2)
+ {m_W^2 \lambda^2 x \cos \beta \over 4 \pi^2 v}
f(m_{{\tilde \chi}_1}^2, \ m_{{\tilde \chi}_2}^2) \cr
& & \cr
& &\mbox{} - {m_W^2 \lambda^2 x \cos \beta \over 8 \pi^2 v}
\log \left(
{m_{{\tilde \chi}_1}^2 m_{{\tilde \chi}_2}^2 \over \Lambda^4} \right)
- {m_W^2 \over 8 \pi^2 v}
{\lambda \cos \beta
(\lambda x \Delta_{{\tilde \chi}_1} + \Delta_{\tilde \chi}) \over
(m_{{\tilde \chi}_2}^2 - m_{{\tilde \chi}_1}^2) }
\log \left ( {m_{{\tilde \chi}_2}^2 \over m_{{\tilde \chi}_1}^2} \right)
 \ , \cr
 & & \cr
M_{15}^{\chi} & = & {m_W^4 \over 4 \pi^2 v}
{M_2 \lambda \cos \beta \sin 2 \beta \sin \phi_c \Delta_{{\tilde \chi}_1}
\over (m_{{\tilde \chi}_2}^2 - m_{{\tilde \chi}_1}^2)^2}
g(m_{{\tilde \chi}_1}^2, \ m_{{\tilde \chi}_2}^2) \cr
& & \cr
& &\mbox{} - {m_W^4 \over 4 \pi^2 v}
{M_2 \lambda \cos \beta \sin 2 \beta \sin \phi_c \over
(m_{{\tilde \chi}_2}^2 - m_{{\tilde \chi}_1}^2)}
\log \left ( {m_{{\tilde \chi}_2}^2 \over m_{{\tilde \chi}_1}^2} \right)
 \ , \cr
 & & \cr
M_{23}^{\chi} & = &\mbox{} - {m_W^4 \over 2 \pi^2 v^2} {M_2
\lambda x \sin \beta \sin \phi_c \Delta_{{\tilde \chi}_2} \over
(m_{{\tilde \chi}_2}^2 - m_{{\tilde \chi}_1}^2)^2} g(m_{{\tilde
\chi}_1}^2, \ m_{{\tilde \chi}_2}^2)  \cr
& & \cr
& &\mbox{} - {m_W^4
\over 2 \pi^2 v^2} {M_2 \lambda x \sin \beta \sin \phi_c \over
(m_{{\tilde \chi}_2}^2 - m_{{\tilde \chi}_1}^2)} \log \left (
{m_{{\tilde \chi}_2}^2 \over m_{{\tilde \chi}_1}^2} \right)
 \ , \cr
 & & \cr
M_{24}^{\chi} & = &\mbox{} - {m_W^2 \over 8 \pi^2 v}
{\lambda \sin \beta \Delta_{{\tilde \chi}_2} \Delta_{\tilde \chi}
\over (m_{{\tilde \chi}_2}^2 - m_{{\tilde \chi}_1}^2)^2}
g(m_{{\tilde \chi}_1}^2, \ m_{{\tilde \chi}_2}^2)
+ {m_W^2 \lambda^2 x \sin \beta \over 4 \pi^2 v}
f(m_{{\tilde \chi}_1}^2, \ m_{{\tilde \chi}_2}^2) \cr
& & \cr
& &\mbox{} -{m_W^2 \lambda^2 x \sin \beta \over 8 \pi^2 v}
\log \left( {m_{{\tilde \chi}_1}^2 m_{{\tilde \chi}_2}^2 \over \Lambda^4}
\right)
- {m_W^2 \over 8 \pi^2 v}
{\lambda \sin \beta
(\lambda x \Delta_{{\tilde \chi}_2} + \Delta_{\tilde \chi}) \over
(m_{{\tilde \chi}_2}^2 - m_{{\tilde \chi}_1}^2)}
\log \left( {m_{{\tilde \chi}_2}^2 \over m_{{\tilde \chi}_1}^2}
\right)   \ ,    \cr
 & & \cr
M_{25}^{\chi} & = & {m_W^4 \over 4 \pi^2 v}
{M_2 \lambda \sin \beta \sin 2 \beta \sin \phi_c \Delta_{{\tilde \chi}_2}
\over (m_{{\tilde \chi}_2}^2 - m_{{\tilde \chi}_1}^2)^2}
g(m_{{\tilde \chi}_1}^2, \ m_{{\tilde \chi}_2}^2)  \cr
& & \cr
& &\mbox{} + {m_W^4 \over 4 \pi^2 v}
{M_2 \lambda \sin \beta \sin 2 \beta \sin \phi_c \over
(m_{{\tilde \chi}_2}^2 - m_{{\tilde \chi}_1}^2)}
\log \left ( {m_{{\tilde \chi}_2}^2 \over m_{{\tilde \chi}_1}^2} \right)
 \ , \cr
 & & \cr
M_{34}^{\chi} & = &\mbox{} - {m_W^2 \over 4 \pi^2 v}
{M_2 \lambda^2 x \sin \phi_c \Delta_{\tilde \chi}
\over (m_{{\tilde \chi}_2}^2 - m_{{\tilde \chi}_1}^2)^2}
g(m_{{\tilde \chi}_1}^2, \ m_{{\tilde \chi}_2}^2)
- {m_W^2 \over 4 \pi^2 v} {M_2 \lambda^3 x^2 \sin \phi_c \over (m_{{\tilde \chi}_2}^2 - m_{{\tilde \chi}_1}^2)}
\log \left ( {m_{{\tilde \chi}_2}^2 \over m_{{\tilde \chi}_1}^2} \right)
 \ , \cr
 & & \cr
M_{35}^{\chi} & = &\mbox{} {m_W^4 \over 2 \pi^2 v}
{M_2^2 \lambda^2 x \sin 2 \beta \sin^2 \phi_c \over (m_{{\tilde \chi}_2}^2 - m_{{\tilde \chi}_1}^2)^2}
g(m_{{\tilde \chi}_1}^2, \ m_{{\tilde \chi}_2}^2)
+ {m_W^2 \over 2 \pi^2 v} M_2 \lambda \cos \phi_c
f(m_{{\tilde \chi}_1}^2, \ m_{{\tilde \chi}_2}^2)
 \ , \cr
 & & \cr
M_{45}^{\chi} & = & {m_W^2 \over 8 \pi^2}
{M_2 \lambda^2 \sin 2 \beta \sin \phi_c \Delta_{\tilde \chi}
\over (m_{{\tilde \chi}_2}^2 - m_{{\tilde \chi}_1}^2)^2}
g(m_{{\tilde \chi}_1}^2, \ m_{{\tilde \chi}_2}^2) \cr
& & \cr
& &\mbox{} + {m_W^2 \over 8 \pi^2}
{M_2 \lambda^3 x \sin 2 \beta \sin \phi_c \over
(m_{{\tilde \chi}_2}^2 - m_{{\tilde \chi}_1}^2)}
\log \left ( {m_{{\tilde \chi}_2}^2 \over m_{{\tilde \chi}_1}^2} \right)
\ ,
\end{eqnarray}
where
\begin{eqnarray}
\Delta_{{\tilde \chi}_1} & = &
M_2^2 +\lambda^2 x^2 + 2 m_W^2 \cos 2 \beta + 2 \lambda x M_2 \tan \beta \cos \phi_c \ , \cr
\Delta_{{\tilde \chi}_2} & = &
M_2^2 +\lambda^2 x^2 - 2 m_W^2 \cos 2 \beta + 2 \lambda x M_2 \cot \beta \cos \phi_c \ , \cr
\Delta_{\tilde \chi} & = &
\lambda^3 x^3 - \lambda x M_2^2 + 2 \lambda x m_W^2 + 2 m_W^2 M_2 \sin 2 \beta \cos \phi_c \ .
\end{eqnarray}
\vskip 0.2 in
\noindent
{\large {\bf Appendix F}}
\vskip 0.2 in
\noindent
The elements for the field-dependent mass matrix of the scalar Higgs bosons at the tree level are
\begin{eqnarray}
{\cal M}_{S_{11}} & = & {3 \over v} m_Z^2 \cos \beta h_1 + {\left (2 v \lambda^2 - {m_Z^2 \over v} \right ) \sin \beta h_2} + {2 x \lambda^2 h_4}
+ {3 m_Z^2 h_1^2 \over 2 v^2} + {\left ( \lambda^2 -{ m_Z^2 \over 2 v^2 } \right) h_2^2} \cr
& &\mbox{}  + {\left ( \lambda^2 \cos^2 \beta  -{ m_Z^2 \cos 2 \beta  \over 2 v^2}\right )h_3^2}
+ {\lambda^2 h_4^2} +{\lambda^2 h_5^2} + {m_Z^2 \cos^2 \beta} \cr
& &\mbox{} + {\lambda x \left ( A_{\lambda} + k x \right ) \tan \beta} \ , \cr
 & & \cr
{\cal M}_{S_{22}} & = & {\left (2 v \lambda^2 - {m_Z^2 \over v} \right ) \cos \beta h_1} + {3 \over v} m_Z^2 \sin \beta h_2 + {2 x \lambda^2 h_4}
+ {\left ( \lambda^2 -{ m_Z^2 \over 2 v^2 } \right) h_1^2}
+ {3 \over 2 v^2} m_Z^2 h_2^2  \cr
& &\mbox{} + {\left (\lambda^2 \sin^2 \beta  + { m_Z^2 \cos 2 \beta  \over 2 v^2} \right ) h_3^2} + {\lambda^2 h_4^2} + {\lambda^2 h_5^2} + {m_Z^2 \sin^2 \beta} \cr
& &\mbox{} + {\lambda x \left ( A_{\lambda} + k x\right ) \cot \beta} \ , \cr
 & & \cr
{\cal M}_{S_{33}} & = & {2 v \lambda \left (\lambda \cos \beta -k \sin \beta \right ) h_1} + {2 v \lambda \left ( \lambda \sin \beta - k \cos \beta \right ) h_2} + {2 k ( 6 k x - A_k) h_4}  + {\lambda^2 h_1^2}  \cr
& &\mbox{} + {\lambda^2 h_2^2} + {\lambda \left( \lambda + k \sin 2 \beta \right) h_3^2} + {6 k^2 h_4^2} + {2 k^2 h_5^2} - {2 k \lambda h_1 h_2}
 + {v^2 \over 2 x} A_{\lambda} \lambda \sin 2 \beta  \cr
& &\mbox{} - {A_k k x} + {4 k^2 x^2} \ , \cr
 & & \cr
{\cal M}_{S_{12}} & = & \left (2 \lambda^2 v -{m_Z^2 \over v} \right) (\sin \beta h_1 + \cos \beta h_2) - {\lambda \left( A_{\lambda} + 2 k x \right) h_4} - {k \lambda h_4^2} + {k \lambda h_5^2} \cr
& &\mbox{} + {\left ( 2 \lambda^2 - {m_Z^2 \over v^2} \right ) h_1 h_2} - {\lambda x (A_{\lambda} + k x) } + {\left (\lambda^2 v^2 - { m_Z^2 \over 2} \right) \sin 2 \beta} \ , \cr
 & & \cr
{\cal M}_{S_{13}} & = & {2 x \lambda^2 h_1} - {\lambda (A_{\lambda} + 2 k x h_2} + {2 v \lambda (\lambda \cos \beta - k \sin \beta) h_4}
+ {2 \lambda^2 h_1 h_4} - {2 k \lambda h_2 h_4} \cr
& &\mbox{}  - {2 k \lambda \cos \beta h_3 h_5} + {2 v x \lambda^2 \cos \beta} - {\lambda v \left(A_\lambda + 2 k x \right) \sin \beta} \ , \cr
 & & \cr
{\cal M}_{S_{23}} & = & - {\lambda (A_{\lambda} + 2 k x) h_1} + {2 x \lambda^2 h_2} + {2 v \lambda (\lambda \sin \beta - k \cos \beta ) h_4}
- {2 k \lambda h_1 h_4} + {2 \lambda^2 h_2 h_4} \cr
& &\mbox{}  - {2 k \lambda \sin \beta h_3 h_5}  + {2 v x \lambda^2 \sin \beta}
-{\lambda v (A_{\lambda} + 2 k x ) \cos \beta}   \ .
\end{eqnarray}
\vskip 0.2 in
\noindent
{\large {\bf Appendix G}}
\vskip 0.2 in
\noindent
The elements for the field-dependent mass matrix of the pseudoscalar Higgs bosons at the tree level are
\begin{eqnarray}
{\cal M}_{P_{44}} & = & {1 \over v} ( 2 \lambda^2 v^2 \cos^2 \beta - m_Z^2 \cos 2 \beta) \cos \beta h_1 + {1 \over v} ( 2 \lambda^2 v^2 \sin^2 \beta + m_Z^2 \sin 2 \beta) \sin \beta h_2 \cr
& &\mbox{} + { \lambda \{ 2 \lambda x ( A_\lambda + 2 k x) \sin 2 \beta \} h_3 } + {1 \over 2 v^2} ( 2 \lambda^2 v^2 \cos^2 \beta - m_Z^2 \cos 2 \beta) \cos \beta h_1^2  \cr
& &\mbox{} + {1 \over 2 v^2} ( 2 \lambda^2 v^2 \sin^2 \beta - m_Z^2 \cos 2 \beta) \cos \beta h_2^2  \cr
& &\mbox{} + {3 \over 4 v^2} \{ 2 \lambda^2 v^2 \sin^2 2\beta + m_Z^2 ( 1 + \cos 4 \beta ) \} h_3^2  + { \lambda ( \lambda + k \sin 2 \beta ) h_4 }  \cr
& &\mbox{} + { \lambda ( \lambda - k \sin 2 \beta ) h_5 } + {2 \lambda x (A_\lambda + k x) \over \sin 2 \beta} \ , \cr
 & & \cr
{\cal M}_{P_{55}} & = & { 2 \lambda ( \lambda v \cos \beta + k v \sin \beta ) h_1 } + { 2 \lambda ( \lambda v \sin \beta + k v \cos \beta ) h_2 } + { 2 k ( A_k + 2 k x) h_4 }  \cr
& &\mbox{} + {\lambda^2 h_1^2} + {\lambda^2 h_2^2} + { \lambda ( \lambda - k \sin 2 \beta) h_3^2 } + {2 k^2 h_4^2} + {6 k^2 h_5^2} + {2 \lambda k h_1 h_2 } + {3 A_k k x} \cr
& &\mbox{} + {\lambda v^2 \left ( 2 k + { A_{\lambda} \over 2 x} \right ) \sin 2 \beta} \ , \cr
 & & \cr
{\cal M}_{P_{45}} & = & { \lambda ( A_\lambda - 2 k x) \cos \beta h_1 } + { \lambda ( A_\lambda - 2 k x) \sin \beta h_2 } - {2 k \lambda v h_4} - {2 k \lambda \cos \beta h_1 h_4} \cr
& &\mbox{} - {2 k \lambda \sin \beta h_2 h_4} + {2 \lambda ( \lambda - k \sin 2 \beta) h_3 h_5} + {\lambda v (A_{\lambda} - 2 k x )} \ .
\end{eqnarray}

\vfil\eject


\vfil\eject
{\bf Figure Captions}

\vskip 0.3 in
\noindent
Fig. 1 : The plot of the neutral Higgs boson masses as functions of $\Lambda$ for $\phi_1 = 10\pi/19$.
We set $\tan \beta$ = 5, $\lambda$ = 0.5, $k$ = 0.4, $A_{\lambda}$ = 590 GeV, $A_k$ = 5 GeV, $x$ = 178 GeV,
$m_Q = m_L$ = 967 GeV, $m_T = m_B = m_E$ = 794 GeV,
$A_t = A_b = A_{\tau}$ = 907 GeV, $M_2$ = 357 GeV, $\phi_t = \phi_b = \phi_{\tau} = \phi_c = 10 \pi/19$.

\vskip 0.3 in
\noindent
Fig. 2  : The plot of the neutral Higgs boson masses as functions of $\phi_1$ for $\Lambda = 300$ GeV.
The remaining parameters have the same values as in Fig. 1.

\vskip 0.3 in
\noindent
Fig. 3a : The plot of mass matrix elements $M_{13}$ (solid curve), $M_{23}$ (dashed curve), and $M_{34}$ (dotted curve)
of the neutral Higgs bosons as functions of $\phi_1$. The values of other parameters are the same as in Fig. 2.

\vskip 0.3 in
\noindent
Fig. 3b : The plot of mass matrix elements $M_{15}$ (solid curve), $M_{25}$ (dashed curve), and $M_{45}$ (dotted curve)
of the neutral HIggs bosons as functions of $\phi_1$. The values of other parameters are the same as in Fig. 2.

\vskip 0.3 in
\noindent
Fig. 4a : The ratios of mass matrix elements $|M_{13}|/|M_{12}|$ (solid curve), $|M_{23}|/|M_{12}|$ (dashed curve),
and $|M_{34}|/|M_{12}|$ (dotted curve) of the neutral Higgs bosons are plotted as functions of $\phi_1$.
The values of other parameters are the same as in Fig. 2.

\vskip 0.3 in
\noindent
Fig. 4b : The ratios of mass matrix elements $|M_{13}|/|M_{14}|$ (solid curve), $|M_{23}|/|M_{14}|$ (dashed curve),
and $|M_{34}|/|M_{14}|$ (dotted curve) of the neutral Higgs bosons are plotted as functions of $\phi_1$.
The values of other parameters are the same as in Fig. 2.

\vskip 0.3 in
\noindent
Fig. 4c : The ratios of mass matrix elements $|M_{13}|/|M_{24}|$ (solid curve), $|M_{23}|/|M_{24}|$ (dashed curve),
and $|M_{34}|/|M_{24}|$ (dotted curve) of the neutral Higgs bosons are plotted as functions of $\phi_1$.
The values of other parameters are the same as in Fig. 2.

\vskip 0.3 in
\noindent
Fig. 4d : The ratios of mass matrix elements $|M_{13}|/|M_{35}|$ (solid curve), $|M_{23}|/|M_{35}|$ (dashed curve),
and $|M_{34}|/|M_{35}|$ (dotted curve) of the neutral Higgs bosons are plotted as functions of $\phi_1$.
The values of other parameters are the same as in Fig. 2.

\vskip 0.3 in
\noindent
Fig. 4e : The ratios of mass matrix elements $|M_{15}|/|M_{12}|$ (solid curve), $|M_{25}|/|M_{12}|$ (dashed curve),
and $|M_{45}|/|M_{12}|$ (dotted curve) of the neutral Higgs bosons are plotted as functions of $\phi_1$.
The values of other parameters are the same as in Fig. 2.

\vskip 0.3 in
\noindent
Fig. 4f : The ratios of mass matrix elements $|M_{15}|/|M_{14}|$ (solid curve), $|M_{25}|/|M_{14}|$ (dashed curve),
and $|M_{45}|/|M_{14}|$ (dotted curve) of the neutral Higgs bosons are plotted as functions of $\phi_1$.
The values of other parameters are the same as in Fig. 2.

\vskip 0.3 in
\noindent
Fig. 4g : The ratios of mass matrix elements $|M_{15}|/|M_{24}|$ (solid curve), $|M_{25}|/|M_{24}|$ (dashed curve),
and $|M_{45}|/|M_{24}|$ (dotted curve) of the neutral Higgs bosons are plotted as functions of $\phi_1$.
The values of other parameters are the same as in Fig. 2.

\vskip 0.3 in
\noindent
Fig. 4h : The ratios of mass matrix elements $|M_{15}|/|M_{35}|$ (solid curve), $|M_{25}|/|M_{35}|$ (dashed curve),
and $|M_{45}|/|M_{35}|$ (dotted curve) of the neutral Higgs bosons are plotted as a function of $\phi_1$.
The values of other parameters are the same as in Fig. 2.

\vskip 0.3 in
\noindent
Fig. 5a : The squares of the elements $O_{13}^2$ (solid curve), $O_{23}^2$ (dashed curve), and $O_{34}^2$ (dotted curve)
of the orthogonal matrix defined in the text are plotted as functions of $\phi_1$.
The values of other parameters are the same as in Fig. 2.

\vskip 0.3 in
\noindent
Fig. 5b : The squares of the elements $O_{15}^2$ (solid curve), $O_{25}^2$ (dashed curve), and $O_{45}^2$ (dotted curve)
of the orthogonal matrix defined in the text are plotted as functions of $\phi_1$.
The values of other parameters are the same as in Fig. 2.

\vskip 0.3 in
\noindent
Fig. 6 : The plot of $\rho$ as a function of $\phi_1$.
The values of other parameters are the same as in Fig. 2.

\vskip 0.3 in
\noindent
Fig. 7 : We plot $\rho_{\rm max}$ as a function of $\phi_1$
for $\Lambda$ = 300 GeV.
The maximum value of  $\rho$ is obtained by randomly searching 20,000 points for given $\phi_1$
in the region of the parameter space whose boundaries are set as
$0 \le \phi_t = \phi_b = \phi_{\tau} = \phi_c < 2 \pi$,
$2 \le \tan \beta \le 40$,
$0 < \lambda, k \le 0.7$,
$0 < A_{\lambda}, A_k, x, M_2 \le 1000 $ GeV,
$0 < A_t = A_b = A_{\tau} \le 1000 $ GeV,
$0 < m_Q = m_L \le 1000 $ GeV,
and $0 < m_T = m_B = m_E \le 1000 $GeV.

\vskip 0.3 in
\noindent
Fig. 8 : The 20,000 points of $R_{h1}$, randomly evaluated in the same parameter region as Fig. 7,
are plotted against $C_3$, under the condition of $(m_{h_{(n+1)}} - m_{h_{(n)}}) > 10$ GeV ($n$ = 1 to 4).

\vfil\eject

\setcounter{figure}{0}
\def\figurename{}{}%
\renewcommand\thefigure{Fig. 1}
\begin{figure}[t]
\epsfxsize=12cm
\hspace*{2.cm}
\epsffile{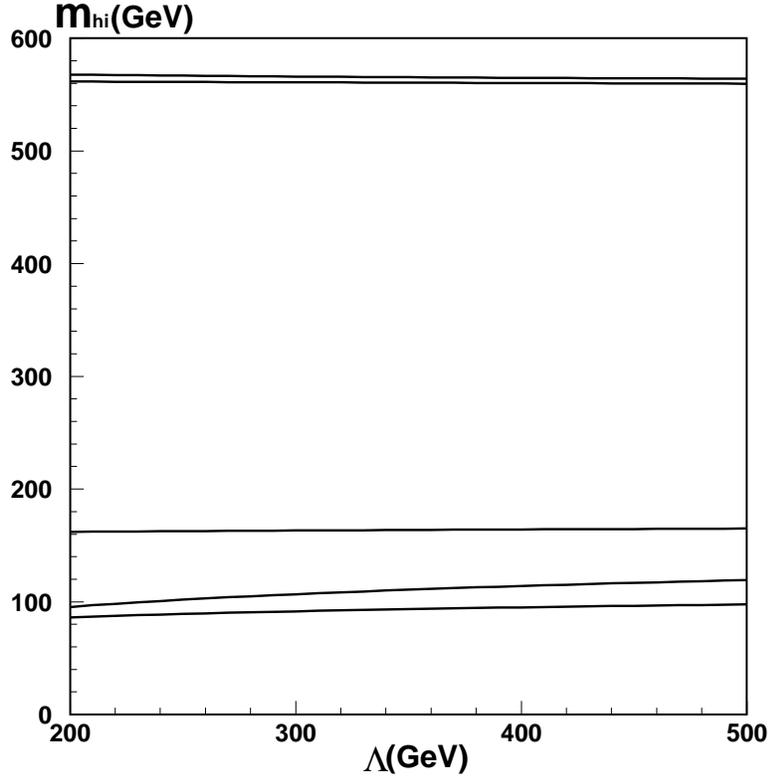}
\caption[plot]
{The plot of the neutral Higgs boson masses as functions of $\Lambda$ for $\phi_1 = 10\pi/19$.
We set $\tan \beta$ = 5, $\lambda$ = 0.5, $k$ = 0.4, $A_{\lambda}$ = 590 GeV, $A_k$ = 5 GeV, $x$ = 178 GeV,
$m_Q = m_L$ = 967 GeV, $m_T = m_B = m_E$ = 794 GeV,
$A_t = A_b = A_{\tau}$ = 907 GeV, $M_2$ = 357 GeV, $\phi_t = \phi_b = \phi_{\tau} = \phi_c = 10 \pi/19$.}
\end{figure}

\renewcommand\thefigure{Fig. 2}
\begin{figure}[t]
\epsfxsize=12cm
\hspace*{2.cm}
\epsffile{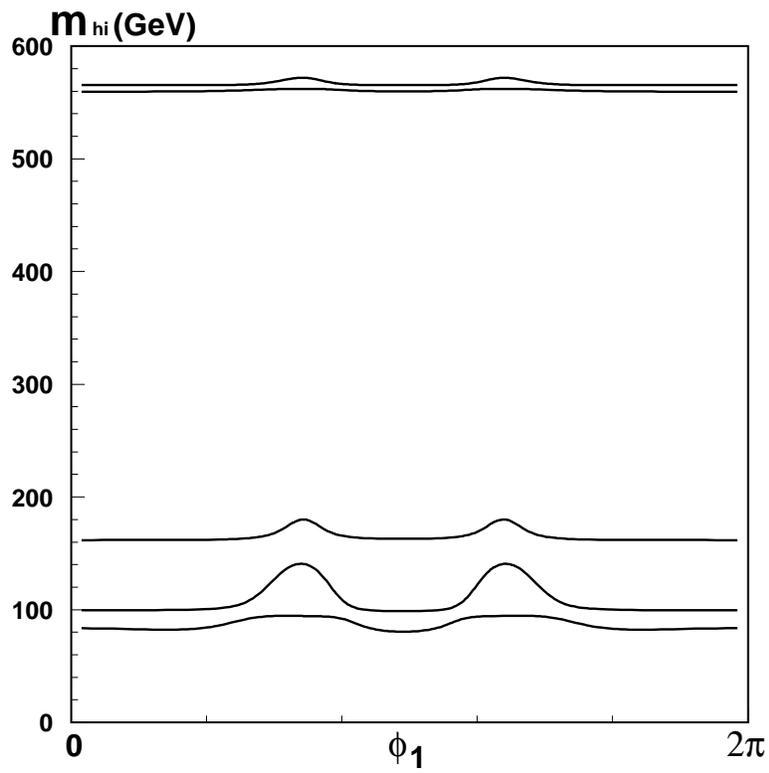}
\caption[plot]
{The plot of the neutral Higgs boson masses as functions of $\phi_1$ for $\Lambda = 300$ GeV.
The remaining parameters have the same values as in Fig. 1.}
\end{figure}

\renewcommand\thefigure{Fig. 3a}
\begin{figure}[t]
\epsfxsize=12cm
\hspace*{2.cm}
\epsffile{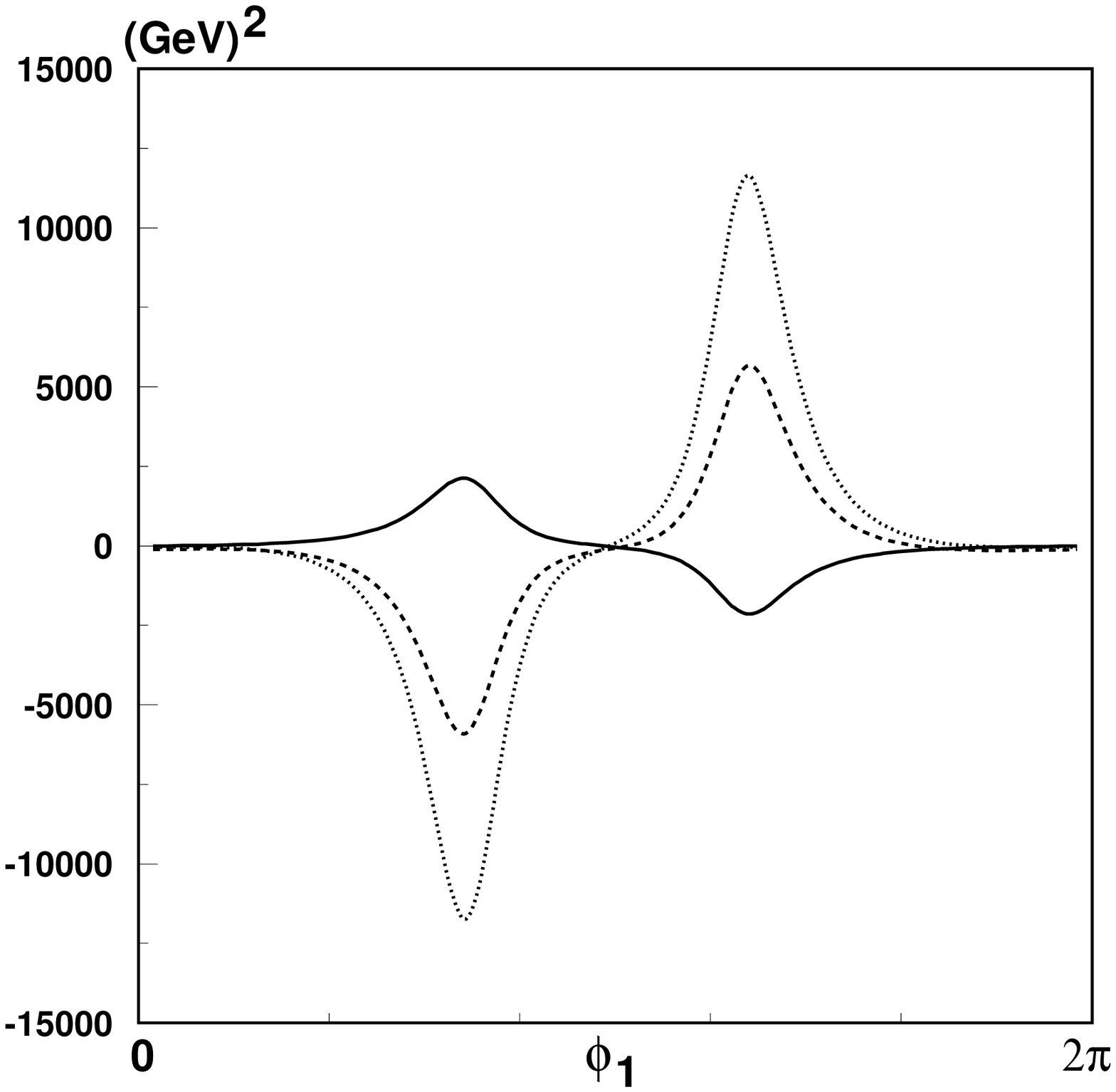}
\caption[plot]
{The plot of mass matrix elements $M_{13}$ (solid curve), $M_{23}$ (dashed curve), and $M_{34}$ (dotted curve)
of the neutral Higgs bosons as functions of $\phi_1$. The values of other parameters are the same as in Fig. 2.}
\end{figure}

\renewcommand\thefigure{Fig. 3b}
\begin{figure}[t]
\epsfxsize=12cm
\hspace*{2.cm}
\epsffile{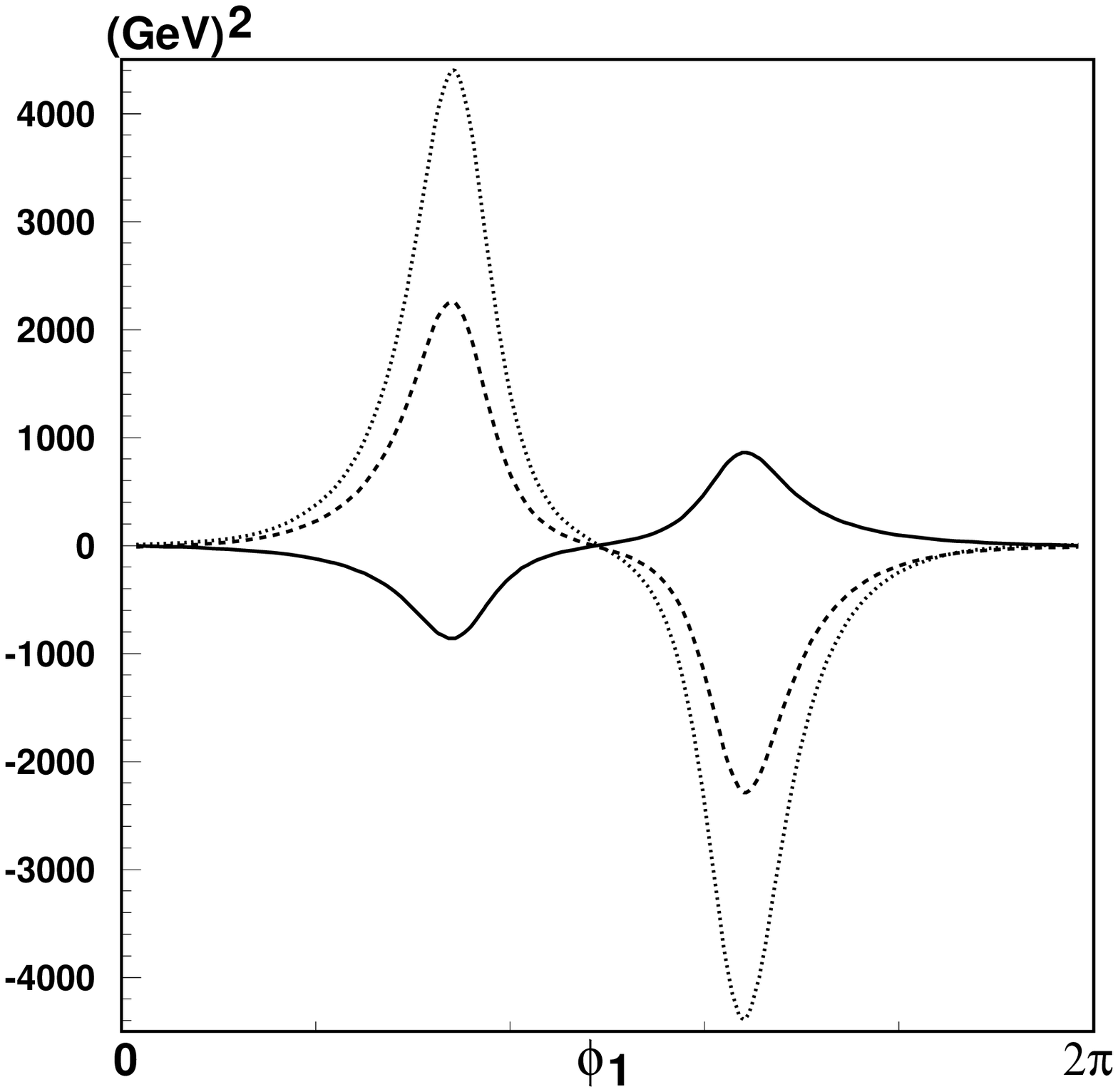}
\caption[plot]
{The plot of mass matrix elements $M_{15}$ (solid curve), $M_{25}$ (dashed curve), and $M_{45}$ (dotted curve)
of the neutral HIggs bosons as functions of $\phi_1$. The values of other parameters are the same as in Fig. 2.}
\end{figure}

\renewcommand\thefigure{Fig. 4a}
\begin{figure}[t]
\epsfxsize=12cm
\hspace*{2.cm}
\epsffile{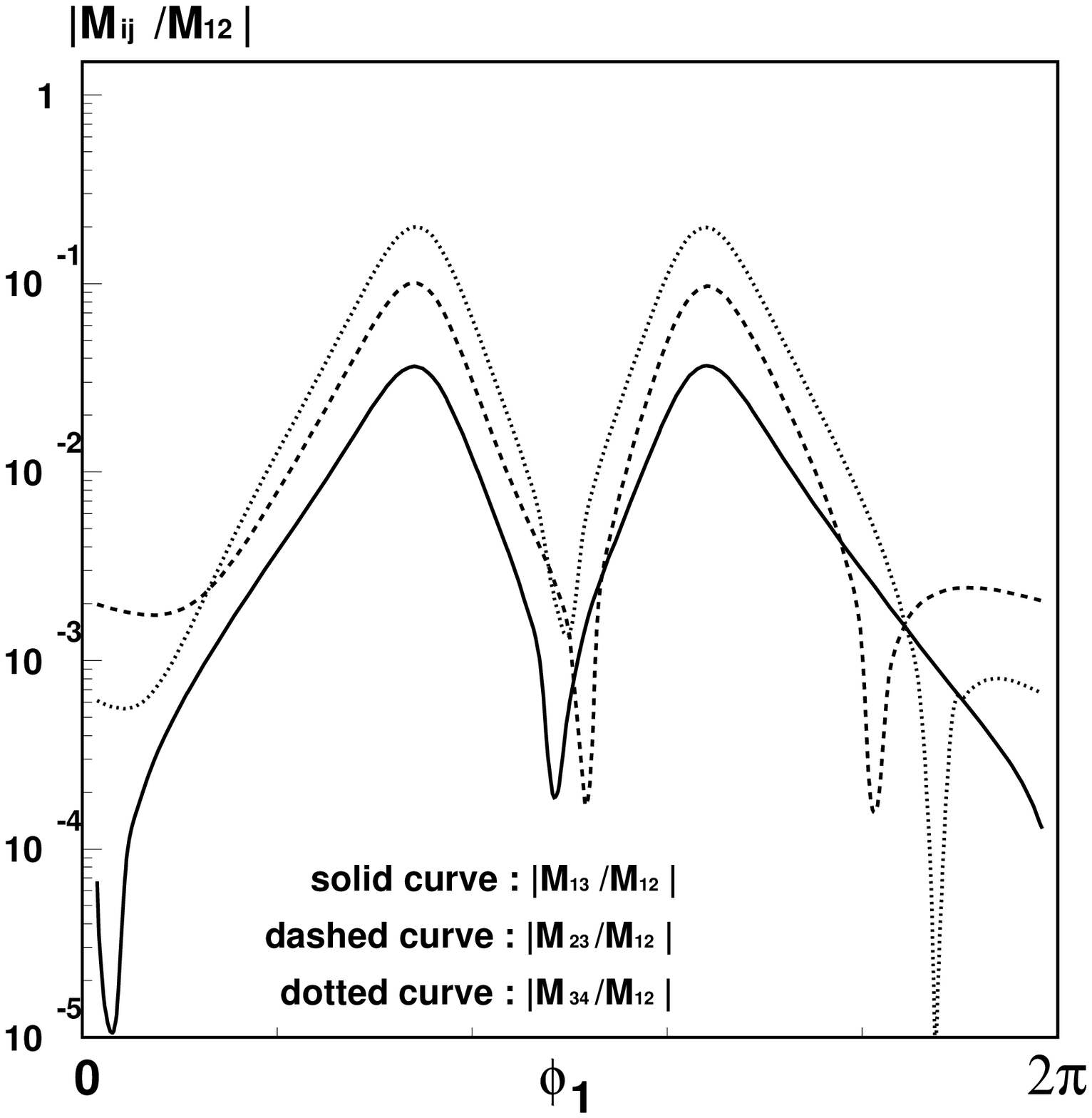}
\caption[plot]
{The ratios of mass matrix elements $|M_{13}|/|M_{12}|$ (solid curve), $|M_{23}|/|M_{12}|$ (dashed curve),
and $|M_{34}|/|M_{12}|$ (dotted curve) of the neutral Higgs bosons are plotted as functions of $\phi_1$.
The values of other parameters are the same as in Fig. 2.}
\end{figure}

\renewcommand\thefigure{Fig. 4b}
\begin{figure}[t]
\epsfxsize=12cm
\hspace*{2.cm}
\epsffile{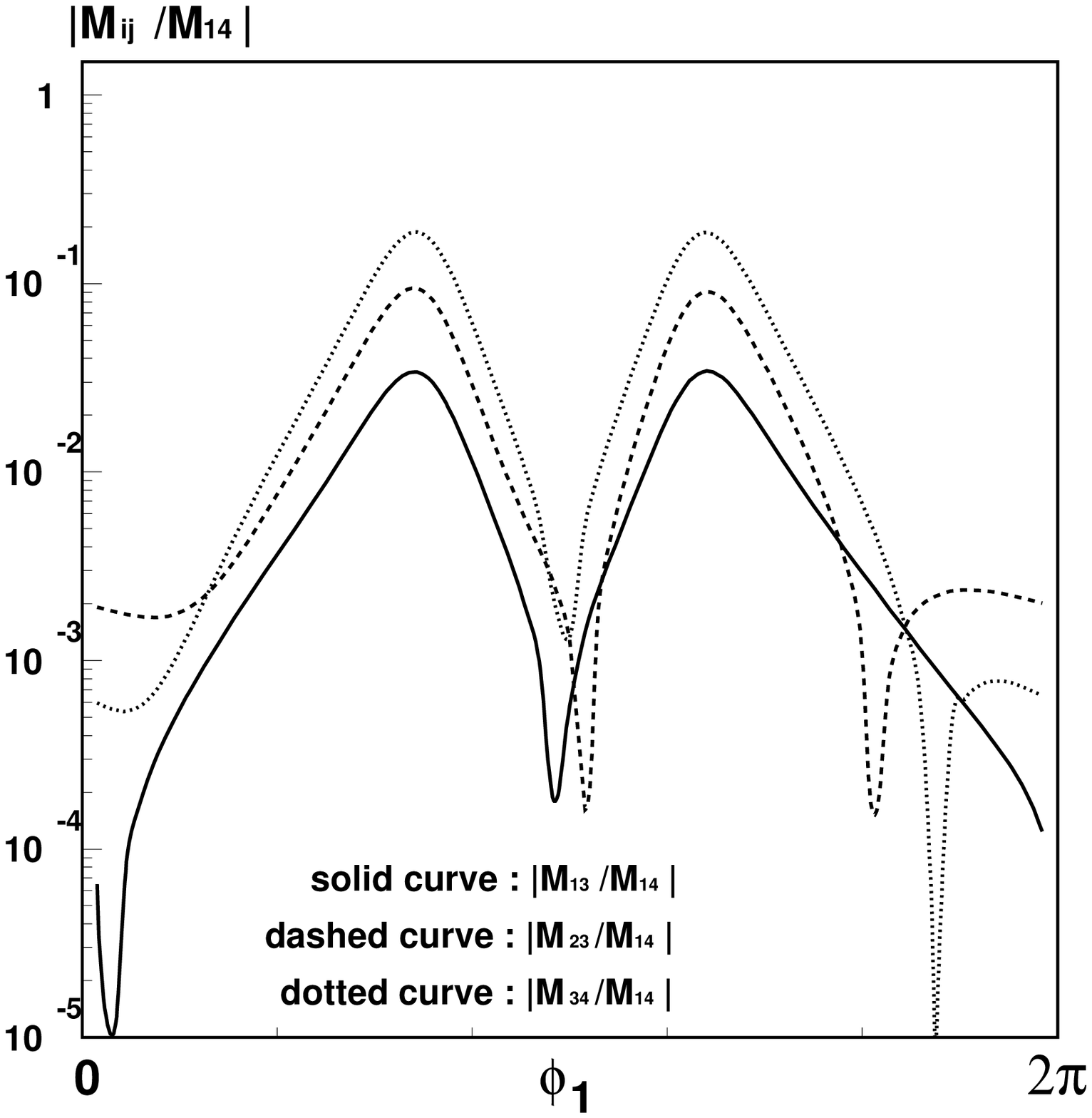}
\caption[plot]
{The ratios of mass matrix elements $|M_{13}|/|M_{14}|$ (solid curve), $|M_{23}|/|M_{14}|$ (dashed curve),
and $|M_{34}|/|M_{14}|$ (dotted curve) of the neutral Higgs bosons are plotted as functions of $\phi_1$.
The values of other parameters are the same as in Fig. 2.}
\end{figure}

\renewcommand\thefigure{Fig. 4c}
\begin{figure}[t]
\epsfxsize=12cm
\hspace*{2.cm}
\epsffile{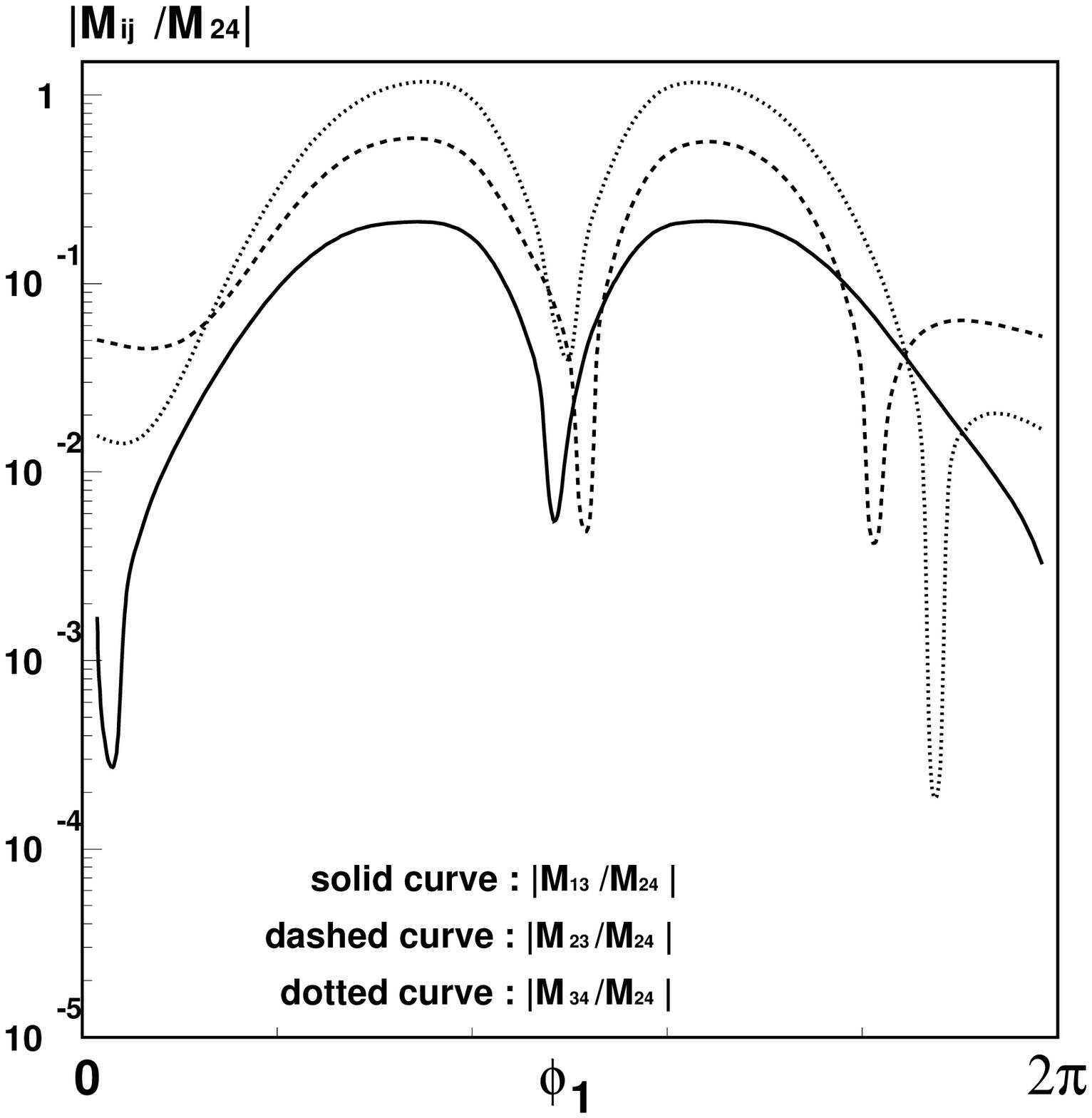}
\caption[plot]
{The ratios of mass matrix elements $|M_{13}|/|M_{24}|$ (solid curve), $|M_{23}|/|M_{24}|$ (dashed curve),
and $|M_{34}|/|M_{24}|$ (dotted curve) of the neutral Higgs bosons are plotted as functions of $\phi_1$.
The values of other parameters are the same as in Fig. 2.}
\end{figure}

\renewcommand\thefigure{Fig. 4d}
\begin{figure}[t]
\epsfxsize=12cm
\hspace*{2.cm}
\epsffile{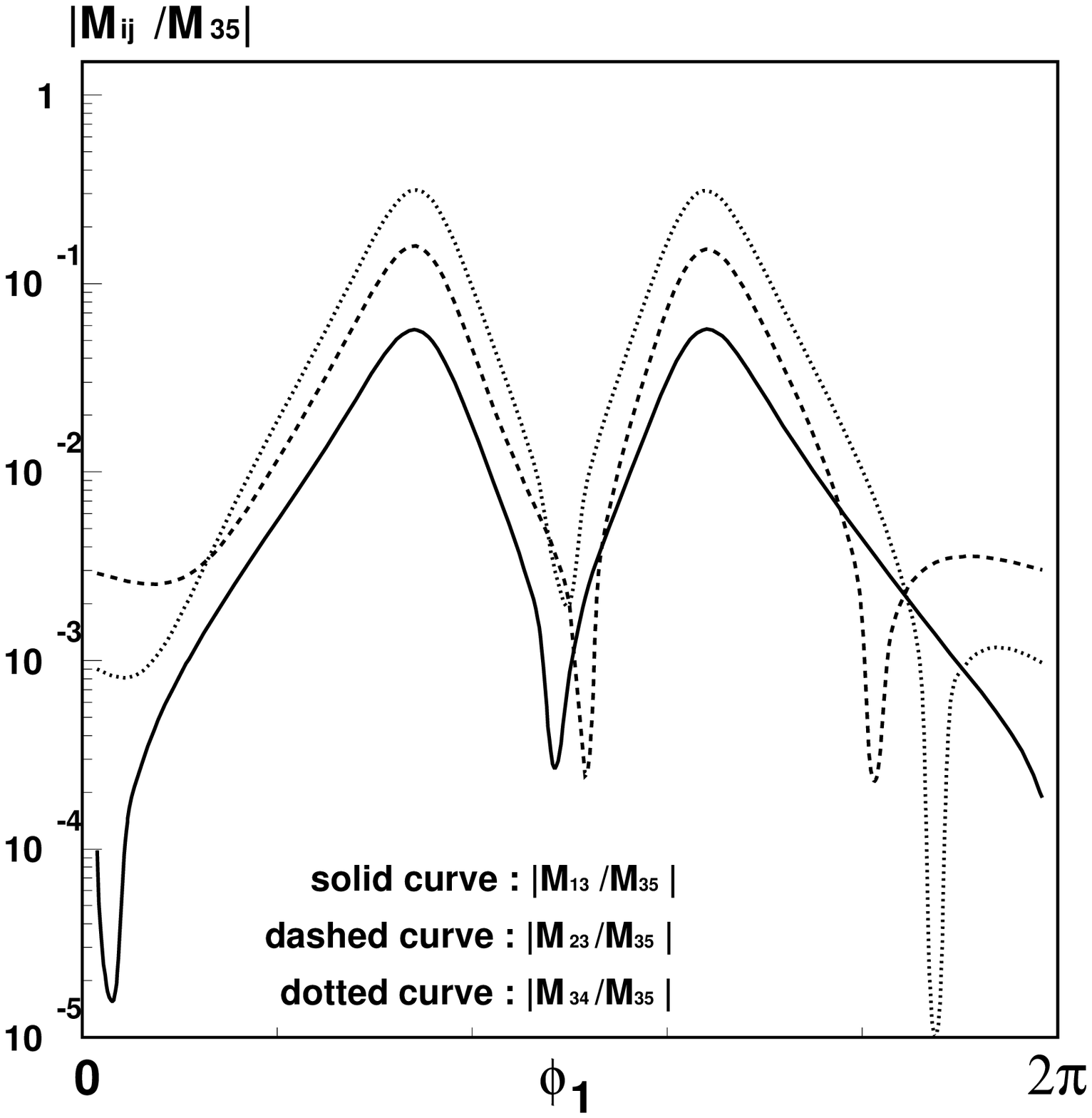}
\caption[plot]
{The ratios of mass matrix elements $|M_{13}|/|M_{35}|$ (solid curve), $|M_{23}|/|M_{35}|$ (dashed curve),
and $|M_{34}|/|M_{35}|$ (dotted curve) of the neutral Higgs bosons are plotted as functions of $\phi_1$.
The values of other parameters are the same as in Fig. 2.}
\end{figure}

\renewcommand\thefigure{Fig. 4e}
\begin{figure}[t]
\epsfxsize=12cm
\hspace*{2.cm}
\epsffile{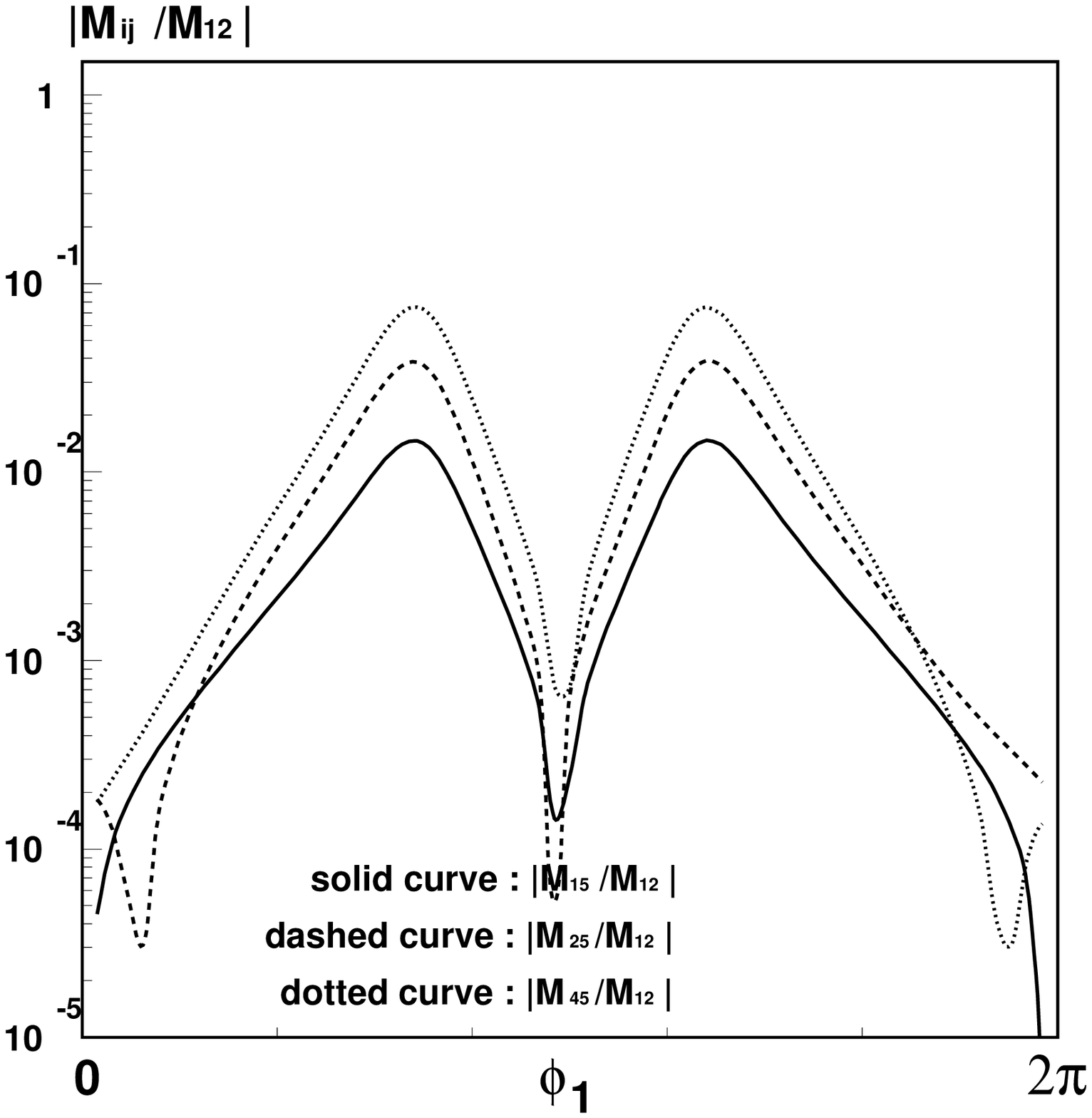}
\caption[plot]
{The ratios of mass matrix elements $|M_{15}|/|M_{12}|$ (solid curve), $|M_{25}|/|M_{12}|$ (dashed curve),
and $|M_{45}|/|M_{12}|$ (dotted curve) of the neutral Higgs bosons are plotted as functions of $\phi_1$.
The values of other parameters are the same as in Fig. 2.}
\end{figure}

\renewcommand\thefigure{Fig. 4f}
\begin{figure}[t]
\epsfxsize=12cm
\hspace*{2.cm}
\epsffile{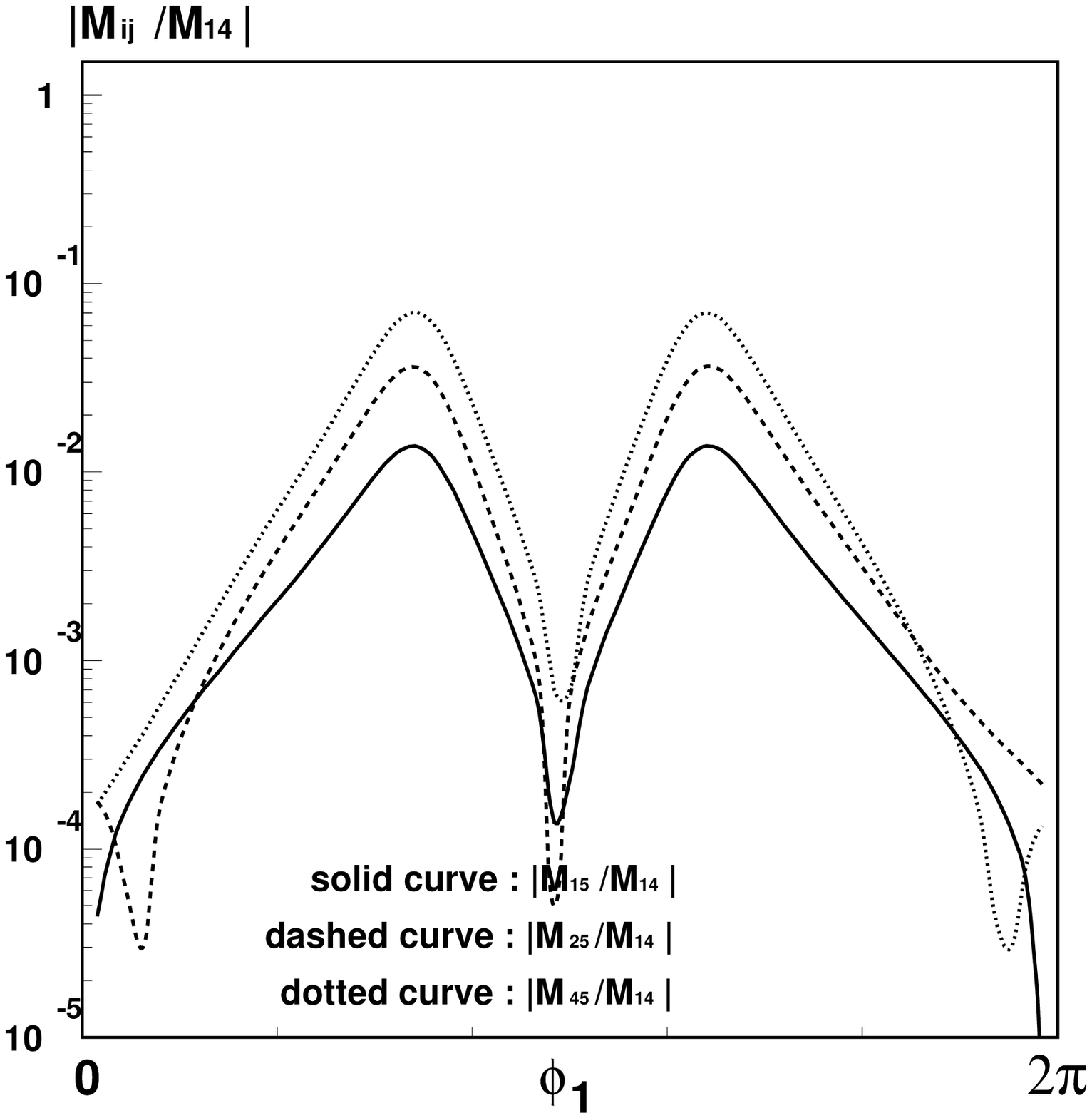}
\caption[plot]
{The ratios of mass matrix elements $|M_{15}|/|M_{14}|$ (solid curve), $|M_{25}|/|M_{14}|$ (dashed curve),
and $|M_{45}|/|M_{14}|$ (dotted curve) of the neutral Higgs bosons are plotted as functions of $\phi_1$.
The values of other parameters are the same as in Fig. 2.}
\end{figure}

\renewcommand\thefigure{Fig. 4g}
\begin{figure}[t]
\epsfxsize=12cm
\hspace*{2.cm}
\epsffile{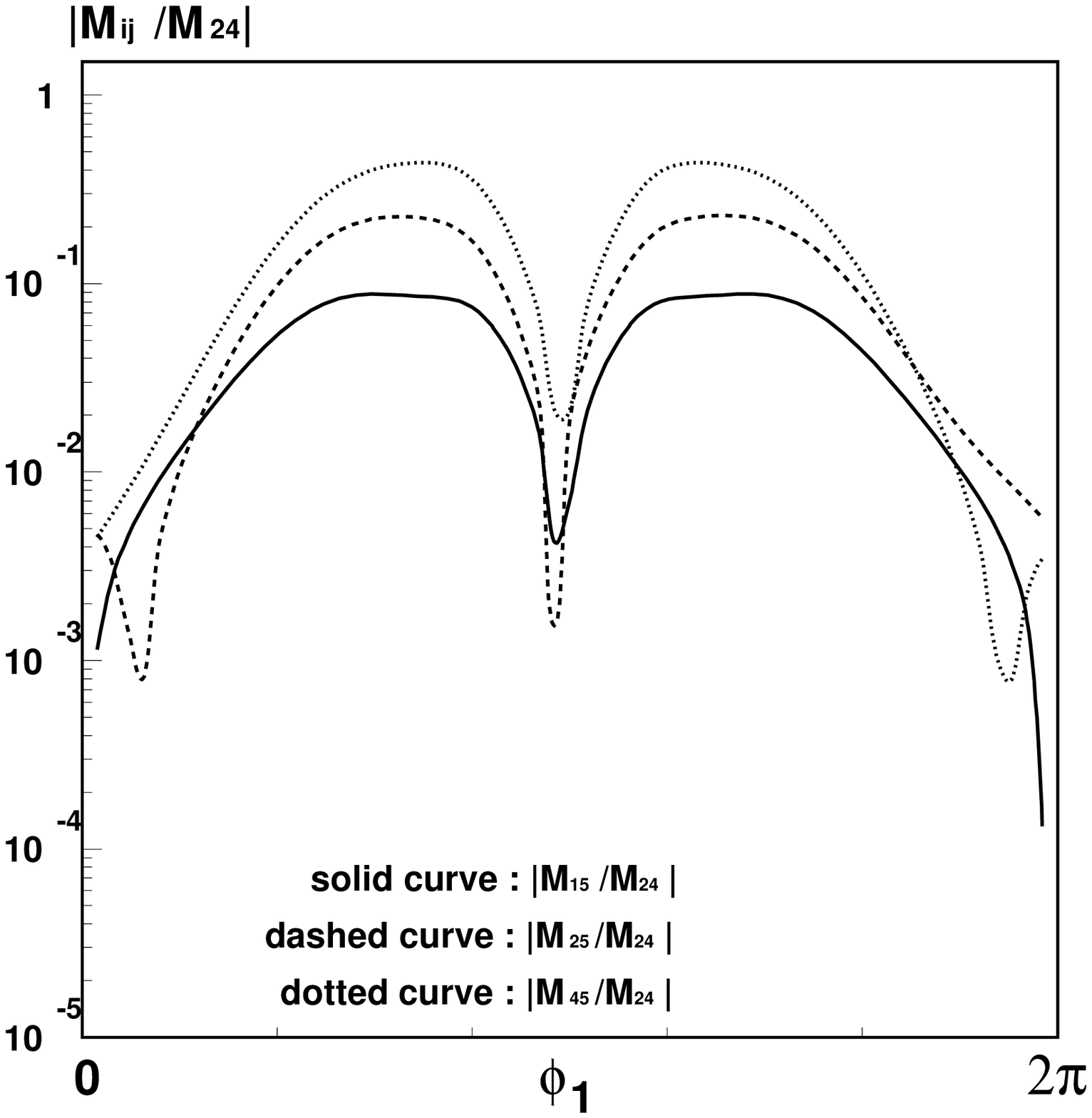}
\caption[plot]
{The ratios of mass matrix elements $|M_{15}|/|M_{24}|$ (solid curve), $|M_{25}|/|M_{24}|$ (dashed curve),
and $|M_{45}|/|M_{24}|$ (dotted curve) of the neutral Higgs bosons are plotted as functions of $\phi_1$.
The values of other parameters are the same as in Fig. 2.}
\end{figure}

\renewcommand\thefigure{Fig. 4h}
\begin{figure}[t]
\epsfxsize=12cm
\hspace*{2.cm}
\epsffile{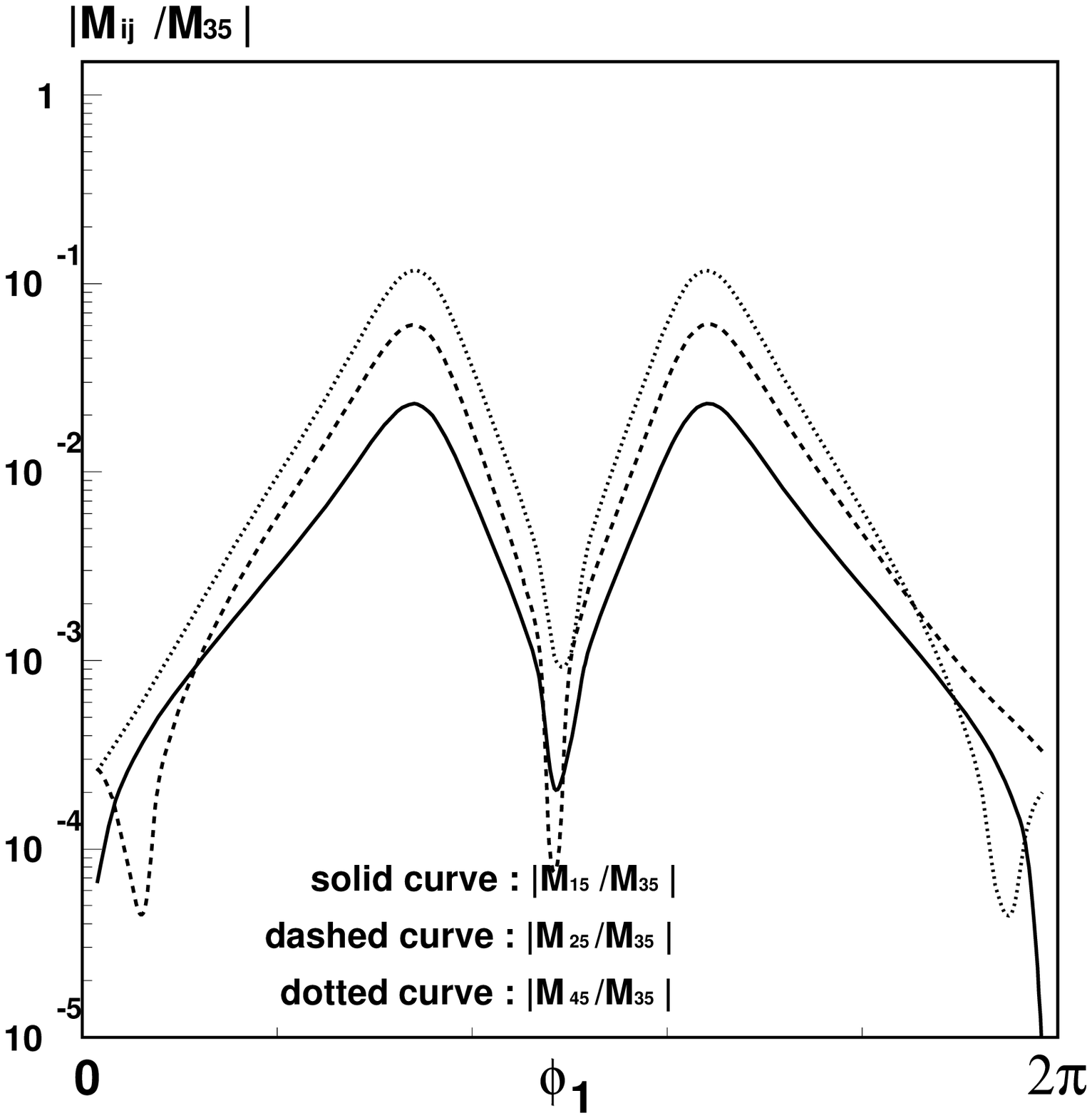}
\caption[plot]
{The ratios of mass matrix elements $|M_{15}|/|M_{35}|$ (solid curve), $|M_{25}|/|M_{35}|$ (dashed curve),
and $|M_{45}|/|M_{35}|$ (dotted curve) of the neutral Higgs bosons are plotted as a function of $\phi_1$.
The values of other parameters are the same as in Fig. 2.}
\end{figure}

\renewcommand\thefigure{Fig. 5a}
\begin{figure}[t]
\epsfxsize=12cm
\hspace*{2.cm}
\epsffile{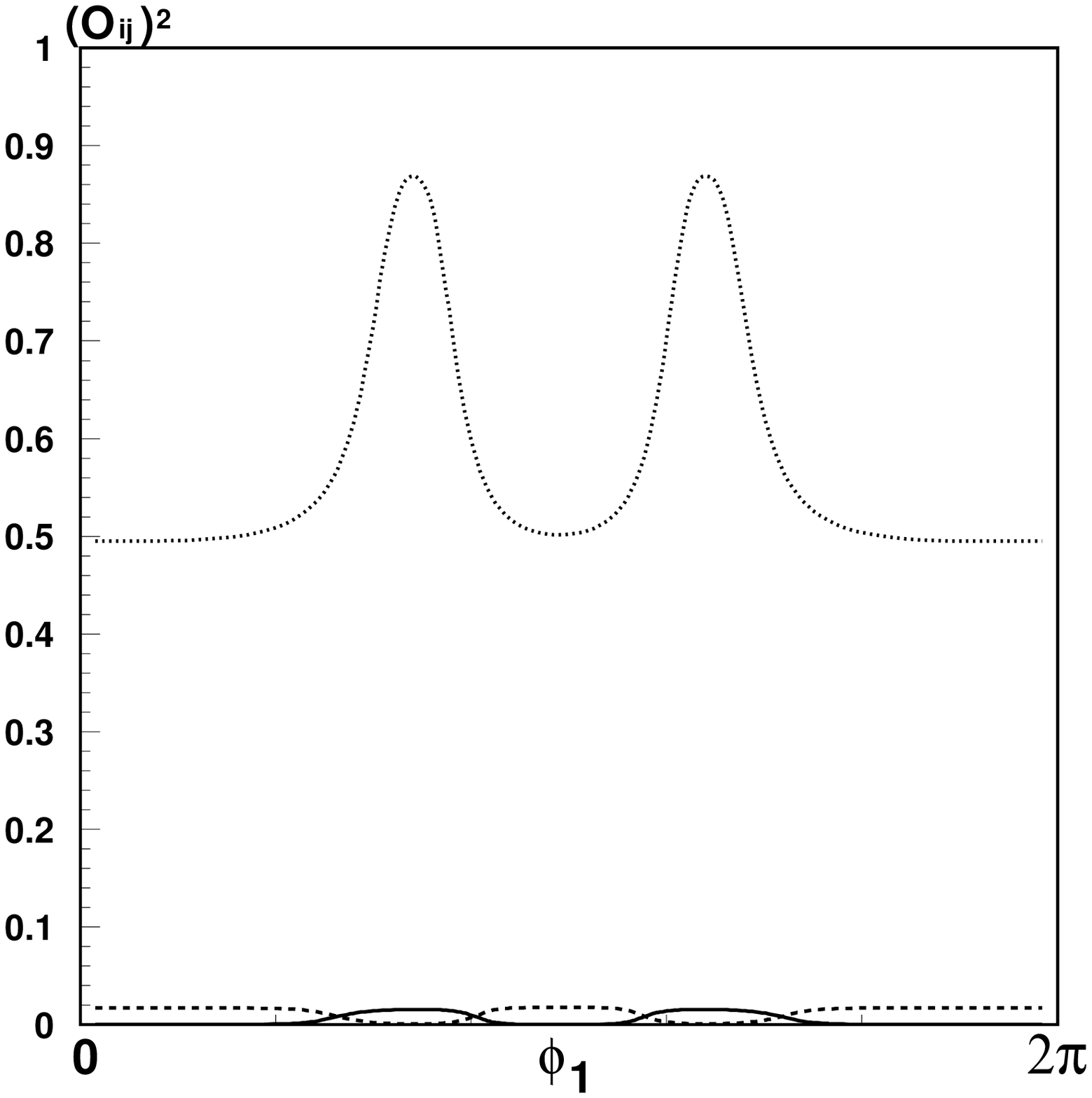}
\caption[plot]
{The squares of the elements $O_{13}^2$ (solid curve), $O_{23}^2$ (dashed curve), and $O_{34}^2$ (dotted curve)
of the orthogonal matrix defined in the text are plotted as functions of $\phi_1$.
The values of other parameters are the same as in Fig. 2.}
\end{figure}

\renewcommand\thefigure{Fig. 5b}
\begin{figure}[t]
\epsfxsize=12cm
\hspace*{2.cm}
\epsffile{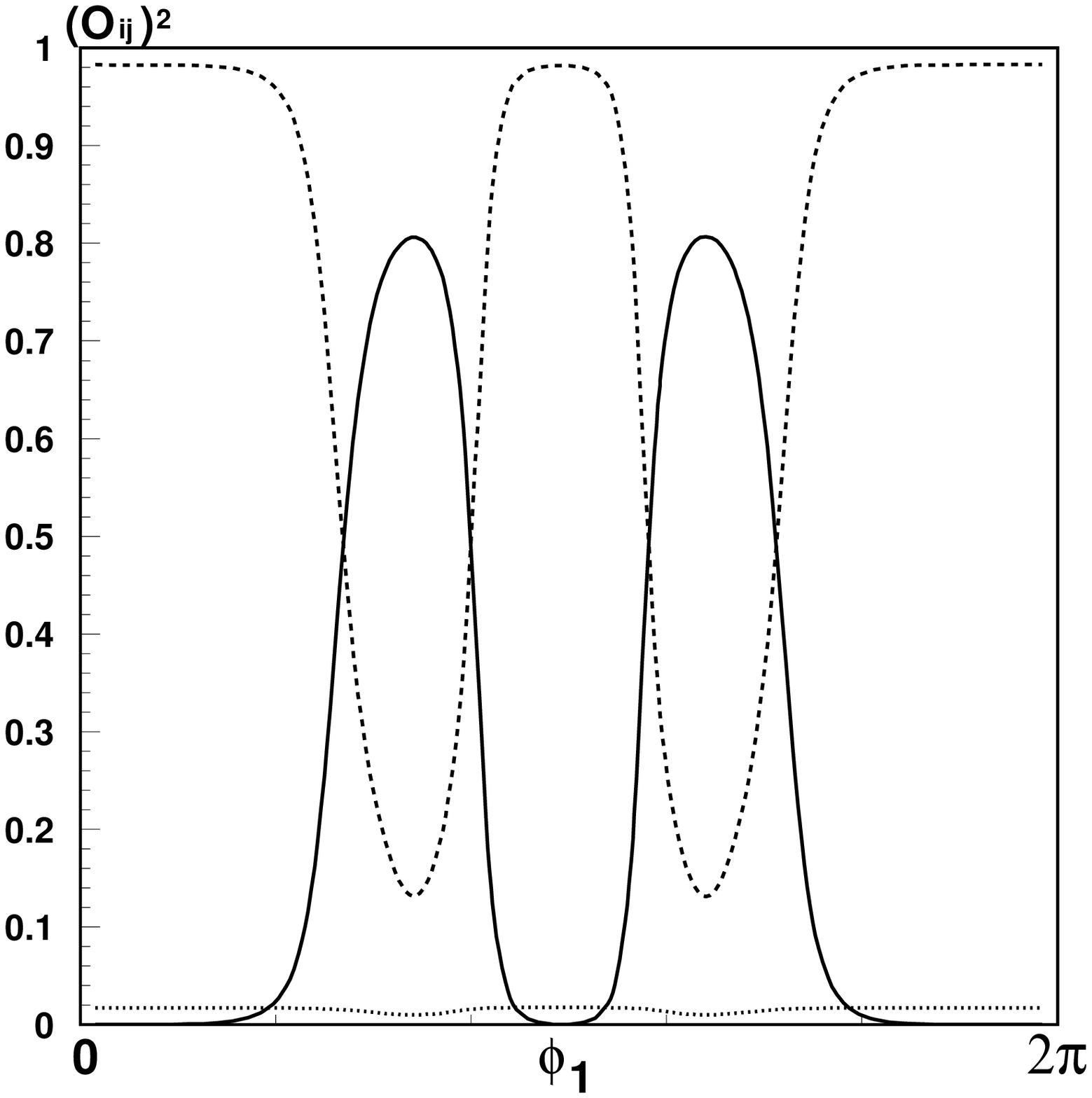}
\caption[plot]
{The squares of the elements $O_{15}^2$ (solid curve), $O_{25}^2$ (dashed curve), and $O_{45}^2$ (dotted curve)
of the orthogonal matrix defined in the text are plotted as functions of $\phi_1$.
The values of other parameters are the same as in Fig. 2.}
\end{figure}

\renewcommand\thefigure{Fig. 6}
\begin{figure}[t]
\epsfxsize=12cm
\hspace*{2.cm}
\epsffile{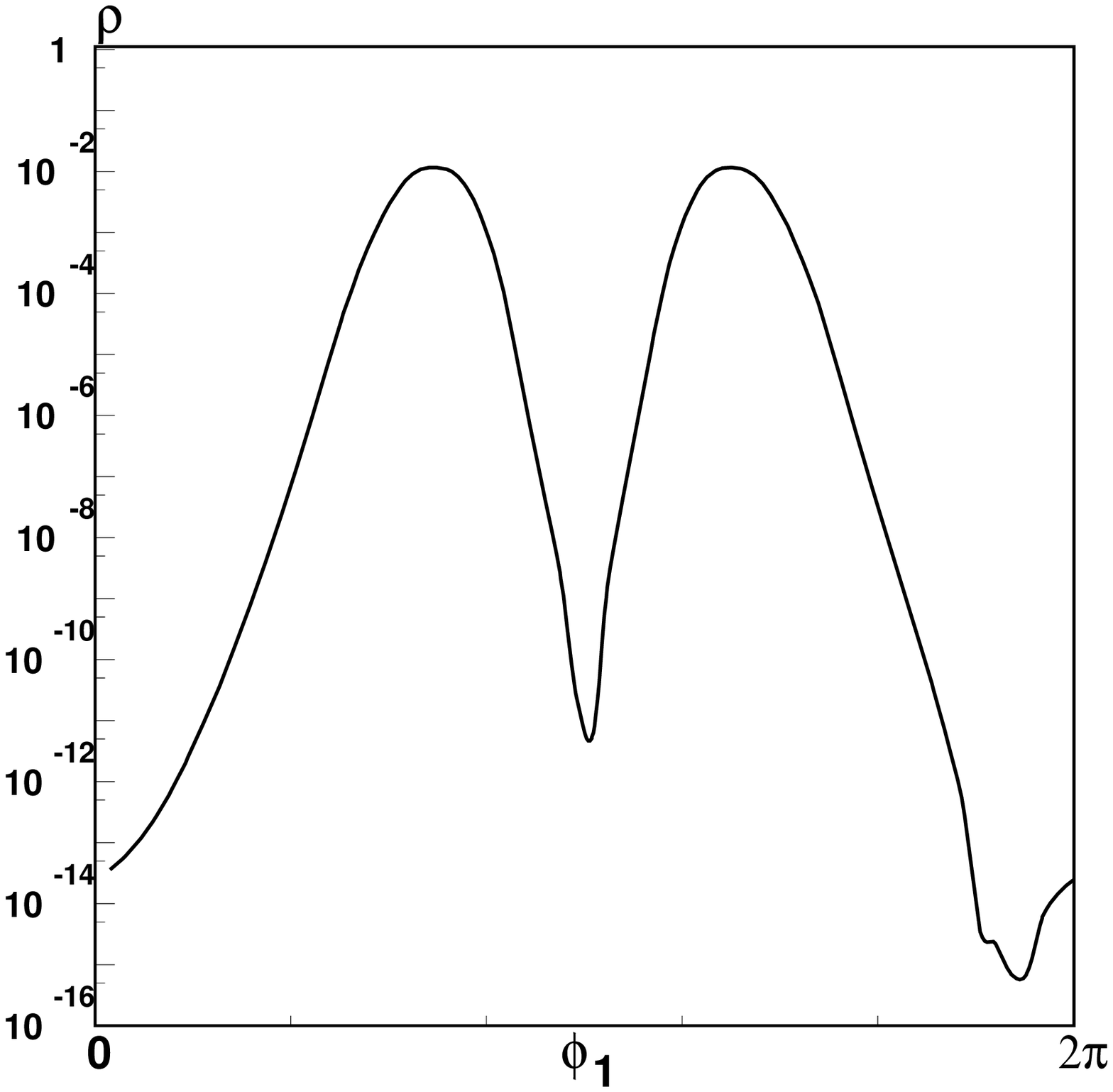}
\caption[plot]
{The plot of $\rho$ as a function of $\phi_1$.
The values of other parameters are the same as in Fig. 2.}
\end{figure}

\renewcommand\thefigure{Fig. 7}
\begin{figure}[t]
\epsfxsize=12cm
\hspace*{2.cm}
\epsffile{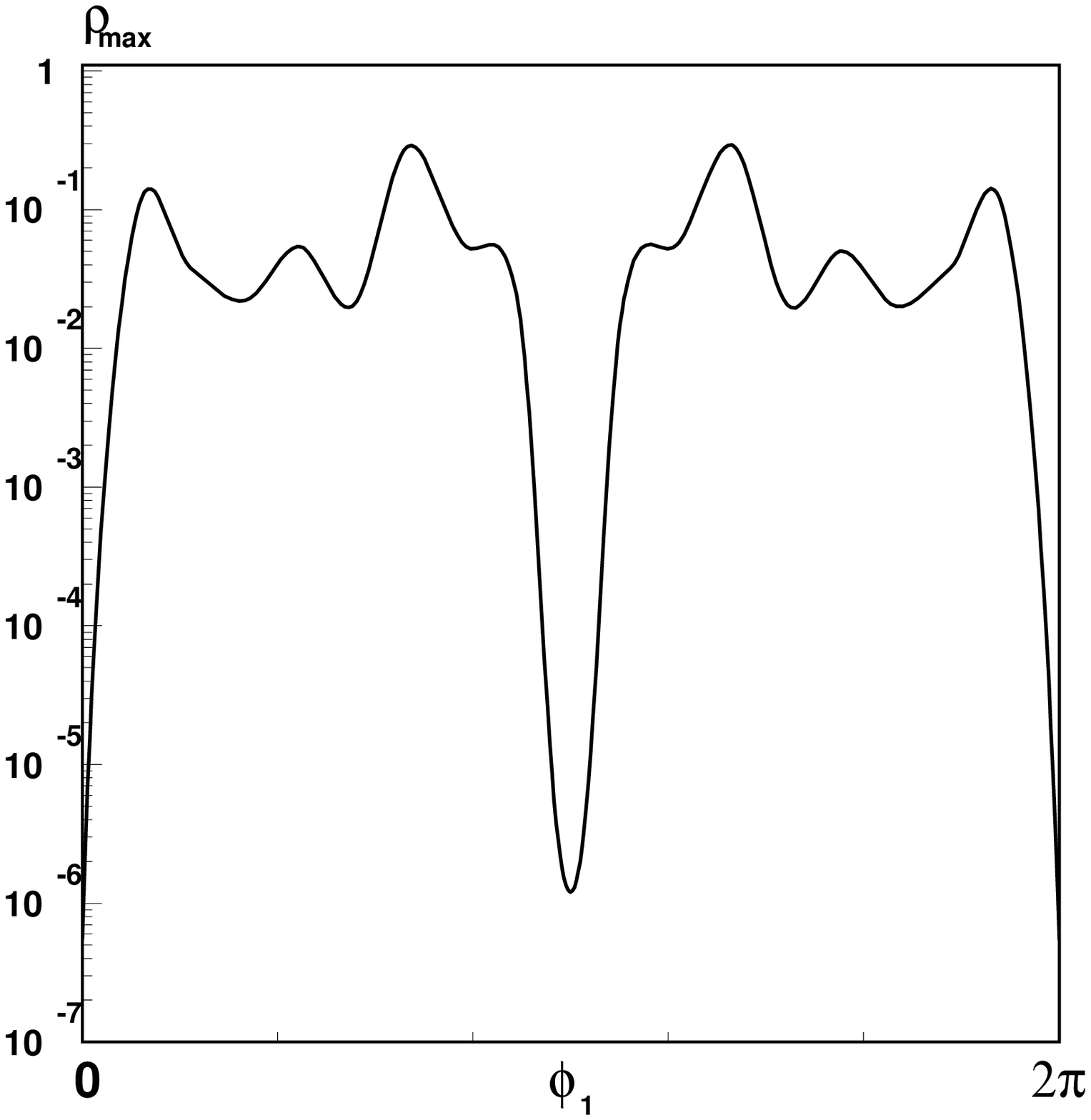}
\caption[plot]
{We plot $\rho_{\rm max}$ as a function of $\phi_1$
for $\Lambda$ = 300 GeV.
The maximum value of  $\rho$ is obtained by randomly searching 20,000 points for given $\phi_1$
in the region of the parameter space whose boundaries are set as
$0 < \phi_t = \phi_b = \phi_{\tau} = \phi_c < 2 \pi$,
$2 \le \tan \beta \le 40$,
$0 < \lambda, k \le 0.7$,
$0 < A_{\lambda}, A_k, x, M_2 \le 1000 $ GeV,
$0 < A_t = A_b = A_{\tau} \le 1000 $ GeV,
$0 < m_Q = m_L \le 1000 $ GeV,
and $0 < m_T = m_B = m_E \le 1000 $GeV.}
\end{figure}

\renewcommand\thefigure{Fig. 8}
\begin{figure}[t]
\epsfxsize=12cm
\hspace*{2.cm}
\epsffile{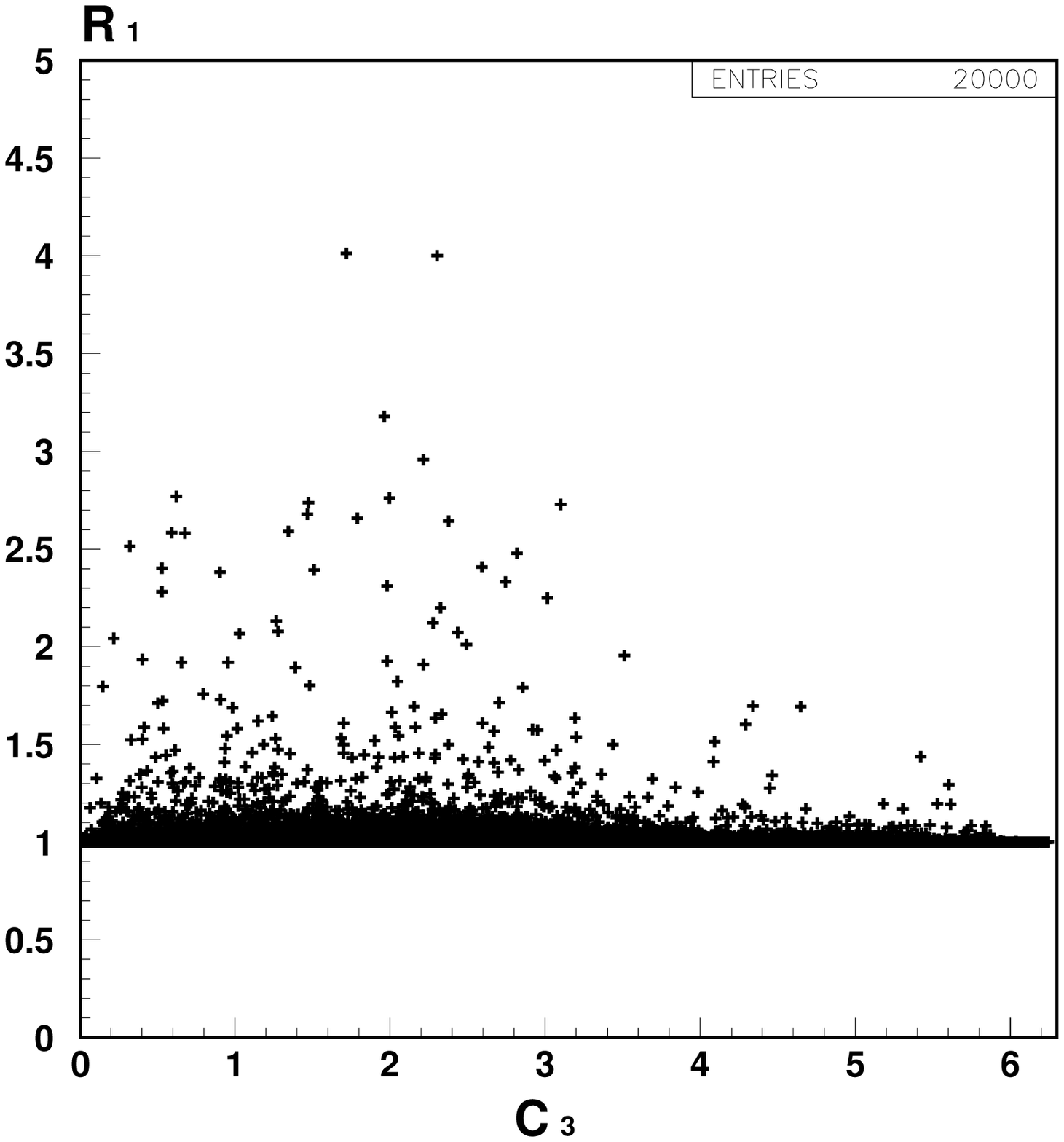}
\caption[plot]
{The 20,000 points of $R_{h1}$, randomly evaluated in the same parameter region as Fig. 7,
are plotted against $C_3$, under the condition of $(m_{h_{(n+1)}} - m_{h_{(n)}}) > 10$ GeV ($n$ = 1 to 4).}
\end{figure}

\end{document}